\documentclass[manuscript]{acmart}
\usepackage{amsmath}
\usepackage{amsfonts}

\usepackage{braket}
\usepackage{fancyhdr}
\usepackage{booktabs}

\usepackage[T1]{fontenc}
\usepackage[utf8]{inputenc}
\usepackage{regexpatch}
\usepackage{letltxmacro}
\usepackage{aligned-overset}

\newcommand{\code}[1]{{\small\texttt{#1}}}

\usepackage{graphicx} % Required for inserting images
\usepackage{outlines}

\usepackage{alltt}
\usepackage{listings,lstcoq, lstegglog}

\usepackage{hyperref}
\usepackage[noabbrev,nameinlink,capitalize]{cleveref} % Import after listings!

\usepackage{enumitem}

\usepackage{xspace}

\usepackage{nowidow}
\usepackage{microtype}

\usepackage{wrapfig}
\usepackage{float}
\usepackage{caption}
\usepackage{subcaption}

\usepackage{threeparttable}

\usepackage{mathpartir}

\newcommand{\revemptyset}{\ensuremath{\text{\reflectbox{$\emptyset$}}}}
\newcommand{\emdash}{\---}
\usepackage{xcolor}
\definecolor{ltblue}{rgb}{0,0.4,0.4}
\definecolor{dkblue}{rgb}{0,0.1,0.6}
\definecolor{dkgreen}{rgb}{0,0.35,0}
\definecolor{dkviolet}{rgb}{0.3,0,0.5}
\definecolor{dkred}{rgb}{0.5,0,0}
\definecolor{dkorange}{HTML}{e87954}
\definecolor{dkyellow}{HTML}{f7cd7a}
\definecolor{dkpink}{HTML}{d152cf}

\newcommand{\zx}[2]{\text{\coqe{ZX #1 #2}}}
\newcommand{\zxold}[2]{\texttt{ZX}\,\,\text{#1}\,\,\text{#2}}
\newcommand{\semantics}[1]{\ensuremath{\llbracket\texttt{#1}\rrbracket}}
\newcommand{\sem}[1]{\llbracket #1 \rrbracket} % for math mode

\newcommand{\sdold}[2]{\texttt{SD}\,\,\text{#1}\,\,\text{#2}}

\newcommand{\oftype}[2]{\text{#1}\,:\,\text{#2}}

\newcommand{\vyzx}{\textsl{Vy}\textsc{ZX}\xspace}
\newcommand{\VyZX}{\textbf{V}erif\textbf{y} the \textbf{ZX}-calculus} % At least give this a descriptive name, dammit!
\newcommand{\vizx}{ZXV\textsc{iz}\xspace}
\newcommand{\sqir}{\textsc{sqir}\xspace}

\newcommand{\qlib}{\texttt{QuantumLib}\xspace}
\newcommand{\pyzx}{PyZX\xspace}

\newcommand{\Coq}{Rocq\xspace}
\newcommand{\vscode}{VSCode\xspace}
\newcommand{\zxcalc}{ZX-calculus\xspace}
\newcommand{\zxdiag}{ZX-diagram\xspace}
\newcommand{\zxdiags}{\zxdiag{}s\xspace}

\newcommand{\chyp}{Chyp\xspace}
\newcommand{\qwire}{\ensuremath{\mathcal{Q}\textsc{wire}}\xspace}
\newcommand{\qbricks}{\ensuremath{\mathcal{Q}\textsc{bricks}}\xspace}

\newcommand{\voqc}{\textsc{voqc}\xspace}
\newcommand{\egraph}{E-Graph\xspace}

\newcommand{\choijam}{Choi-Jamiołkowski\xspace}

\newcommand{\R}{\mathbb{R}}
\newcommand{\C}{\mathbb{C}}
\newcommand{\N}{\mathbb{N}}

\usepackage{etoolbox} % replaces ifthen package
\newtoggle{comments}
\togglefalse{comments}

\iftoggle{comments}{
  \newcommand{\fixme}[1]{\textbf{\textcolor{red}{[ Fixme: #1]}}}
  \newcommand{\todo}[1]{\textbf{\textcolor{green}{[ TODO: #1 ]}}}
  \newcommand{\rnr}[1]{\textbf{\textcolor{blue}{[ Robert: #1 ]}}}
  \newcommand{\adrian}[1]{\textbf{\textcolor{olive}{[ Adrian: #1 ]}}}
  \newcommand{\ael}[1]{\adrian{#1}}
  \newcommand{\ben}[1]{\textbf{\textcolor{orange}{[ Ben: #1 ]}}}
  \newcommand{\bhakti}[1]{\textbf{\textcolor{purple}{[ Bhakti: #1 ]}}}
  \newcommand{\wjbs}[1]{\textbf{\textcolor{purple}{[ William: #1 ]}}}
}{
  \newcommand{\fixme}[1]{}
  \newcommand{\todo}[1]{}
  \newcommand{\rnr}[1]{}
  \newcommand{\adrian}[1]{}  
  \newcommand{\ael}[1]{}  
  \newcommand{\ben}[1]{}
  \newcommand{\bhakti}[1]{}
  \newcommand{\wjbs}[1]{}
}

\usepackage{tikz}

\usetikzlibrary{backgrounds}
\usetikzlibrary{arrows}
\usetikzlibrary{shapes,shapes.geometric,shapes.misc}
\usetikzlibrary{decorations.pathmorphing}
\usetikzlibrary{decorations.pathreplacing}
\usetikzlibrary{decorations.markings}

\tikzstyle{every picture}=[baseline=-0.25em,scale=0.5]

\pgfkeys{/tikz/tikzit fill/.initial=0}
\pgfkeys{/tikz/tikzit draw/.initial=0}
\pgfkeys{/tikz/tikzit shape/.initial=0}
\pgfkeys{/tikz/tikzit category/.initial=0}

\newcommand{\tikzfig}[1]{%
\IfFileExists{#1.tikz}
  {\input{#1.tikz}}
  {%
    \IfFileExists{./figures/#1.tikz}
      {\input{./figures/#1.tikz}}
      {\tikz[baseline=-0.5em]{\node[draw=red,font=\color{red},fill=red!10!white] {\textit{#1}};}}%
  }%
}
\newcommand{\ctikzfig}[1]{%
\begin{center}\rm
  \input{#1.tikz}
\end{center}}

\newcommand{\hmaxtikzfig}[1]{\centering\resizebox{\textwidth}{!}{\input{#1.tikz}}}

\pgfdeclarelayer{edgelayer}
\pgfdeclarelayer{nodelayer}
\pgfsetlayers{background,edgelayer,nodelayer,main}
\tikzstyle{none}=[inner sep=0mm]
\tikzstyle{every loop}=[]
\tikzstyle{mark coordinate}=[inner sep=0pt,outer sep=0pt,minimum size=3pt,fill=black,circle]
\input{zh.tikzdefs}
\tikzstyle{dot}=[inner sep=0.3mm, minimum width=2mm, minimum height=2mm, draw, shape=circle, font={\footnotesize}, tikzit fill=magenta, fill=white]
\tikzstyle{white dot}=[dot, fill={rgb,255: red,232; green,165; blue,165}, text depth=-0.2mm, tikzit category=ZH-pf, draw=black]
\tikzstyle{white phase dot}=[minimum size=5mm, font={\footnotesize\boldmath}, shape=rectangle, rounded corners=2mm, inner sep=0.2mm, outer sep=-2mm, scale=0.8, tikzit shape=circle, draw=black, fill={rgb,255: red,232; green,165; blue,165}, tikzit category=ZH-pf, tikzit draw=blue]
\tikzstyle{gray dot}=[dot, fill={rgb,255: red,216; green,248; blue,216}, text depth=-0.2mm, tikzit category=ZH-pf]
\tikzstyle{gray phase dot}=[white phase dot, tikzit shape=circle, tikzit draw=blue, fill={rgb,255: red,216; green,248; blue,216}, font={\footnotesize\boldmath}]
\tikzstyle{hadamard}=[fill=white, draw, inner sep=0.6mm, minimum height=1.5mm, minimum width=1.5mm, shape=rectangle, tikzit shape=rectangle, tikzit category=ZH-pf]
\tikzstyle{small hadamard}=[hadamard]
\tikzstyle{lambda}=[hadamard, fill={rgb,255: red,180; green,180; blue,180}, tikzit shape=rectangle]
\tikzstyle{halfscalar}=[star, fill=black, draw=black, minimum size=8pt, inner sep=0pt]
\tikzstyle{box}=[shape=rectangle, text height=1.5ex, text depth=0.25ex, fill=white, draw=black, minimum height=3mm, minimum width=5mm, font={\small}]
\tikzstyle{Z dot}=[inner sep=0mm, minimum size=2mm, shape=circle, draw=black, fill={zx_green}, tikzit fill=green]
\tikzstyle{Z phase dot}=[minimum size=5mm, font={\footnotesize\boldmath}, shape=rectangle, rounded corners=2mm, inner sep=0.2mm, outer sep=-2mm, scale=0.8, tikzit shape=circle, draw=black, fill={zx_green}, tikzit draw=blue, tikzit fill=green]
\tikzstyle{X dot}=[Z dot, shape=circle, draw=black, fill={zx_red}, tikzit fill=red]
\tikzstyle{X phase dot}=[Z phase dot, tikzit shape=circle, tikzit draw=blue, fill={zx_red}, font={\footnotesize\color{black}\boldmath}, tikzit fill=red]
\tikzstyle{H box}=[hadamard]
\tikzstyle{st}=[star, star points=5, fill=white, draw=black, inner sep=1.2pt, line width=1.2pt, tikzit fill=blue, tikzit draw=red, tikzit category=ZH-pf]
\tikzstyle{triangle}=[regular polygon, regular polygon sides=3, fill=white, draw=black, inner sep=0pt, minimum width=1em, tikzit draw=blue, tikzit category=ZH-pf, tikzit fill=cyan]
\tikzstyle{not}=[fill={rgb,255: red,180; green,180; blue,180}, draw=black, shape=circle, font={$\neg$}, dot]
\tikzstyle{vertex}=[inner sep=0mm, minimum size=1mm, shape=circle, draw=black, fill=black]
\tikzstyle{vertex set}=[inner sep=0mm, minimum size=1mm, shape=circle, draw=black, fill=white, font={\footnotesize\boldmath}]
\tikzstyle{wide point}=[fill=white, draw, shape=isosceles triangle, shape border rotate=-90, isosceles triangle stretches=true, inner sep=0pt, minimum width=1.5cm, minimum height=6.12mm, yshift=-0.0mm]
\tikzstyle{medium gray box}=[semilarge box, fill={rgb,255: red,180; green,180; blue,180}]
\tikzstyle{small box}=[rectangle, inline text, fill=white, draw, minimum height=5mm, yshift=-0.5mm, minimum width=5mm, font={\small}]
\tikzstyle{small gray box}=[small box, fill={rgb,255: red,180; green,180; blue,180}]
\tikzstyle{medium box}=[rectangle, inline text, fill=white, draw, minimum height=5mm, yshift=-0.5mm, minimum width=8mm, font={\small}]
\tikzstyle{ddot}=[line width=1.6pt, inner sep=0mm, minimum width=2.5mm, minimum height=2.5mm, draw, shape=circle]
\tikzstyle{dd white}=[ddot, fill=white, tikzit draw=green]
\tikzstyle{dd white phase}=[white phase dot, line width=1.6pt, tikzit draw=yellow]
\tikzstyle{dd gray}=[ddot, fill={rgb,255: red,180; green,180; blue,180}, tikzit draw=green]
\tikzstyle{dd gray phase}=[gray phase dot, line width=1.6pt, tikzit draw=yellow]
\tikzstyle{cnotbot}=[fill=white, draw=black, shape=circle]
\tikzstyle{new edge style 0}=[->-]
\tikzstyle{Inner arrow}=[->-]
\tikzstyle{State Prep}=[fill=white, draw=black, shape=rounded rectangle, rounded rectangle east arc=0pt, tikzit shape=rectangle, tikzit category=Measurement, tikzit fill={rgb,255: red,247; green,0; blue,255}]
\tikzstyle{Measure}=[fill=white, draw=black, shape=rounded rectangle, rounded rectangle west arc=0pt, tikzit category=Measurement, tikzit fill=cyan, tikzit shape=rectangle]
\tikzstyle{Gate}=[fill=white, draw=black, shape=rectangle, tikzit category=Measurement, tikzit shape=rectangle]
\tikzstyle{Vertex Highlight}=[fill=orange, draw=orange, shape=circle]
\tikzstyle{medium box}=[fill=white, draw=black, shape=rectangle, minimum width=0.75cm, minimum height=1cm]

\tikzstyle{simple}=[-, fill=none]
\tikzstyle{hadamard edge}=[-, dashed, dash pattern=on 2pt off 1pt, thick, draw=blue]
\tikzstyle{gray}=[-, draw={blue!60!white}, tikzit draw=blue]
\tikzstyle{blue}=[-, draw={blue!60!white}, tikzit draw=blue]
\tikzstyle{brace edge}=[-, tikzit draw=blue, decorate, decoration={brace,amplitude=1mm,raise=-1mm}]
\tikzstyle{diredge}=[->]
\tikzstyle{not edge}=[-, dashed, dash pattern=on 2pt off 1.5pt, thick, draw={rgb,255: red,255; green,68; blue,68}]
\tikzstyle{double edge}=[-, double, shorten <=-1mm, shorten >=-1mm, double distance=2pt]
\tikzstyle{boldedge}=[-, line width=1.6pt, shorten <=-0.17mm, shorten >=-0.17mm, tikzit draw=blue]
\tikzstyle{separatediagrams}=[-, dashed, dash pattern=on 2pt off 1.5pt, thick, draw=black]
\tikzstyle{x stabilizer}=[-, double=black, draw={zx_red}, line width=1pt]
\tikzstyle{z stabilizer}=[-, double=black, draw={zx_green}, line width=1 pt]
\tikzstyle{zx stabilizer}=[-, double=black, draw={zx_yellow}, line width=1 pt]
\tikzstyle{measurement}=[-, tikzit fill=cyan, double]

\newcommand{\pngfig}[2]{%
    \IfFileExists{#1.pdf}{\includegraphics[width=#2]{#1.pdf}}
    {
        \IfFileExists{./figures/#1.pdf}{\includegraphics[width=#2]{./figures/#1.pdf}}
        {
            \IfFileExists{#1.png}
              {\PackageWarning{pngfig}{#1.png is a PNG; slows down compile time; consider converting to pdf}
                \includegraphics[width=#2]{#1.png}}
              {
                \IfFileExists{./figures/#1.png}
                  {\PackageWarning{pngfig}{./figures/#1.png is a PNG; slows down compile time; consider converting to pdf}
                  \includegraphics[width=#2]{./figures/#1.png}}
                  {\tikz[baseline=-0.5em]{\node[draw=red,font=\color{red},fill=red!10!white] {\textit{#1}};}}%
              }
        }
    }
}

\newcommand{\pnghfig}[2]{%
    \IfFileExists{#1.pdf}{\includegraphics[height=#2]{#1.pdf}}
    {
        \IfFileExists{./figures/#1.pdf}{\includegraphics[height=#2]{./figures/#1.pdf}}
        {
            \IfFileExists{#1.png}
              {\PackageWarning{pnghfig}{#1.png is a PNG; slows down compile time; consider converting to pdf}
                \includegraphics[height=#2]{#1.png}}
              {
                \IfFileExists{./figures/#1.png}
                  {\PackageWarning{pnghfig}{./figures/#1.png is a PNG; slows down compile time; consider converting to pdf}
                  \includegraphics[height=#2]{./figures/#1.png}}
                  {\tikz[baseline=-0.5em]{\node[draw=red,font=\color{red},fill=red!10!white] {\textit{#1}};}}%
              }
        }
    }
}

\newcommand\blfootnote[1]{%
  \begingroup
  \renewcommand\thefootnote{}\footnotetext{#1}%
  \endgroup
}

\newcommand{\id}{\texttt{id}}

\newtheorem{dfn}{Definition}

\newcommand{\lstlineref}[2]{\hyperref[#2]{\Cref*{#1}:\ref*{#2}}}
\newcommand{\lstlinerangeref}[3]{\hyperref[#2]{\Cref*{#1}:\ref*{#2}-\ref*{#3}}}
\crefname{lem}{Lemma}{Lemmas}
\Crefname{lem}{Lemma}{Lemmas}
\crefname{cor}{Corrolary}{Corrolaries}
\Crefname{cor}{Corrolary}{Corrolaries}
\crefname{thm}{Theorem}{Theorems}
\Crefname{thm}{Theorem}{Theorems}
\crefname{def}{Definition}{Definitions}
\Crefname{def}{Definition}{Definitions}
\crefname{ex}{Exercise}{Exercises}
\Crefname{ex}{Exercise}{Exercises}
\crefname{exa}{Example}{Examples}
\Crefname{exa}{Example}{Examples}

\usepackage{stmaryrd}

\title{\vyzx: Formal Verification of a Graphical Quantum Language}

\author{Adrian Lehmann*}
\affiliation{%
\institution{University of Chicago}
\country{USA}
}
\author{Ben Caldwell*}
\affiliation{%
\institution{University of Chicago}
\country{USA}
}
\author{Bhakti Shah}
\affiliation{%
\institution{University of St Andrews}
\country{UK}
}
\author{William Spencer}
\affiliation{%
\institution{University of Chicago}
\country{USA}
}
\author{Robert Rand}
\affiliation{%
\institution{University of Chicago}
\country{USA}
}
\date{April 2025}

\ccsdesc[500]{Theory of computation~Program verification}
\ccsdesc[500]{Theory of computation~Equational logic and rewriting}
\ccsdesc[500]{Theory of computation~Quantum information theory}

\begin{document}

\begin{abstract}

%Mathematical representations of graphs often resemble adjacency matrices or lists, representations that facilitate whiteboard reasoning and algorithm design. 
Graphical languages are a convenient shorthand to represent computation, with rewrite rules relating one graph to another. 
In contrast, proof assistants rely heavily on inductive datatypes, particularly when giving semantics to embedded languages.
This creates obstacles to formally reasoning about graphical languages, since imposing an inductive structure obfuscates the diagrammatic nature of graphical languages, along with their corresponding equational theories.
To address this gap, we present \vyzx, a verified library for reasoning about inductively defined graphical languages.
These inductive constructs arise naturally from category-theoretic definitions.
We developed \vyzx to \VyZX, a graphical language for reasoning about quantum computation.
The \zxcalc comes with a collection of diagrammatic rewrite rules that preserve the graph's semantic interpretation.
We show how inductive graphs in \vyzx are used to prove the soundness of the \zxcalc rewrite rules and apply them in practice using standard proof assistant techniques.
We also provide an IDE-integrated visualizer for proof engineers to directly reason about diagrams in graphical form.
\end{abstract}

\maketitle

\blfootnote{* Equal contribution}

\section{Introduction}

\begin{wrapfigure}{R}{.2\textwidth}
    \centering
        \begin{tikzpicture}
	\begin{pgfonlayer}{nodelayer}
		\node [style=none] (1) at (0, 0) {};
		\node [style=none] (2) at (0, 2) {};
		\node [style=white dot] (3) at (2, 0) {};
		\node [style=gray dot] (4) at (2, 2) {};
		\node [style=none] (5) at (4, 0) {};
		\node [style=none] (6) at (4, 2) {};  
	\end{pgfonlayer}
	\begin{pgfonlayer}{edgelayer}
		\draw (1) to (3);
		\draw (2) to (4);
		\draw (3) to (4);
		\draw (3) to (5);
		\draw (4) to (6);
	\end{pgfonlayer}
\end{tikzpicture}
    \caption{\centering A \zxdiag representing a CNOT gate}
    \label{fig:cnot}
\end{wrapfigure} %
How do we formally verify a graphical language? Formal verification involves giving rigid semantics to syntactic objects, while graphical languages represent computations as graphs, allowing us to easily visualize %computational 
processes.
%such as the  \zxcalc, a lightweight language for quantum computing? 
Of special interest are graphical languages that follow the principle that \emph{only connectivity matters}~\cite{coecke2017picturing}: We care about the connections in a graph, not the position of any given node. \Cref{fig:cnot} shows a simple graphical program, in which the green and red nodes\footnote{We use the accessible shades of green and red from \href{https://zxcalculus.com/accessibility.html}{zxcalculus.com/accessibility.html} for this paper.} represent processes that are each connected to one input, one output, and each other.

In this example, we can connect the green and red nodes vertically without worrying about their left-to-right ordering in the graph. 
However, when giving \emph{semantics} to a graph, we first have to fix an ordering corresponding to the flow of data in the program. In this paper, we will focus on the \zxcalc, a lightweight language for quantum computing, in which nodes correspond to matrices and their connections correspond to matrix multiplication. To translate \Cref{fig:cnot} into matrix products, we will need to choose whether the green (top) or red (bottom) node comes first, even though the full program's semantics will be the same regardless of our choice.
The standard approach here is to reposition the nodes in the diagram slightly, so that any two connected nodes have a left-to-right ordering. We see one of the two ways of ordering the green and red nodes in \Cref{fig:cnot-slide}. %
\begin{wrapfigure}{R}{.2\textwidth}
    \centering
        \begin{tikzpicture}
	\begin{pgfonlayer}{nodelayer}
		\node [style=none] (1) at (0, 0) {};
		\node [style=none] (2) at (0, 2) {};
		\node [style=white dot] (3) at (2.75, 0) {};
		\node [style=gray dot] (4) at (1.25, 2) {};
		\node [style=none] (5) at (4, 0) {};
		\node [style=none] (6) at (4, 2) {};  
		\node [style=none] (7) at (2, -0.25) {};
		\node [style=none] (8) at (2, 2.25) {};
		\node [style=none] (9) at (-0.25, 1) {};
		\node [style=none] (10) at (4.25, 1) {};
 \end{pgfonlayer}
	\begin{pgfonlayer}{edgelayer}
		\draw (1) to (3);
		\draw (2) to (4);
		\draw (3) to (4);
		\draw (3) to (5);
		\draw (4) to (6);
		\draw[densely dotted] (7) to (8);
		\draw[densely dotted] (9) to (10);
            
 \end{pgfonlayer}
\end{tikzpicture}
    \caption{\centering A CNOT adjusted for ordering.}
    \label{fig:cnot-slide}
\end{wrapfigure} 
Unfortunately, while decomposing a \zxdiag into an ordered composition of nodes allows us to give it semantics, it complicates reasoning about the diagram. If we now wish to shift the green node past the red one, we will need to prove that this is semantics-preserving. De facto, we are dealing with a far more rigid structure than the graphical diagram we first saw, and manipulating it in the expected ways will prove difficult.

This paper is about bridging this gap in the \Coq prover~\cite{Rocq}, formerly known as Coq. It shows how we can represent a \zxdiag in terms of its compositional structure in a proof assistant, while regaining key rules that pertain to graphs, including ``only connectivity matters''. Moreover, since \zxcalc is a \emph{calculus} endowed with a range of semantics-preserving rewrite rules, we prove the soundness of these rules as well, allowing users to graphically transform one \zxdiag into an equivalent one.

We use these insights to present a formally verified \zxcalc library called \vyzx. By verified, we mean that our transformations, particularly the complete equational theory for the \zxcalc, have been proven sound with respect to the linear-algebraic semantics of \zxdiags. 

% Besides this central contribution,
%of representing graphical structures inductively, endowing them with semantics, and proving the soundness of a range of rewrite rules, 
% \vyzx provides features to make it a practical tool for \zxcalc reasoning:
In pursuit of practical reasoning about \zxdiags, \vyzx presents several key contributions:

\textbf{\vyzx diagrams are inductively defined and parametric} (\Cref{sec:vyzx-diags}). % Section 3
\zxdiags can have a variable number of inputs, connections, and outputs. Practically, \zxcalc rules are often written parametrically, to allow application in diverse contexts.
\vyzx's inductive structure allows for variables for all inputs, connections, and outputs, allowing us to state and prove rules in their most general form.
As our structure is inductively defined, we are able to do inductive proof over the structure of diagrams in a natural way.
We can also reason over the structure of diagrams with variable placeholders for arbitrary diagrams.

\textbf{\vyzx allows for graphical rewriting on top of the denotational semantics} (\Cref{sec:vizx}). %Section 4
\vyzx allows for inductive reasoning about graphs via denotational semantics. Semantics are given in the form of \qlib matrices, but not all proofs require the user to appeal to those semantics.
This is because \vyzx verifies a complete set of standard \zxcalc rewrite rules, allowing the user to prove equivalences between \zxdiags purely diagrammatically. 

We also provide a visualizer for graphical structures. The \vizx rocq-lsp plugin integrates graphical proof states into the user's proof writing environment. % Section 4
This addresses a fundamental limitation of existing graphical libraries, including \vyzx, that graphs are hard to represent and reason about textually.
In \vyzx, we frequently see examples with a deeply nested structure that obfuscates the \zxdiags they represent. 
A human-readable graphical visualization of the proof state makes \vyzx proofs significantly more approachable.

\textbf{\vyzx implements a complete set of rewrite rules} (\cref{sec:vyzx}). We show how to prove the complete equational theory of Jeandel et al.~\cite{jeandel2019completeness} using \vyzx. This allows us to move from structural proof to diagrammatic proofs, as a complete equational theory on diagrams obviates the need to reason about the denotation of \zxdiags.

\textbf{\vyzx verifies the universality of \zxdiags} (\cref{sec:universality}). Extrapolating from a proof sketch by van de Wetering~\cite{vandewetering2020zxcalculus}, we show how to construct arbitrary scalars as \zxdiags. We then show how to graphically take the sum of two ZX-diagrams, following Jeandel et al.~\cite{Jeandel2024AdditionZX}. Combining these two results gives us an elegant proof that we can represent any linear map as a \zxdiag.

\textbf{\vyzx works with other semantic models} (\cref{sec:circuits}). % Section 7
Other works in verified quantum computing, such as \sqir~\cite{hietala2021sqir} and \voqc~\cite{hietala2021voqc}, use a matrix-based \Coq quantum computing library  ``\qlib''~\cite{QuantumLib} to define their semantics.
We use \qlib to establish interoperability with \sqir circuits
% %
and show how these tools can interact through quantum circuit ingestion to \vyzx diagrams.

Although \vyzx was developed for the \zxcalc, the same principles apply to any symmetric monoidal category, which can represent a broad range of mathematical structures, all of which can be visualized as graphs~\cite{Selinger2010}.

\section{The ZX Calculus}\label{sec:zx}

\begin{figure}
    \centering
    \begin{subfigure}{.45\textwidth}
        \centering
        \begin{tikzpicture}
	\begin{pgfonlayer}{nodelayer}
		\node [style=gray dot] (0) at (0, 0) {$\alpha$};
		\node [style=none] (1) at (1.5, 1.25) {};
		\node [style=none] (2) at (1.5, -1.25) {};
		\node [style=none] (3) at (-1.5, 1.25) {};
		\node [style=none] (4) at (-1.5, -1.25) {};
		\node [style=none] (5) at (1.5, 0) {};
		\node [style=none] (7) at (1.5, 0.25) {\vdots};
		\node [style=none] (8) at (-1.5, 0.25) {\vdots};
		\node [style=none] (9) at (-2.5, 0.1) {$n$};
		\node [style=none] (10) at (2.5, 0.1) {$m$};
		\node [style=none] (11) at (-1.5, 0.75) {};
		\node [style=none] (12) at (-1.5, -0.75) {};
		\node [style=none] (13) at (1.5, -0.75) {};
		\node [style=none] (14) at (1.5, 0.75) {};
	\end{pgfonlayer}
	\begin{pgfonlayer}{edgelayer}
		\draw [in=180, out=30, looseness=1.25] (0) to (1.center);
		\draw [in=180, out=-30, looseness=1.25] (0) to (2.center);
		\draw [in=150, out=0, looseness=1.25] (3.center) to (0);
		\draw [in=-150, out=0, looseness=1.25] (4.center) to (0);
		\draw [in=165, out=0, looseness=1.50] (11.center) to (0);
		\draw [in=-165, out=0, looseness=1.50] (12.center) to (0);
		\draw [in=-180, out=15, looseness=1.50] (0) to (14.center);
		\draw [in=-180, out=-15, looseness=1.50] (0) to (13.center);
	\end{pgfonlayer}
\end{tikzpicture}
        \par\bigskip
        \(\ket{0}^{\otimes m}  \bra{0}^{\otimes n} + e^{i\alpha} \ket{1}^{\otimes m} \bra{1}^{\otimes n}\)
    \end{subfigure}
    \begin{subfigure}{.45\textwidth}
        \centering
        \begin{tikzpicture}
	\begin{pgfonlayer}{nodelayer}
		\node [style=white dot] (0) at (0, 0) {$\alpha$};
		\node [style=none] (1) at (1.5, 1.25) {};
		\node [style=none] (2) at (1.5, -1.25) {};
		\node [style=none] (3) at (-1.5, 1.25) {};
		\node [style=none] (4) at (-1.5, -1.25) {};
		\node [style=none] (5) at (1.5, 0.25) {};
		\node [style=none] (6) at (1.5, 0.25) {};
		\node [style=none] (7) at (1.5, 0.25) {\vdots};
		\node [style=none] (8) at (-1.5, 0.25) {\vdots};
		\node [style=none] (9) at (-2.5, 0.1) {$n$};
		\node [style=none] (10) at (2.5, 0.1) {$m$};
		\node [style=none] (11) at (-1.5, 0.75) {};
		\node [style=none] (12) at (-1.5, -0.75) {};
		\node [style=none] (13) at (1.5, -0.75) {};
		\node [style=none] (14) at (1.5, 0.75) {};
	\end{pgfonlayer}
	\begin{pgfonlayer}{edgelayer}
		\draw [in=180, out=30, looseness=1.25] (0) to (1.center);
		\draw [in=180, out=-30, looseness=1.25] (0) to (2.center);
		\draw [in=150, out=0, looseness=1.25] (3.center) to (0);
		\draw [in=-150, out=0, looseness=1.25] (4.center) to (0);
		\draw [in=165, out=0, looseness=1.50] (11.center) to (0);
		\draw [in=-165, out=0, looseness=1.50] (12.center) to (0);
		\draw [in=-180, out=15, looseness=1.50] (0) to (14.center);
		\draw [in=-180, out=-15, looseness=1.50] (0) to (13.center);
	\end{pgfonlayer}
\end{tikzpicture}
        \par\bigskip
        \(\ket{+}^{\otimes m}  \bra{+}^{\otimes n} + e^{i\alpha} \ket{-}^{\otimes m} \bra{-}^{\otimes n}\)   
    \end{subfigure}
    \caption{\centering Z and X spiders with their standard bra-ket semantics.}
    \label{fig:eqnXZ}
\end{figure}

In most existing quantum software, like IBM's Qiskit~\cite{Qiskit} or Google's Cirq~\cite{cirq}, quantum programs are written as \emph{quantum circuits}~\cite{deutsch1989circuits}, a quantum analog to classical circuits that generally features from three to over a dozen distinct basic gates.\footnote{For an accessible and thorough introduction to quantum computing, we encourage the reader to consult the standard textbook in the area~\cite{nielsen2010}.}
By contrast, \zxdiags are expressible with only two kinds of nodes and a more flexible graphical structure. 
In addition to their comparative simplicity, \zxdiags have proven useful for creating quantum circuit optimizers~\cite{kissinger2020Pyzx,cowtan2020opt,debeaudrap2020opt}, simulating quantum circuits~\cite{kissinger2022simulating}, implementing error-correcting codes~\cite{zxlattice,chancellor2018graphical}, and reasoning about measurement-based quantum computing~\cite{zx-mbqc,mcelvanney2023flowpreserving}.

Fundamentally, \zxdiags are graphs with green and red nodes, called Z and X \emph{spiders}, with $n$ inputs and $m$ outputs, along with a rotation angle $\alpha \in [0,2\pi)$. 
Spiders without an explicit rotation parameter are considered to have a $0$ rotation.
The semantics of Z and X spiders are shown in \Cref{fig:eqnXZ}.
%
% While this paper doesn't aim to explain quantum computing in general,\footnote{The interested reader is encouraged to consult the standard textbook in the area~\cite{nielsen2010}.} we will briefly explain the notation above.
%
Here, $\ket{0}^{\otimes n}$ with $n$ repeated zeros represents (in Dirac's \emph{bra-ket} notation) a $2^n$-length basis vector with a $1$ in the first position and zeros elsewhere, while $\bra{0}^{\otimes n}$ represents its transpose. 
Similarly, $\ket{1}^{\otimes n}$ is a $2^n$-length vector with a $1$ in the last position. 
The intuition behind these spiders is that they take in a quantum state, preserve the $\ket{0}^{\otimes n}$ vectors, and rotate the $\ket{1}^{\otimes n}$ vectors by $\alpha$, postselecting on the two cases above.
%and \emph{postselect} on those parts of the state that aren't entirely $0$ or $1$. 
%
The red X spiders act equivalently, but using the X basis.
From a linear algebraic standpoint, Z and X spiders correspond to complex-valued matrices.
The number of inputs and outputs to a spider determines the dimensions of the matrix it represents.
Combined, the $Z$ and $X$ spiders are sufficient to represent any quantum computation (see \Cref{sec:gates}).

The \emph{\zxcalc}~\cite{CoeckeDuncan2011,coecke2017picturing} uses \zxdiags with a set of rewrite rules to translate between equivalent quantum operations.
We show a sample of common rewrite rules in \Cref{sec:zx-rules}.
Note that instead of denoting \zxdiags as equal, we denote them as proportional (\(\propto\)), meaning they are equal up to a non-zero scalar factor.
Considering equality up to scalar factors is a convention in \zxcalc, as any non-zero scalar can be written as a \zxdiag with no inputs or outputs.

This presentation of \zxcalc draws upon van de Wetering's \cite{vandewetering2020zxcalculus} extensive survey.
We also recommend Coecke's introduction to the \zxcalc~\cite{coecke2023basic}.

\subsection{Meta-Rules}

\paragraph{Colorswapping}

We define a color-swapped \zxdiag as a \zxdiag with the same structure but changing every spider from Z to X and X to Z while preserving the angle.
It can be shown that if a rule can be applied to a \zxdiag \(\text{zx}_1\) transforming it into \(\text{zx}_2\), then it can be applied to a color-swapped version of \(\text{zx}_1\) transforming it into a color-swapped \(\text{zx}_2\).
With this in mind, we show any rule only for one color configuration, understanding that it applies to the color-swapped version.

\paragraph{Only connectivity matters}

\begin{figure}
    \centering
    \begin{tikzpicture}
	\begin{pgfonlayer}{nodelayer}
		\node [style=white dot] (0) at (-8, 0) {};
		\node [style=gray dot] (1) at (-8, -1) {};
		\node [style=gray dot] (2) at (-8, 1) {};
		\node [style=white dot] (3) at (-5, -1) {};
		\node [style=white dot] (4) at (-5, 1) {};
		\node [style=none] (5) at (-9, 1) {};
		\node [style=none] (6) at (-9, 0) {};
		\node [style=none] (7) at (-9, -1) {};
		\node [style=gray dot] (8) at (-2, 1) {};
		\node [style=gray dot] (10) at (-2, -1) {};
		\node [style=none] (11) at (-1, 1) {};
		\node [style=none] (13) at (-1, -1) {};
		\node [style=white dot] (14) at (2, -1) {};
		\node [style=gray dot] (15) at (2, 0) {};
		\node [style=gray dot] (16) at (2, 1) {};
		\node [style=white dot] (17) at (7, 0) {};
		\node [style=white dot] (18) at (5, 1) {};
		\node [style=none] (19) at (1, 1) {};
		\node [style=none] (20) at (1, 0) {};
		\node [style=none] (21) at (1, -1) {};
		\node [style=gray dot] (22) at (8, 1) {};
		\node [style=gray dot] (23) at (8, 0) {};
		\node [style=none] (24) at (9, 1) {};
		\node [style=none] (25) at (9, -1) {};
		\node [style=none] (26) at (0, 0) {$=$};
	\end{pgfonlayer}
	\begin{pgfonlayer}{edgelayer}
		\draw (2) to (4);
		\draw [in=195, out=15] (0) to (4);
		\draw [in=165, out=-15] (0) to (3);
		\draw (1) to (3);
		\draw (4) to (8);
		\draw (10) to (3);
		\draw [bend left=75, looseness=1.25] (4) to (3);
		\draw (8) to (11.center);
		\draw (10) to (13.center);
		\draw (6.center) to (0);
		\draw (5.center) to (2);
		\draw (1) to (7.center);
		\draw (16) to (18);
		\draw [in=-165, out=15] (14) to (18);
		\draw [in=-165, out=0] (14) to (17);
		\draw (15) to (17);
		\draw (18) to (22);
		\draw (23) to (17);
		\draw [in=165, out=-15] (18) to (17);
		\draw [in=180, out=0] (22) to (24.center);
		\draw [in=165, out=-15, looseness=1.50] (23) to (25.center);
		\draw [in=-180, out=0, looseness=0.75] (20.center) to (14);
		\draw (19.center) to (16);
		\draw [in=0, out=-180, looseness=0.75] (15) to (21.center);
	\end{pgfonlayer}
\end{tikzpicture}
    \caption{\centering Two equivalent \zxdiags, where the right diagram is deformed. Connections and qubit order are maintained.}\label{fig:deform}
\end{figure}

\begin{figure}
    \centering
    \begin{tikzpicture}
	\begin{pgfonlayer}{nodelayer}
		\node [style=none] (39) at (-14, 0) {$\propto$};
		\node [style=none] (40) at (-15, 1) {};
		\node [style=none] (41) at (-16, 1) {};
		\node [style=none] (42) at (-16, 0) {};
		\node [style=none] (43) at (-16, -1) {};
		\node [style=none] (44) at (-17, -1) {};
		\node [style=none] (45) at (-13, 0) {};
		\node [style=none] (46) at (-12, 0) {};
		\node [style=none] (47) at (-11, 0) {$\propto$};
		\node [style=none] (55) at (-4, 0.5) {};
		\node [style=none] (57) at (-4, -0.5) {};
		\node [style=none] (59) at (-5, 0.5) {};
		\node [style=none] (60) at (-6, -0.5) {};
		\node [style=none] (61) at (-5, -0.5) {};
		\node [style=none] (62) at (-6, 0.5) {};
		\node [style=none] (63) at (-1, 0.5) {};
		\node [style=none] (64) at (-2, 0.5) {};
		\node [style=none] (65) at (-1, -0.5) {};
		\node [style=none] (66) at (-2, -0.5) {};
		\node [style=none] (67) at (-3, 0) {$\propto$};
		\node [style=none] (68) at (1.5, 0.5) {};
		\node [style=none] (69) at (1.5, -0.5) {};
		\node [style=none] (72) at (2.5, -0.5) {};
		\node [style=none] (73) at (2.5, 0.5) {};
		\node [style=none] (74) at (3.5, 0) {$\propto$};
		\node [style=none] (76) at (6, -0.5) {};
		\node [style=none] (78) at (6, 0.5) {};
		\node [style=none] (79) at (5, 0.5) {};
		\node [style=none] (80) at (5, -0.5) {};
		\node [style=none] (132) at (-12.5, -4) {};
		\node [style=none] (133) at (-12.5, -3) {};
		\node [style=none] (134) at (-10, -4.5) {};
		\node [style=none] (135) at (-11, -5.5) {};
		\node [style=none] (136) at (-12.5, -5) {};
		\node [style=gray dot] (137) at (-11, -4.5) {};
		\node [style=none] (138) at (-10, -5.5) {};
		\node [style=none] (139) at (-8, -4) {};
		\node [style=none] (140) at (-5.5, -5.5) {};
		\node [style=none] (141) at (-5.5, -4.5) {};
		\node [style=none] (142) at (-7, -3) {};
		\node [style=none] (143) at (-8, -5) {};
		\node [style=gray dot] (144) at (-6.75, -4.5) {};
		\node [style=none] (145) at (-8, -3) {};
		\node [style=none] (146) at (-9, -4.5) {$\propto$};
		\node [style=none] (147) at (-1, -4.5) {};
		\node [style=none] (148) at (1, -4) {};
		\node [style=none] (149) at (3, -4.5) {};
		\node [style=none] (150) at (1, -5) {};
		\node [style=gray dot] (151) at (2, -4.5) {};
		\node [style=none] (152) at (0, -4.5) {$\propto$};
		\node [style=gray dot] (153) at (-2, -4.5) {};
		\node [style=none] (154) at (-3.5, -5) {};
		\node [style=none] (155) at (-3.5, -4) {};
		\node [style=none] (156) at (-3, -4.5) {};
		\node [style=none] (157) at (5, -4) {};
		\node [style=none] (158) at (7, -4.5) {};
		\node [style=none] (159) at (6, -5.5) {};
		\node [style=gray dot] (160) at (6, -4.5) {};
		\node [style=none] (161) at (11, -4) {};
		\node [style=none] (162) at (9, -4.5) {};
		\node [style=none] (163) at (11, -5) {};
		\node [style=gray dot] (164) at (10, -4.5) {};
		\node [style=none] (165) at (8, -4.5) {$\propto$};
		\node [style=none] (166) at (7, -5.5) {};
		\node [style=none] (167) at (-8, -1) {};
		\node [style=none] (168) at (-9, -1) {};
		\node [style=none] (169) at (-9, 0) {};
		\node [style=none] (170) at (-9, 1) {};
		\node [style=none] (171) at (-10, 1) {};
		\node [style=none] (172) at (9, 0.5) {};
		\node [style=none] (173) at (9, -0.5) {};
		\node [style=none] (174) at (8, -0.5) {};
		\node [style=none] (175) at (8, 0.5) {};
		\node [style=none] (176) at (10.5, 0) {$\propto$};
		\node [style=none] (177) at (11.5, -0.5) {};
		\node [style=none] (178) at (11.5, 0.5) {};
		\node [style=none] (179) at (12.5, 0.5) {};
		\node [style=none] (180) at (12.5, -0.5) {};
	\end{pgfonlayer}
	\begin{pgfonlayer}{edgelayer}
		\draw [in=0, out=180, looseness=1.25] (40.center) to (41.center);
		\draw [bend right=90, looseness=1.75] (41.center) to (42.center);
		\draw [bend left=90, looseness=1.75] (42.center) to (43.center);
		\draw [in=0, out=180, looseness=1.25] (43.center) to (44.center);
		\draw (46.center) to (45.center);
		\draw [in=0, out=180, looseness=1.25] (59.center) to (60.center);
		\draw [in=0, out=180, looseness=1.25] (61.center) to (62.center);
		\draw [in=0, out=180, looseness=1.25] (55.center) to (61.center);
		\draw [in=0, out=180, looseness=1.25] (57.center) to (59.center);
		\draw (64.center) to (63.center);
		\draw (66.center) to (65.center);
		\draw [in=180, out=180, looseness=1.75] (68.center) to (69.center);
		\draw [in=180, out=0, looseness=1.25] (69.center) to (73.center);
		\draw [in=180, out=0, looseness=1.25] (68.center) to (72.center);
		\draw [in=180, out=180, looseness=1.75] (79.center) to (80.center);
		\draw (76.center) to (80.center);
		\draw (78.center) to (79.center);
		\draw [in=180, out=0, looseness=1.25] (133.center) to (135.center);
		\draw [in=-150, out=0] (136.center) to (137);
		\draw [in=150, out=0] (132.center) to (137);
		\draw (134.center) to (137);
		\draw (135.center) to (138.center);
		\draw [in=0, out=180, looseness=1.25] (140.center) to (142.center);
		\draw [in=-150, out=0] (143.center) to (144);
		\draw [in=150, out=0] (139.center) to (144);
		\draw (141.center) to (144);
		\draw (142.center) to (145.center);
		\draw [in=-150, out=0] (150.center) to (151);
		\draw [in=150, out=0] (148.center) to (151);
		\draw (149.center) to (151);
		\draw (147.center) to (153);
		\draw [bend left=45, looseness=1.25] (156.center) to (153);
		\draw (156.center) to (154.center);
		\draw (156.center) to (155.center);
		\draw [bend right=45, looseness=1.25] (156.center) to (153);
		\draw [in=180, out=180, looseness=1.75] (159.center) to (160);
		\draw [in=150, out=0] (157.center) to (160);
		\draw (158.center) to (160);
		\draw [in=-30, out=180] (163.center) to (164);
		\draw [in=30, out=180] (161.center) to (164);
		\draw (162.center) to (164);
		\draw (166.center) to (159.center);
		\draw [in=0, out=-180, looseness=1.25] (167.center) to (168.center);
		\draw [bend left=90, looseness=1.75] (168.center) to (169.center);
		\draw [bend right=90, looseness=1.75] (169.center) to (170.center);
		\draw [in=0, out=-180, looseness=1.25] (170.center) to (171.center);
		\draw [in=0, out=0, looseness=1.75] (172.center) to (173.center);
		\draw [in=0, out=180, looseness=1.25] (173.center) to (175.center);
		\draw [in=0, out=180, looseness=1.25] (172.center) to (174.center);
		\draw [in=0, out=0, looseness=1.75] (179.center) to (180.center);
		\draw (177.center) to (180.center);
		\draw (178.center) to (179.center);
	\end{pgfonlayer}
\end{tikzpicture}
    \caption{\centering The Only Connectivity Matters rules~\cite[Section 2.2]{jeandel2018complete}}\label{fig:ocm-rules}
\end{figure}

Perhaps the most important meta-rule in the \zxcalc is that only connectivity matters (OCM).
This means that wires can be arbitrarily deformed as long as the input and output order to the overall diagram is maintained. 
Note that this especially means that the overall diagram's matrix semantics maintain their dimensions while we allow diagram components to change their matrix dimensions.
This rule is very powerful as it allows us to move diagrams into the most convenient form to apply rules to.
Another important corollary of OCM is that it doesn't matter whether a wire is an input or output to a node.
We convert inputs to outputs (and vice versa) by moving the wires over the spider and adding a cup ($\subset$) or cap ($\supset$) as appropriate.
From a quantum information perspective, this corresponds to exchanging inputs and outputs by adding appropriate Bell states or Bell measurements.
We show an example of OCM deformations in \Cref{fig:deform}, where we can observe a Bell measurement in the left diagram becoming a simple wire between the two central red spiders.
The full set of rules generating OCM are given in \Cref{fig:ocm-rules}.

\subsection{Rewrite Rules}\label{sec:zx-rules}

\begin{figure}[h]
    \centering
    \begin{subfigure}{.32\textwidth}
        \centering
        \begin{tikzpicture}
	\begin{pgfonlayer}{nodelayer}
		\node [style=white dot] (0) at (0, 0) {$\alpha$};
		\node [style=none] (3) at (1, -0.25) {};
		\node [style=none] (4) at (-1, -0.25) {};
		\node [style=none] (5) at (-1, -0.5) {};
		\node [style=none] (6) at (-1, -0.5) {.};
		\node [style=none] (7) at (1, -0.5) {.};
		\node [style=none] (8) at (1, -0.75) {};
		\node [style=none] (9) at (1, -0.75) {};
		\node [style=none] (10) at (-1, -0.75) {.};
		\node [style=none] (11) at (-1, -1) {};
		\node [style=none] (12) at (1, -1) {};
		\node [style=white dot] (13) at (4, 0) {$\alpha$};
		\node [style=none] (14) at (5, -0.25) {};
		\node [style=none] (15) at (3, -0.25) {};
		\node [style=none] (16) at (3, -0.5) {};
		\node [style=none] (17) at (3, -0.5) {.};
		\node [style=none] (18) at (5, -0.5) {.};
		\node [style=none] (19) at (5, -0.75) {};
		\node [style=none] (20) at (5, -0.75) {.};
		\node [style=none] (21) at (3, -0.75) {.};
		\node [style=none] (22) at (3, -1) {};
		\node [style=none] (23) at (5, -1) {};
		\node [style=none] (24) at (2, 0) {$\propto$};
	\end{pgfonlayer}
	\begin{pgfonlayer}{edgelayer}
		\draw [in=195, out=0] (4.center) to (0);
		\draw [in=120, out=60, loop] (0) to ();
		\draw [in=-180, out=-15, looseness=1.25] (0) to (3.center);
		\draw [in=180, out=-30, looseness=0.75] (0) to (12.center);
		\draw [in=0, out=-150, looseness=0.75] (0) to (11.center);
		\draw [in=195, out=0] (15.center) to (13);
		\draw [in=-180, out=-15, looseness=1.25] (13) to (14.center);
		\draw [in=180, out=-30, looseness=0.75] (13) to (23.center);
		\draw [in=0, out=-150, looseness=0.75] (13) to (22.center);
	\end{pgfonlayer}
\end{tikzpicture}
        \caption{\centering Self-loop Removal}\label{fig:self-loop-removal}
    \end{subfigure}
    \begin{subfigure}{.32\textwidth}
        \centering
        \begin{tikzpicture}
	\begin{pgfonlayer}{nodelayer}
		\node [style=white dot] (0) at (-2, 0) {};
		\node [style=gray dot] (1) at (-3, 0) {};
		\node [style=none] (2) at (-1, 0.75) {};
		\node [style=none] (6) at (-1, -0.75) {};
		\node [style=none] (9) at (0, 0) {$\propto$};
		\node [style=none] (12) at (1, -0.75) {};
		\node [style=none] (13) at (1, 0.75) {};
		\node [style=white dot] (14) at (1.5, -0.75) {};
		\node [style=white dot] (15) at (1.5, 0.75) {};
		\node [style=gray dot] (16) at (3.5, 0.75) {};
		\node [style=gray dot] (17) at (3.5, -0.75) {};
		\node [style=none] (18) at (4, 0.75) {};
		\node [style=none] (19) at (4, -0.75) {};
		\node [style=none] (20) at (-4, 0.75) {};
		\node [style=none] (24) at (-4, -0.75) {};
	\end{pgfonlayer}
	\begin{pgfonlayer}{edgelayer}
		\draw (13.center) to (15);
		\draw (15) to (16);
		\draw [in=165, out=-15] (15) to (17);
		\draw [in=-165, out=15, looseness=1.25] (14) to (16);
		\draw (16) to (18.center);
		\draw (17) to (19.center);
		\draw (14) to (17);
		\draw (12.center) to (14);
		\draw [in=165, out=0] (20.center) to (1);
		\draw (1) to (0);
		\draw [in=-180, out=15] (0) to (2.center);
		\draw [in=-180, out=-15] (0) to (6.center);
		\draw [in=0, out=-180, looseness=0.75] (1) to (24.center);
	\end{pgfonlayer}
\end{tikzpicture}
        \caption{\centering Bialgebra rule}\label{fig:bi-alg}
    \end{subfigure}
    \begin{subfigure}{.32\textwidth}
        \centering
        \begin{tikzpicture}
	\begin{pgfonlayer}{nodelayer}
		\node [style=white dot] (0) at (-3.25, 0) {};
		\node [style=gray dot] (1) at (-1.75, 0) {};
		\node [style=none] (2) at (-1, 0) {};
		\node [style=none] (3) at (-4, 0) {};
		\node [style=white dot] (4) at (1.75, 0) {};
		\node [style=gray dot] (5) at (3.25, 0) {};
		\node [style=none] (6) at (4, 0) {};
		\node [style=none] (7) at (1, 0) {};
		\node [style=none] (8) at (0, 0) {$\propto$};
	\end{pgfonlayer}
	\begin{pgfonlayer}{edgelayer}
		\draw (3.center) to (0);
		\draw [bend left=15, looseness=1.25] (0) to (1);
		\draw (1) to (2.center);
		\draw [in=-165, out=-15] (0) to (1);
		\draw (7.center) to (4);
		\draw (5) to (6.center);
	\end{pgfonlayer}
\end{tikzpicture}
        \caption{\centering Hopf rule}\label{fig:hopf}
    \end{subfigure}
    \begin{subfigure}{.32\textwidth} 
        \centering
        \begin{tikzpicture}
	\begin{pgfonlayer}{nodelayer}
		\node [style=none] (1) at (-1, 1) {};
		\node [style=white dot] (2) at (0, 0) {$-\alpha$};
		\node [style=none] (3) at (-1, 2) {};
		\node [style=none] (4) at (-1, -1) {};
		\node [style=none] (5) at (1, -1) {};
		\node [style=none] (6) at (1, 1) {};
		\node [style=none] (7) at (1, 2) {};
		\node [style=none] (8) at (1, 0.25) {.};
		\node [style=none] (9) at (1, 0) {.};
		\node [style=none] (10) at (1, -0.25) {.};
		\node [style=none] (11) at (-1, -0.25) {.};
		\node [style=none] (12) at (-1, 0) {.};
		\node [style=none] (13) at (-1, 0.25) {.};
		\node [style=gray dot] (14) at (1.5, 2) {$\pi$};
		\node [style=gray dot] (15) at (1.5, 1) {$\pi$};
		\node [style=gray dot] (16) at (1.5, -1) {$\pi$};
		\node [style=gray dot] (17) at (-1.5, -1) {$\pi$};
		\node [style=gray dot] (18) at (-1.5, 1) {$\pi$};
		\node [style=gray dot] (19) at (-1.5, 2) {$\pi$};
		\node [style=none] (20) at (2, 2) {};
		\node [style=none] (21) at (2, 1) {};
		\node [style=none] (22) at (2, -1) {};
		\node [style=none] (23) at (-2, 2) {};
		\node [style=none] (24) at (-2, 1) {};
		\node [style=none] (25) at (-2, -1) {};
		\node [style=none] (26) at (-6.25, 1) {};
		\node [style=white dot] (27) at (-5.25, 0) {$\alpha$};
		\node [style=none] (28) at (-6.25, 2) {};
		\node [style=none] (29) at (-6.25, -1) {};
		\node [style=none] (30) at (-4.25, -1) {};
		\node [style=none] (31) at (-4.25, 1) {};
		\node [style=none] (32) at (-4.25, 2) {};
		\node [style=none] (33) at (-4.25, 0.25) {.};
		\node [style=none] (34) at (-4.25, 0) {.};
		\node [style=none] (35) at (-4.25, -0.25) {.};
		\node [style=none] (36) at (-6.25, -0.25) {.};
		\node [style=none] (37) at (-6.25, 0) {.};
		\node [style=none] (38) at (-6.25, 0.25) {.};
		\node [style=none] (39) at (-3, 0.5) {$\propto$};
		\node [style=none] (40) at (-4, 2) {};
		\node [style=none] (41) at (-4, 1) {};
		\node [style=none] (42) at (-4, -1) {};
		\node [style=none] (43) at (-6.5, -1) {};
		\node [style=none] (44) at (-6.5, 1) {};
		\node [style=none] (45) at (-6.5, 2) {};
	\end{pgfonlayer}
	\begin{pgfonlayer}{edgelayer}
		\draw [in=135, out=0, looseness=0.50] (3.center) to (2);
		\draw [in=-180, out=45, looseness=0.50] (2) to (7.center);
		\draw [in=-180, out=15, looseness=0.50] (2) to (6.center);
		\draw [in=165, out=0, looseness=0.75] (1.center) to (2);
		\draw [in=15, out=-165, looseness=0.50] (2) to (4.center);
		\draw [in=-180, out=-15, looseness=0.75] (2) to (5.center);
		\draw (19) to (3.center);
		\draw (18) to (1.center);
		\draw (17) to (4.center);
		\draw (16) to (5.center);
		\draw (15) to (6.center);
		\draw (14) to (7.center);
		\draw (23.center) to (19);
		\draw (24.center) to (18);
		\draw (14) to (20.center);
		\draw (15) to (21.center);
		\draw (16) to (22.center);
		\draw (17) to (25.center);
		\draw [in=135, out=0, looseness=0.50] (28.center) to (27);
		\draw [in=-180, out=45, looseness=0.50] (27) to (32.center);
		\draw [in=-180, out=15, looseness=0.50] (27) to (31.center);
		\draw [in=165, out=0, looseness=0.75] (26.center) to (27);
		\draw [in=15, out=-165, looseness=0.50] (27) to (29.center);
		\draw [in=-180, out=-15, looseness=0.75] (27) to (30.center);
		\draw (45.center) to (28.center);
		\draw (44.center) to (26.center);
		\draw (43.center) to (29.center);
		\draw (42.center) to (30.center);
		\draw (41.center) to (31.center);
		\draw (40.center) to (32.center);
	\end{pgfonlayer}
\end{tikzpicture}
        \caption{\centering Bi-\( \pi \) rule}\label{fig:bipi}
    \end{subfigure}
    \begin{subfigure}{.32\textwidth} 
        \centering
        \begin{tikzpicture}
	\begin{pgfonlayer}{nodelayer}
		\node [style=gray dot] (0) at (-3.5, 0) {$k\pi$};
		\node [style=white phase dot] (1) at (-2, 0) {$\alpha$};
		\node [style=none] (2) at (-1, 1) {};
		\node [style=none] (3) at (-1, 0) {};
		\node [style=none] (4) at (-1, -2) {};
		\node [style=none] (5) at (-1, -0.75) {$\vdots$};
		\node [style=none] (6) at (0, 0) {$\propto$};
		\node [style=gray phase dot] (7) at (1, 1) {$k\pi$};
		\node [style=gray phase dot] (8) at (1, 0) {$k\pi$};
		\node [style=gray phase dot] (10) at (1, -2) {$k\pi$};
		\node [style=none] (11) at (2.5, 1) {};
		\node [style=none] (12) at (2.5, 0) {};
		\node [style=none] (13) at (2.0, -0.75) {$\vdots$};
		\node [style=none] (14) at (2.5, -2) {};
	\end{pgfonlayer}
	\begin{pgfonlayer}{edgelayer}
		\draw (1) to (0);
		\draw [in=-180, out=15, looseness=0.75] (1) to (2.center);
		\draw (1) to (3.center);
		\draw [in=-180, out=-30, looseness=0.75] (1) to (4.center);
		\draw (7) to (11.center);
		\draw (8) to (12.center);
		\draw (10) to (14.center);
	\end{pgfonlayer}
\end{tikzpicture}
        \caption{\centering State Copy rule}\label{fig:state-copy}
    \end{subfigure}
    \begin{subfigure}{.32\textwidth}
        \centering
        \begin{tikzpicture}
	\begin{pgfonlayer}{nodelayer}
		\node [style=white dot] (0) at (-1, 1) {$\alpha$};
		\node [style=white dot] (1) at (1, -1) {$\beta$};
		\node [style=none] (3) at (2, -1.25) {};
		\node [style=none] (5) at (2, -2) {};
		\node [style=none] (7) at (-2, 2) {};
		\node [style=none] (8) at (-2, 1.25) {};
		\node [style=none] (10) at (0, 1.25) {};
		\node [style=none] (12) at (0, 2) {};
		\node [style=none] (13) at (0, -2) {};
		\node [style=none] (15) at (0, -1.25) {};
		\node [style=none] (17) at (0, 1.5) {.};
		\node [style=none] (21) at (0, 0) {..};
		\node [style=none] (22) at (0, 1.75) {};
		\node [style=none] (23) at (0, 1.75) {.};
		\node [style=none] (24) at (-2, 1.5) {.};
		\node [style=none] (26) at (-2, 1.75) {.};
		\node [style=none] (27) at (0, -1.75) {};
		\node [style=none] (28) at (0, -1.75) {.};
		\node [style=none] (29) at (0, -1.5) {.};
		\node [style=none] (31) at (2, -1.75) {.};
		\node [style=none] (34) at (2, -1.5) {.};
		\node [style=none] (35) at (3, 0) {$\propto$};
		\node [style=white dot] (37) at (5.5, 0) {$\alpha + \beta$};
		\node [style=none] (39) at (4, 1) {};
		\node [style=none] (40) at (4, 0.25) {};
		\node [style=none] (41) at (4, 0.5) {.};
		\node [style=none] (42) at (4, 0.75) {.};
		\node [style=none] (43) at (4, -0.25) {};
		\node [style=none] (44) at (4, -1) {};
		\node [style=none] (45) at (4, -0.75) {.};
		\node [style=none] (46) at (4, -0.5) {.};
		\node [style=none] (47) at (7, 1) {};
		\node [style=none] (48) at (7, 0.25) {};
		\node [style=none] (49) at (7, 0.5) {.};
		\node [style=none] (50) at (7, 0.75) {.};
		\node [style=none] (51) at (7, -0.25) {};
		\node [style=none] (52) at (7, -1) {};
		\node [style=none] (53) at (7, -0.75) {.};
		\node [style=none] (54) at (7, -0.5) {.};
	\end{pgfonlayer}
	\begin{pgfonlayer}{edgelayer}
		\draw [in=120, out=-30] (0) to (1);
		\draw [in=150, out=-60] (0) to (1);
		\draw [in=-180, out=0, looseness=1.25] (1) to (3.center);
		\draw [in=180, out=-30] (1) to (5.center);
		\draw [in=0, out=-180] (1) to (15.center);
		\draw [in=0, out=-135] (1) to (13.center);
		\draw [in=180, out=0, looseness=1.25] (0) to (10.center);
		\draw [in=180, out=30] (0) to (12.center);
		\draw [in=0, out=135, looseness=1.25] (0) to (7.center);
		\draw [in=0, out=165] (0) to (8.center);
		\draw [in=-180, out=30] (37) to (47.center);
		\draw [in=-180, out=15] (37) to (48.center);
		\draw [in=180, out=-15] (37) to (51.center);
		\draw [in=180, out=-30] (37) to (52.center);
		\draw [in=0, out=-150] (37) to (44.center);
		\draw [in=0, out=-165, looseness=1.25] (37) to (43.center);
		\draw [in=0, out=150, looseness=0.75] (37) to (39.center);
		\draw [in=0, out=165] (37) to (40.center);
	\end{pgfonlayer}
\end{tikzpicture}
        \caption{\centering Spider Fusion}\label{fig:spider-fusion}
    \end{subfigure}
    \begin{subfigure}{.32\textwidth} 
        \centering
        \begin{tikzpicture}
	\begin{pgfonlayer}{nodelayer}
		\node [style=none] (39) at (0, 0) {$\propto$};
		\node [style=gray dot] (45) at (-2, 0) {};
		\node [style=none] (46) at (1, 0) {};
		\node [style=none] (47) at (3, 0) {};
		\node [style=none] (48) at (-1, 0) {};
		\node [style=none] (49) at (-3, 0) {};
	\end{pgfonlayer}
	\begin{pgfonlayer}{edgelayer}
		\draw (49.center) to (45);
		\draw (45) to (48.center);
		\draw (46.center) to (47.center);
	\end{pgfonlayer}
\end{tikzpicture}
        \caption{\centering Identity removal}\label{fig:identity-removal}
    \end{subfigure}
    \begin{subfigure}{.32\textwidth} 
        \centering
        \begin{tikzpicture}
	\begin{pgfonlayer}{nodelayer}
		\node [style=none] (1) at (-3.75, 1) {};
		\node [style=white dot] (2) at (-2.75, 0) {$\alpha$};
		\node [style=none] (4) at (-3.75, -1) {};
		\node [style=none] (5) at (-1.75, -1) {};
		\node [style=none] (6) at (-1.75, 1) {};
		\node [style=none] (8) at (-1.75, 0.25) {.};
		\node [style=none] (9) at (-1.75, 0) {.};
		\node [style=none] (10) at (-1.75, -0.25) {.};
		\node [style=none] (11) at (-3.75, -0.25) {.};
		\node [style=none] (12) at (-3.75, 0) {.};
		\node [style=none] (13) at (-3.75, 0.25) {.};
		\node [style=hadamard] (15) at (-1.25, 1) {};
		\node [style=hadamard] (16) at (-1.25, -1) {};
		\node [style=hadamard] (17) at (-4.25, -1) {};
		\node [style=hadamard] (18) at (-4.25, 1) {};
		\node [style=none] (21) at (-0.75, 1) {};
		\node [style=none] (22) at (-0.75, -1) {};
		\node [style=none] (24) at (-4.75, 1) {};
		\node [style=none] (25) at (-4.75, -1) {};
		\node [style=none] (26) at (1.75, 1) {};
		\node [style=gray dot] (27) at (2.75, 0) {$\alpha$};
		\node [style=none] (29) at (1.75, -1) {};
		\node [style=none] (30) at (3.75, -1) {};
		\node [style=none] (31) at (3.75, 1) {};
		\node [style=none] (33) at (3.75, 0.25) {.};
		\node [style=none] (34) at (3.75, 0) {.};
		\node [style=none] (35) at (3.75, -0.25) {.};
		\node [style=none] (36) at (1.75, -0.25) {.};
		\node [style=none] (37) at (1.75, 0) {.};
		\node [style=none] (38) at (1.75, 0.25) {.};
		\node [style=none] (39) at (0, 0) {$\propto$};
		\node [style=none] (41) at (4, 1) {};
		\node [style=none] (42) at (4, -1) {};
		\node [style=none] (43) at (1.5, -1) {};
		\node [style=none] (44) at (1.5, 1) {};
	\end{pgfonlayer}
	\begin{pgfonlayer}{edgelayer}
		\draw [in=-180, out=15, looseness=0.50] (2) to (6.center);
		\draw [in=165, out=0, looseness=0.75] (1.center) to (2);
		\draw [in=15, out=-165, looseness=0.50] (2) to (4.center);
		\draw [in=-180, out=-15, looseness=0.75] (2) to (5.center);
		\draw (18) to (1.center);
		\draw (17) to (4.center);
		\draw (16) to (5.center);
		\draw (15) to (6.center);
		\draw (24.center) to (18);
		\draw (15) to (21.center);
		\draw (16) to (22.center);
		\draw (17) to (25.center);
		\draw [in=-180, out=15, looseness=0.50] (27) to (31.center);
		\draw [in=165, out=0, looseness=0.75] (26.center) to (27);
		\draw [in=15, out=-165, looseness=0.50] (27) to (29.center);
		\draw [in=-180, out=-15, looseness=0.75] (27) to (30.center);
		\draw (44.center) to (26.center);
		\draw (43.center) to (29.center);
		\draw (42.center) to (30.center);
		\draw (41.center) to (31.center);
	\end{pgfonlayer}
\end{tikzpicture}
        \caption{\centering Color change}\label{fig:color-change}
    \end{subfigure}
    \begin{subfigure}{.32\textwidth}
        \centering
        \begin{tikzpicture}
	\begin{pgfonlayer}{nodelayer}
		\node [style=none] (39) at (0, 0) {$\propto$};
		\node [style=none] (46) at (1, 0) {};
		\node [style=none] (48) at (-1, 0) {};
		\node [style=none] (49) at (-3, 0) {};
		\node [style=hadamard] (50) at (-2, 0) {};
		\node [style=none] (55) at (5, 0) {};
		\node [style=gray phase dot] (57) at (2, 0) {$\frac{\pi}{2}$};
		\node [style=gray phase dot] (58) at (3, 1) {$\frac{-\pi}{2}$};
		\node [style=gray phase dot] (59) at (4, 0) {$\frac{\pi}{2}$};
		\node [style=white dot] (60) at (3, 0) {};
	\end{pgfonlayer}
	\begin{pgfonlayer}{edgelayer}
		\draw (49.center) to (50);
		\draw (50) to (48.center);
		\draw (58) to (60);
		\draw (60) to (57);
		\draw (60) to (59);
		\draw (59) to (55.center);
		\draw (57) to (46.center);
	\end{pgfonlayer}
\end{tikzpicture}
        \caption{\centering Hadamard decomposition}\label{fig:box-decomp}
    \end{subfigure}
    \caption{\centering Some rules of the \zxcalc, where \(\alpha, \beta \in \mathbb{R},k \in \mathbb{N}\). 
        Together, they form a (non-minimal) complete set of rules for the Clifford fragment of \zxdiags~\cite{jeandel2019completeness}. }
    \label{fig:zx-rules}
\end{figure}

The \zxcalc comprises \textit{rewrite rules} that can be used to transform diagrams by replacing a certain subdiagram with another.
A rewrite rule is called \textit{sound} if the matrix semantics of the diagram before the application of the rule are the same as those after the application of the rule. 
Whenever we refer to a rewrite rule, we implicitly mean a rewrite rule which is sound.
\Cref{fig:zx-rules} lists certain rewrite rules of the \zxcalc which are commonly used. 
For each of these rules, there is also a color-swapped version of the rule which changes green spiders to red spiders and red spiders to green spiders.
Note that some of these, such as the Hopf rule (\Cref{fig:hopf}), are usually called derived rules to distinguish them from some set of axioms. 
In \vyzx, all rules are proven sound directly without appeal to axioms, so we make no such distinction.

One of the most critical \zxcalc rules is that spiders connected by an arbitrary (non-zero) number of edges can be fused into a single node with the angles added together.
This rule is shown in \Cref{fig:spider-fusion}. 
Applying this rule twice to \Cref{fig:deform} would yield a diagram with a single red spider. 
The reverse is also true: any spider can be split such that the two new spiders add up to the original angle.
A corollary of this is that spiders with $0$ angle can be split off any input or output, so long as they have the color of the original spider.
Further, with the spider fusion rule, we can manipulate the number of inputs and outputs for all other rules by adding a fusible node to said input or output.

Another simplification is that we can remove self-loops, as shown in \Cref{fig:self-loop-removal}~\cite{vandewetering2020zxcalculus}. 
Intuitively, this rule says that a node cannot provide more information about itself.
%A self-loop is an edge that goes from an edge to itself. \rnr{not needed}
%
The remaining rules are fairly self-explanatory, for an intuitive description of their behavior, see van de Wetering~\cite{vandewetering2020zxcalculus}.

\begin{figure}[h]
    \centering
    \begin{tikzpicture}
	\begin{pgfonlayer}{nodelayer}
		\node [style=none] (4) at (-4, 0.5) {};
		\node [style=none] (5) at (-2, 0.5) {};
		\node [style=gray phase dot] (19) at (-3, 0.5) {$\pi$};
		\node [style=none] (27) at (-3, -1) {Z};
		\node [style=none] (29) at (-1, 0.5) {};
		\node [style=none] (30) at (1, 0.5) {};
		\node [style=none] (31) at (2, 0.5) {};
		\node [style=none] (32) at (5, 0.5) {};
		\node [style=white phase dot] (33) at (0, 0.5) {$\pi$};
		\node [style=none] (34) at (0, -1) {X};
		\node [style=none] (35) at (3.5, -1) {Y};
		\node [style=white phase dot] (36) at (2.75, 0.5) {$\pi$};
		\node [style=gray phase dot] (37) at (4.25, 0.5) {$\pi$};
		\node [style=none] (38) at (6, 0.5) {};
		\node [style=none] (39) at (8, 0.5) {};
		\node [style=gray phase dot] (40) at (7, 0.5) {$\alpha$};
		\node [style=none] (44) at (12, 1) {};
		\node [style=none] (45) at (12, 0) {};
		\node [style=none] (46) at (14, 1) {};
		\node [style=none] (47) at (14, 0) {};
		\node [style=white dot] (48) at (13.5, 0) {};
		\node [style=gray dot] (49) at (12.5, 1) {};
		\node [style=none] (50) at (7, -1) {Rz($\alpha$)};
		\node [style=none] (51) at (13, -1) {CNOT};
		\node [style=none] (58) at (15, 0.5) {};
		\node [style=none] (59) at (19, 0.5) {};
		\node [style=hadamard] (60) at (17, 0.5) {};
		\node [style=none] (61) at (17, -1) {H};
		\node [style=none] (62) at (11, 0.5) {};
		\node [style=none] (63) at (9, 0.5) {};
		\node [style=none] (65) at (10, -1) {T};
		\node [style=gray phase dot] (66) at (10, 0.5) {};
		\node [style=gray phase dot] (67) at (10, 0.5) {$\frac{\pi}{4}$};
	\end{pgfonlayer}
	\begin{pgfonlayer}{edgelayer}
		\draw (32.center) to (37);
		\draw (37) to (36);
		\draw (36) to (31.center);
		\draw (29.center) to (33);
		\draw (33) to (30.center);
		\draw (4.center) to (19);
		\draw (19) to (5.center);
		\draw (38.center) to (40);
		\draw (40) to (39.center);
		\draw (44.center) to (49);
		\draw (49) to (46.center);
		\draw (48) to (47.center);
		\draw (45.center) to (48);
		\draw [in=150, out=-30, looseness=0.75] (49) to (48);
		\draw (58.center) to (60);
		\draw (60) to (59.center);
		\draw (63.center) to (67);
		\draw (67) to (62.center);
	\end{pgfonlayer}
\end{tikzpicture}
    \caption{\centering Common gates represented in the \zxcalc.}\label{fig:zx-gates}
\end{figure}

\subsection{Representing Unitary Gates}\label{sec:gates}

We translate some standard unitary gates from the quantum circuit model into \zxcalc in \Cref{fig:zx-gates}. Z, T, and Rz($\alpha$) represent Z-axis rotations by the given angles, while X and Y rotate qubits by $\pi$ around the X and Y axes. The H (or Hadamard) gate switches between the Z and X bases; following van de Wetering~ \cite[Section 3.6]{vandewetering2020zxcalculus} it can be decomposed into a sequence of red and green nodes. Finally, CNOT applies an X rotation (or NOT) on the second qubit contingent on the first being $\ket{1}$. 
This set of unitary gates corresponds to the RzQ gate set~\cite{nam2018automated}, a universal gate set for quantum circuits (i.e. one capable of implementing any quantum computation).
%Thinking of wires as transporting information between graph nodes, 
% Note that OCM (perhaps counterintuitively) tells us that we can slide the green ``control'' past the red NOT without changing the semantics of the diagram, even though this switches the inputs and outputs of the spiders. 

Having translations from gates into ZX allows us to convert quantum circuits to diagrams.
Once we translate circuits into diagrams, we can use the \zxcalc rules without being bound by the rigid structure of the circuit model.
While several equational theories have recently been proposed for quantum circuits, these are generally more complicated than those of the \zxcalc~\cite{clement2023complete,clement2024minimal}.

Due to its broad applicability to quantum computing, including circuit optimizations~\cite{kissinger2020Pyzx,cowtan2020opt} and error correction~\cite{zxlattice,chancellor2018graphical}, as well as its array of extensions (such as the ZH~\cite{backens2019zh} and ZW~\cite{hadzihasanovic2017zw} calculi), the \zxcalc is a prime target for formal verification. However, achieving this in practice requires representing these graphical structures inductively, which will prove challenging.

\section{Inductive \zxdiags}\label{sec:vyzx-diags}

Given the \zxcalc is a symmetric monoidal category (the interested reader may find more information about the category theoretical foundations in \Cref{sec:inductive}) generated by the Z and X \textit{spiders}, we can naturally translate the \zxcalc into an inductive datatype for \Coq, as shown in \Cref{fig:blockconstructors}.
Note that for convenience we add a box (\(\square\)), which corresponds to the conventional Hadamard box in \zxcalc, which is technically not a core part of the \zxcalc.
We define additional constructs for convenience.
Relevant constructs include \coqe{n_stack1 : forall n : nat, ZX 1 1 -> ZX n n}, written \coqe{n ↑ zx}, which stacks a ZX diagram with a single input and output \(n\) times upwards. 
We also create a shorthand for \(n\) stacked wires called \coqe{n_wire n}.

To assign meaning to our syntactic constructs, we construct a semantic evaluation function \(\semantics{.} : \zx{n}{m} \to \texttt{Matrix } 2^m \times 2^n\);
a \zxdiag with $n$ inputs and $m$ outputs semantically evaluates to a complex-valued matrix of size $2^m$ by $2^n$.
In practice, all our semantics are built using QuantumLib's~\cite{QuantumLib} matrices and complex numbers.
\begin{figure}[b]
    \centering
    \begin{align*}
        \inferrule
        {\oftype{in out}{\(\N\)} \\ \oftype{\(\alpha\)}{\(\R\)}}
        {\oftype{\texttt{Z} in out \(\alpha\)}{\zxold{in}{out}}}
        \hspace{1.5em}
        \inferrule{ }{\oftype{\(\supset\)}{\zxold{0}{2}}}
        \hspace{.75em}&\hspace{.75em}
        \inferrule{ }{\oftype{\(\subset\)}{\zxold{2}{0}}}
        \hspace{1.5em}
        \inferrule
        {\oftype{in out}{\(\N\)} \\ \oftype{\(\alpha\)}{\(\R\)}}
        {\oftype{\texttt{X} in out \(\alpha\)}{\zxold{in}{out}}}
        \\
        \inferrule{ }{\oftype{\(\--\)}{\zxold{1}{1}}}
        \hspace{1.5em}
        \inferrule{ }{\oftype{\(\square\)}{\zxold{1}{1}}}
        \hspace{.75em}&\hspace{.75em}
        \inferrule{ }{\oftype{\(\times\)}{\zxold{2}{2}}}
        \hspace{1.5em}
        \inferrule{ }{\oftype{\(\revemptyset\)}{\zxold{0}{0}}}
        \\
        \inferrule
        {\oftype{\texttt{zx\textsubscript{0}}}{\zxold{in}{mid}} \\ \oftype{\texttt{zx\textsubscript{1}}}{\zxold{mid}{out}}}
        {\oftype{\texttt{zx\textsubscript{0}}  \(\leftrightarrow\) \texttt{zx\textsubscript{1}}}{\zxold{in}{out}}}
        \hspace{.75em}&\hspace{.75em}
        \inferrule
        {\oftype{\texttt{zx\textsubscript{0}}}{\zxold{in\textsubscript{0}}{out\textsubscript{0}}} \\ \oftype{\texttt{zx\textsubscript{1}}}{\zxold{in\textsubscript{1}}{out\textsubscript{1}}}}
        {\oftype{\texttt{zx\textsubscript{0}} \(\updownarrow\) \texttt{zx\textsubscript{1}}}{\zxold{(in\textsubscript{0} + in\textsubscript{1})}{(out\textsubscript{0} + out\textsubscript{1})}}}
    \end{align*}
    \caption{\centering The inductive constructors for block representation \zxdiags}\label{fig:blockconstructors}
\end{figure}

\begin{figure}[b]
    \centering
    \begin{align*}
    \semantics{\(\revemptyset\)} = I_{1 \times 1} \hspace{1.5em} \semantics{\(\--\)} = I_{2 \times 2} \hspace{.75em}&\hspace{.75em} \semantics{\(\square\)} = H \hspace{1.5em} \semantics{\(\subset\)} = [1,0,0,1]^T\\
    \semantics{\(\times\)} = |00\rangle\langle 00|+|11\rangle\langle 11|+|01\rangle\langle 10|+|10\rangle\langle 01| \hspace{.75em}&\hspace{.75em} \semantics{\(\supset\)} = [1,0,0,1] \\
    \semantics{Z n m \(\alpha\)} = \ket{0}^{\otimes m}  \bra{0}^{\otimes n} + e^{i\alpha} \ket{1}^{\otimes m} \bra{1}^{\otimes n} \hspace{.75em}&\hspace{.75em} \semantics{X n m \(\alpha\)} = H^{\otimes m} \times \semantics{Z n m \(\alpha\)} \times H^{\otimes n}\\
    \semantics{zx\textsubscript{0} \(\leftrightarrow\) zx\textsubscript{1}} = \semantics{zx\textsubscript{1}} \times \semantics{zx\textsubscript{0}} \hspace{.75em}&\hspace{.75em} \semantics{zx\textsubscript{0} \(\updownarrow\) zx\textsubscript{1}} = \semantics{zx\textsubscript{0}} \otimes \semantics{zx\textsubscript{1}}
    \end{align*}
    \caption{\centering\vyzx semantics
    %using the notations defined in \Cref{tab:cat-to-ind}
    }
    \label{fig:semantics}
\end{figure}
We implement \semantics{.} as shown in \Cref{fig:semantics}, with the minor difference that instead of generating the matrices for Z spiders through the composition of vectors, we directly build the resulting matrix.
The semantic definition of the X spider is also constructed using the equation seen in \Cref{fig:color-change}.
Some constructs such as the wire ($\--$), Hadamard box ($\square$),  cup ($\subset$), and cap ($\supset$), could be equivalently created out of various compositions of spiders;
however, given that they will appear frequently in definitions and lemmas, it is useful to have them as their own constructors.

We define the equivalence relation \emph{proportionality} to state that two diagrams are equal up to a scalar factor:
\[\forall (\texttt{zx\textsubscript{0}}, \texttt{zx\textsubscript{1}} : \zx{n}{m}), \texttt{zx\textsubscript{0}} \propto \texttt{zx\textsubscript{1}} \coloneqq \exists c \in \C, \semantics{zx\textsubscript{0}} = c \cdot \semantics{zx\textsubscript{1}} \wedge c \neq 0\]
We show within \Coq that this is an equivalence relation and that \texttt{Stack} (\(\updownarrow\)) and \texttt{Compose} (\(\leftrightarrow\)) are \emph{parametric morphisms}~\cite{SozeauGeneralizedRewriteCoq}, meaning that we can safely rewrite using proportionality within \zxdiags.
Using the definition of proportionality, we can then prove facts about the \zxcalc.

This proportionality judgment is sufficient for most of our use-cases, such as for optimizing unitary circuits~\cite{kissinger2020Pyzx} (see also \cref{sec:circuits}).
For some applications, however, such as reasoning about probabilistic quantum circuits, it is important to track the values of the associated constants.
To facilitate those applications, we define refinements of implicit proportionality (\texttt{\(\propto\)}) with explicit proportionality constants, and prove each lemma with the strictest possible relation. 
We define explicit proportionality (\coqe{\propto[c]}) denoting proportionality by $c$, which must be nonzero:
\[\forall (\texttt{zx\textsubscript{0}}, \texttt{zx\textsubscript{1}} : \zx{n}{m})\ (\texttt{c} : C), \texttt{zx\textsubscript{0}} \propto\!\texttt{[c] }\texttt{zx\textsubscript{1}} \coloneqq \semantics{zx\textsubscript{0}} = c \cdot \semantics{zx\textsubscript{1}} \wedge c \neq 0\]
Using this relation, the explicit constant is captured.
We also define a stricter version \texttt{\(\propto\)=}, which captures semantic equality, i.e., \texttt{\(\propto\)[1]}:
\[\forall (\texttt{zx\textsubscript{0}}, \texttt{zx\textsubscript{1}} : \zx{n}{m}), \texttt{zx\textsubscript{0}} \propto= \texttt{zx\textsubscript{1}} \coloneqq \semantics{zx\textsubscript{0}} = \semantics{zx\textsubscript{1}} \]
In practice, most results hold up to strict semantic equality.
Moreover, we can add a scalar gadget to diagrams to convert statements of \texttt{\(\propto\)[c]} to statements of \texttt{\(\propto\)=} following the technique in van de Wetering~\cite[Section 3.4]{vandewetering2020zxcalculus}.
Concretely, the notation \texttt{to\_gadget} uses tactics to convert a lemma ending in a statement $ \texttt{zx\textsubscript{0}} \propto\!\!\texttt{[c] }\texttt{zx\textsubscript{1}} $ to a lemma ending $\texttt{zx\textsubscript{0}} \propto= \texttt{zx\_of\_const c} \updownarrow \texttt{zx\textsubscript{1}}$, where $\texttt{zx\_of\_const c}$ is a gadget with semantics \coqe{[c] : Matrix 1 1}, a 1 by 1 matrix whose only element is $c$.
This allows us to reason with only \texttt{\(\propto\)=} when we want to prove strict equality of \zxdiags up to their semantic interpretations.
By using the subrelation feature of \Coq's setoid rewriting, the stricter proportionality relations \texttt{\(\propto\)=} and \texttt{\(\propto\)[c]} can be directly rewritten in weaker contexts, allowing us to consistently state lemmas in their strictest forms without loss of functionality.
% TODO: Do we want to say something like this?
For simplicity, we state lemmas in this paper using \texttt{\(\propto\)}, while in the development they are stated in their stricter forms, either \texttt{\(\propto\)=} or \texttt{\(\propto\)[c]}.
In addition to reasoning about proportionality, we often have to reason about the composition of \zxdiags.
Here, a common challenge in dependently typed programming shows up, as we require precise equality of dimensions across proportionality and the composition constructor.
While the proof of the associativity of \texttt{Compose} is trivial, we encounter an issue with \texttt{Stack}.
Consider the diagrams \(\texttt{zx\textsubscript{0}} : \zx{n}{m}, \texttt{zx\textsubscript{1}} : \zx{n'}{m'}, \texttt{zx\textsubscript{2}} : \zx{n''}{m''}\).
Here the diagrams \((\texttt{zx\textsubscript{0}} \updownarrow \texttt{zx\textsubscript{1}}) \updownarrow \texttt{zx\textsubscript{2}} \) and \(\texttt{zx\textsubscript{0}} \updownarrow (\texttt{zx\textsubscript{1}} \updownarrow \texttt{zx\textsubscript{2}}) \) have the incompatible types \zx{(n+n')+n''}{(m+m')+m''} and \zx{n+(n'+n'')}{m+(m'+m'')}, respectively, as we see from the typing rules in \Cref{fig:blockconstructors}.
To bridge this gap, we define a \coqe{cast} function with the following type: 
\begin{coq}    
cast (n m:nat) {n' m':nat} (pfn: n=n') (pfm: m=m') (zx: ZX n' m') : ZX n m.
\end{coq}%
\texttt{cast} uses a proof that \(n = n'\) and \(m = m'\) to change the explicit dimensions of a \zxdiag.
Casting functions have been used in similar contexts, like the Lean matrix library mathlib3's \coqe{matrix.reindex}~\cite{leanmathlib2020} or Rocq vector library's \coqe{cast}~\cite{Rocq}.
Note that \Coq can infer \(n'\) and \(m'\) of \coqe{cast}.
Casts introduce more implicit structure into our already highly structured diagrams.
For example, in proving associativity, we can decide which side of the proportionality judgment is cast to the other's dimensions.
In the implementation, we define numerous lemmas to move casts through our structure under type restrictions appropriately.
In practice, it turns out that the added structure and required moves are the only proof overhead caused.
Potential obligations from creating/manipulating casts can commonly be resolved using a linear arithmetic solver such as \Coq's \coqe{lia}~\cite{besson2007lia}.

With proof irrelevance for natural number equality (see Hedberg's theorem~\cite{hedbergthm}), we then in theory can rewrite \texttt{cast}s without looking at their specific proof terms, as proofs of natural number equality are interchangeable.
Unfortunately, in their current states, \Coq's \coqe{rewrite} and \coqe{setoid_rewrite} tactics cannot perform such rewrites, even if the proofs are provably irrelevant.
However, \vyzx handles most straightforward cases of cast automatically, as we discuss in \Cref{sec:proof-auto}.

We show the \Coq code for stack associativity as an example illustrating the use of \coqe{cast} in \Cref{lst:stack-assoc}.
Here, \texttt{cast} serves to unify the dimensions of the proportionality statement.
Notice that while we could provide an exact proof to cast (\texttt{Nat.add\_assoc}), we instead keep the proof parameterized.
This circumvents the limitations of the \Coq rewrite engine with respect to proof irrelevance by proving the lemma for an arbitrary proof, allowing it to be rewritten both forward and backward in any context.
If used in forward rewrites, having parameterized proofs creates additional obligations, which can either be immediately resolved with automation or deferred to the end, where all remaining obligations generated by cast operations can be resolved using a linear arithmetic solver.
Since our lemmas are parametric over equality proofs, we can backwards rewrite with any arbitrary proof.

For key results, it is prudent to restate the theorem with explicit rather than parametric proof lemmas, exhibiting a witness for the \coqe{cast} equalities. 
These explicit lemmas can be trivially solved using automation that uses the parametric proof with an arithmetic solver.
For example, let us look at \Cref{lst:stack-assoc}.
We can define an explicit version shown in \Cref{lst:stack-assoc-exists}, which also demonstrates the tactic to solve these kinds of problems.
Note that the statements in \Cref{lst:stack-assoc,lst:stack-assoc-exists} are almost identical, except that the latter states the existence of proofs for \texttt{cast}.
Therefore, our approach allows us to provide a lightweight, \Coq rewrite engine compatible, encapsulation of proof irrelevance at the cost of automatically resolvable goals.
\begin{figure}[b]
\begin{lstlisting}[caption={Example of the use of \texttt{cast}: Proving the associativity of stack. Found in \texttt{src/CoreRules/StackRules.v}}, label={lst:stack-assoc}]
Lemma stack_assoc : forall {n0 n1 n2 m0 m1 m2 : nat} 
(zx0 : ZX n0 m0) (zx1 : ZX n1 m1) (zx2 : ZX n2 m2) prfn prfm,
(zx0 ↕ zx1) ↕ zx2 ∝ cast _ _ prfn prfm (zx0 ↕ (zx1 ↕ zx2)). 
\end{lstlisting}
\end{figure}

\begin{figure}[b]
\begin{lstlisting}[caption={Example of the explicit \texttt{cast}: Proving the associativity of stack. Not found in repository.}, label={lst:stack-assoc-exists}]
Global Tactic Notation "crush_explicit_cast" constr(lemma) := 
    unshelve (do 2 eexists; apply lemma); lia.

Lemma stack_assoc_non_param :  forall {n0 n1 n2 m0 m1 m2} 
(zx0 : ZX n0 m0) (zx1 : ZX n1 m1) (zx2 : ZX n2 m2), exists prfn prfm,
(zx0 ↕ zx1) ↕ zx2 ∝ cast _ _ prfn prfm (zx0 ↕ (zx1 ↕ zx2)).
Proof.
    crush_explicit_cast stack_assoc.
Qed.

\end{lstlisting}
\end{figure}

\section{\vizx}\label{sec:vizx}

With all the structure we introduce for \zxdiags in \vyzx, our textual, inductive representation of \zxdiags can be deeply nested and tricky to parse.
To address this, we exploit the inherently visual nature of our diagrams to provide a human-readable diagrammatic representation using \vizx.
\begin{figure}
    \centering
    \begin{subfigure}{0.75\textwidth}
    \begin{subfigure}[t]{.18\textwidth}
        \centering
        \includegraphics[width=0.8\linewidth]{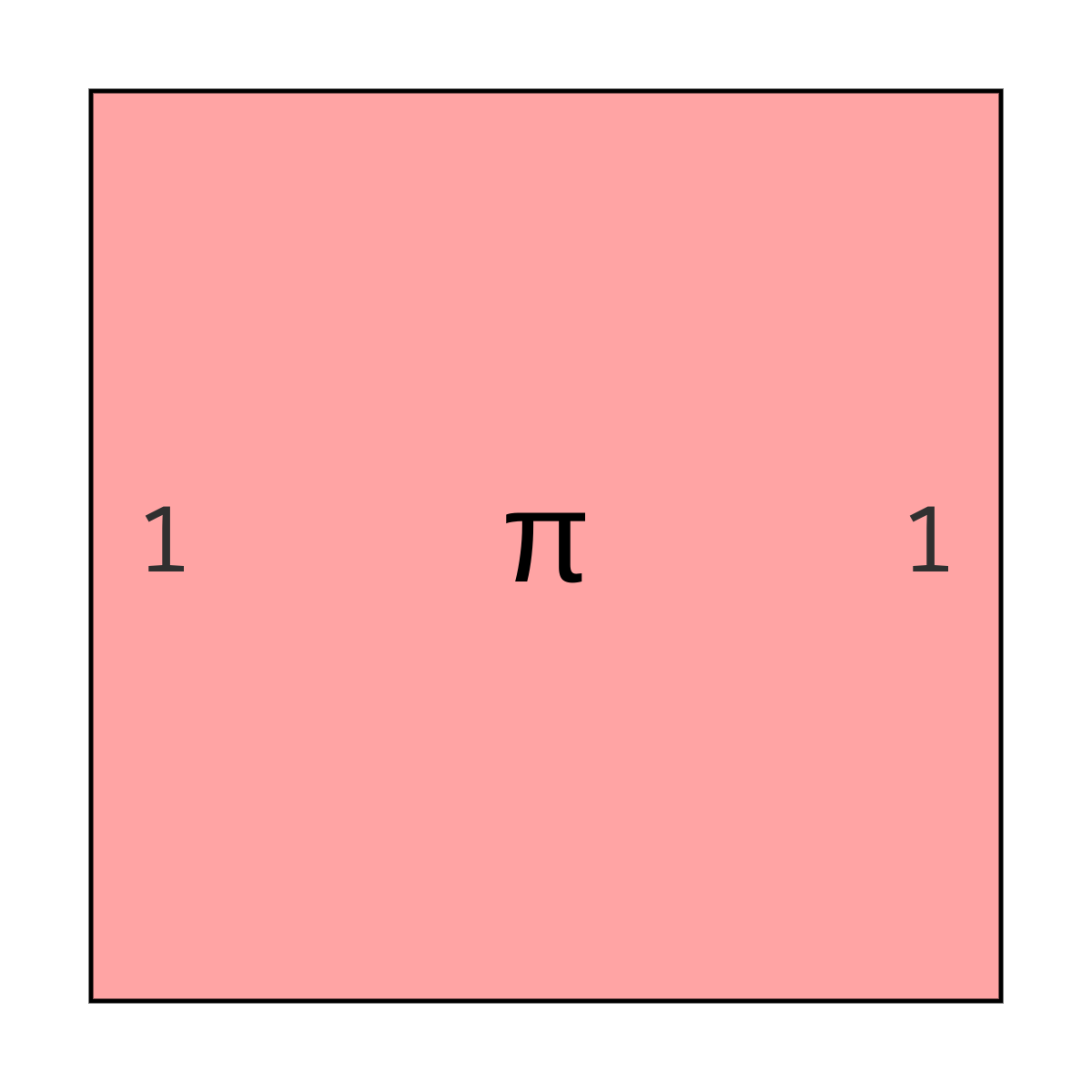}
        \caption{\centering A $\pi$ X spider.}
        \label{fig:vizx-x-spider-long}
    \end{subfigure}
    \begin{subfigure}[t]{.29\textwidth}
        \centering
        \includegraphics[width=0.8\linewidth]{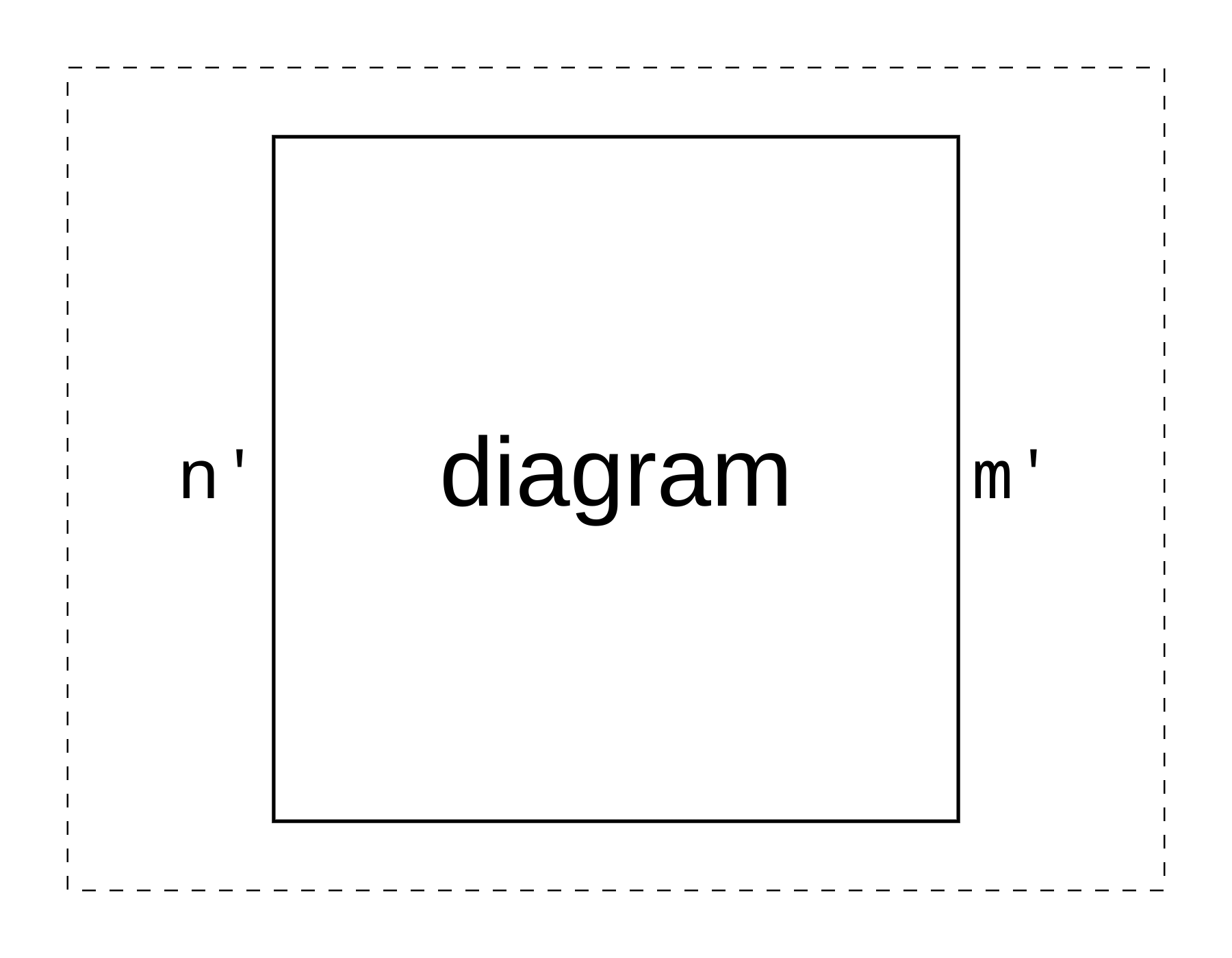}
        \caption{\centering A diagram cast to \coqe{n'} inputs and \coqe{m'} outputs.}
        \label{fig:vizx-cast}
    \end{subfigure}
    \begin{subfigure}[t]{.33\textwidth}
        \centering
            \includegraphics[width=0.8\linewidth]{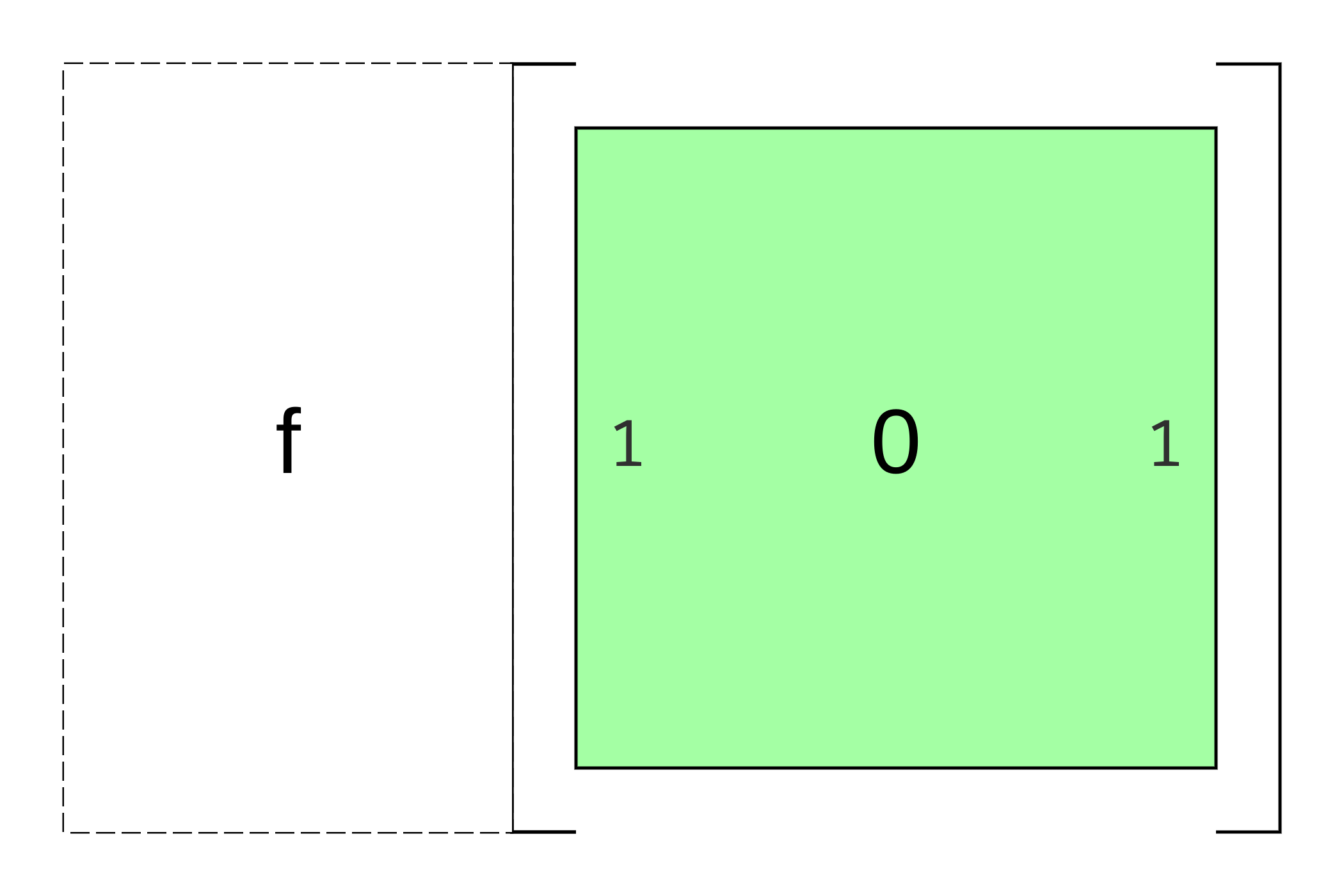}
            \caption{\centering A function \coqe{f} applied to a Z spider.}
            \label{fig:vizx-arbitrary-function}
    \end{subfigure}
    \begin{subfigure}[t]{.28\textwidth}
            \centering
            \includegraphics[width=0.5\linewidth]{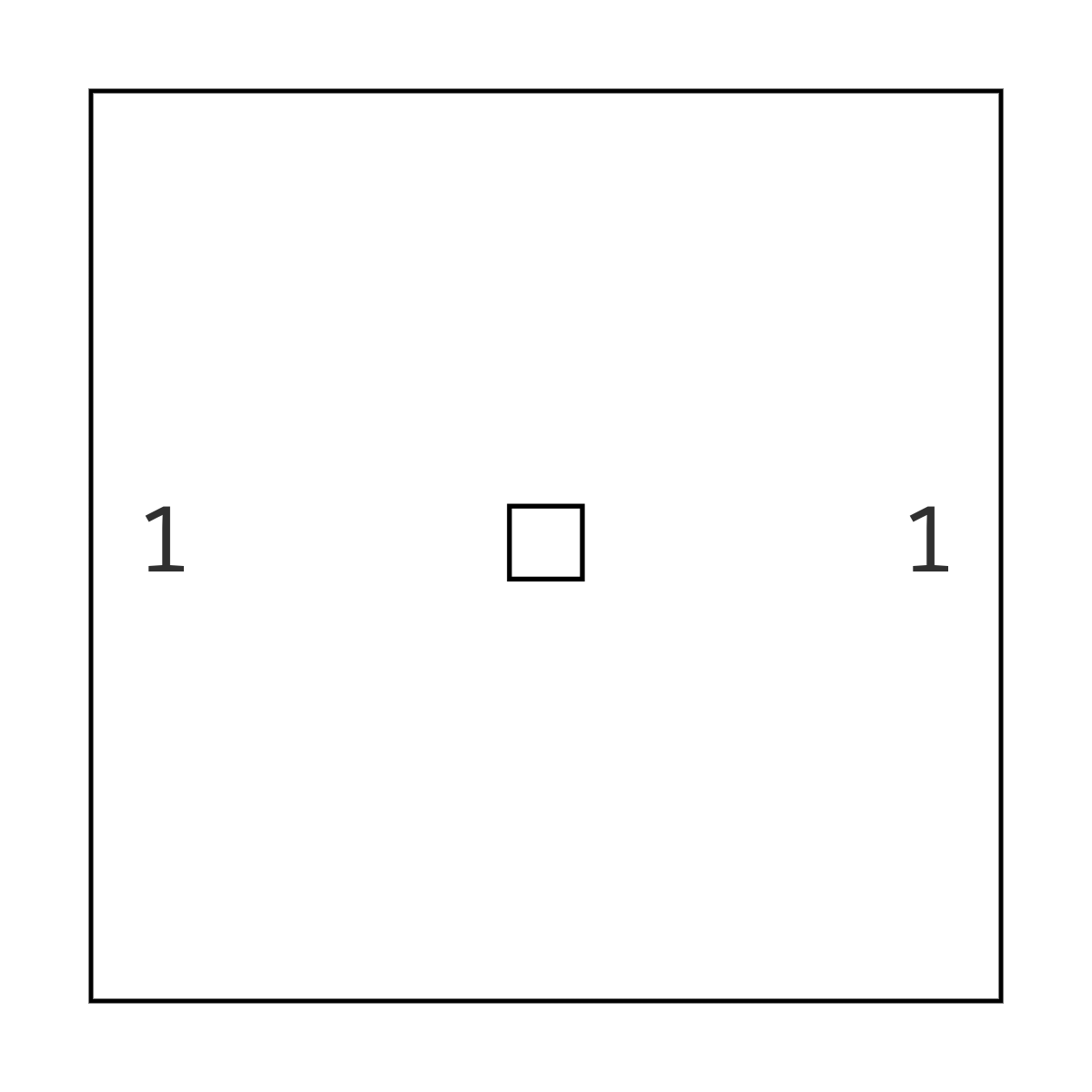}
            \caption{\centering Hadamard box.}
            \label{fig:vizx-box}
    \end{subfigure}\hspace{0.8cm}
    \begin{subfigure}[t]{.2\textwidth}
            \centering
            \includegraphics[width=0.75\linewidth]{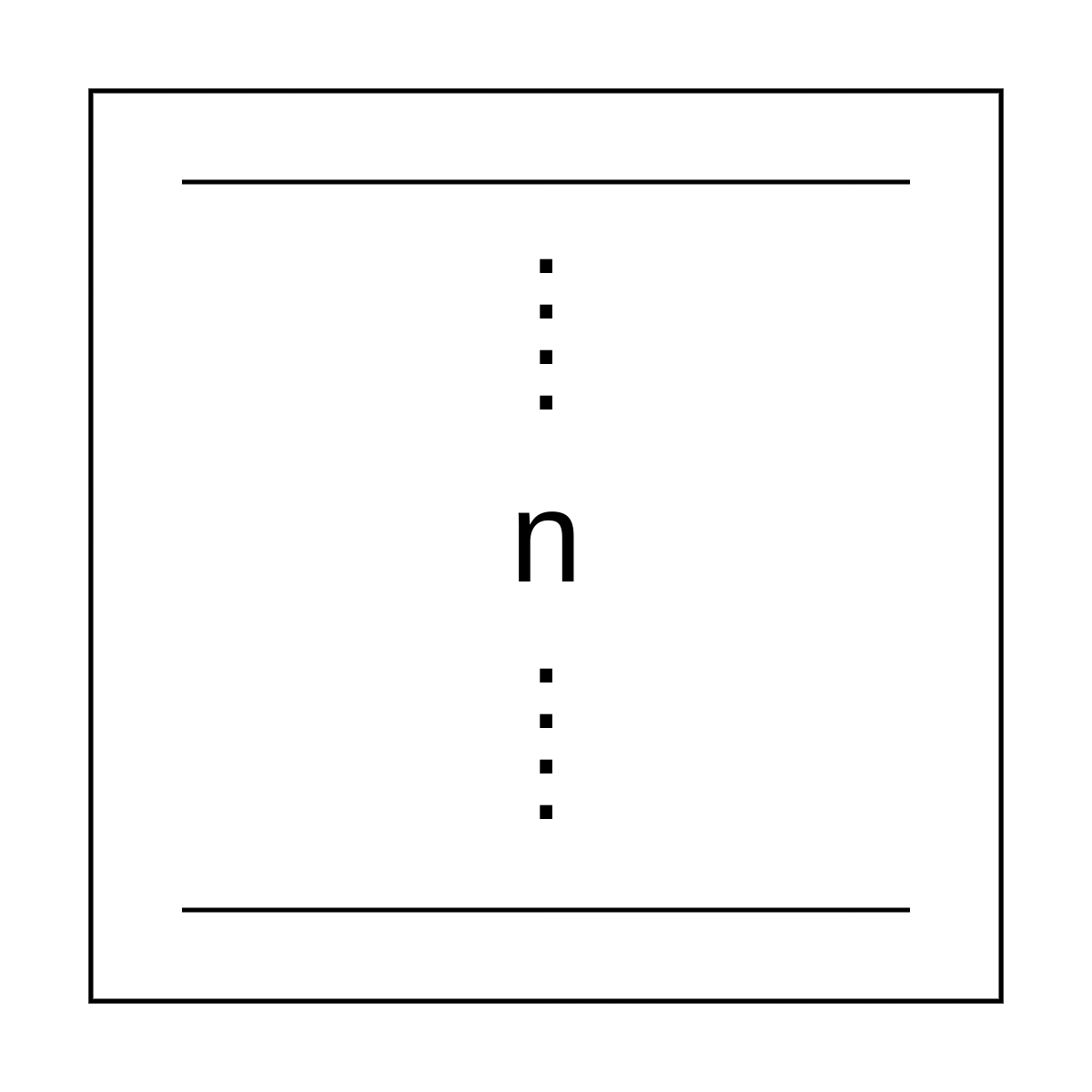}
            \caption{\centering \coqe{n_wire n}}
            \label{fig:vizx-n-wire}\hspace{0.8cm}
    \end{subfigure}\hspace{0.8cm}
    \begin{subfigure}[t]{0.30\textwidth}
        \centering
        \includegraphics[width=\linewidth]{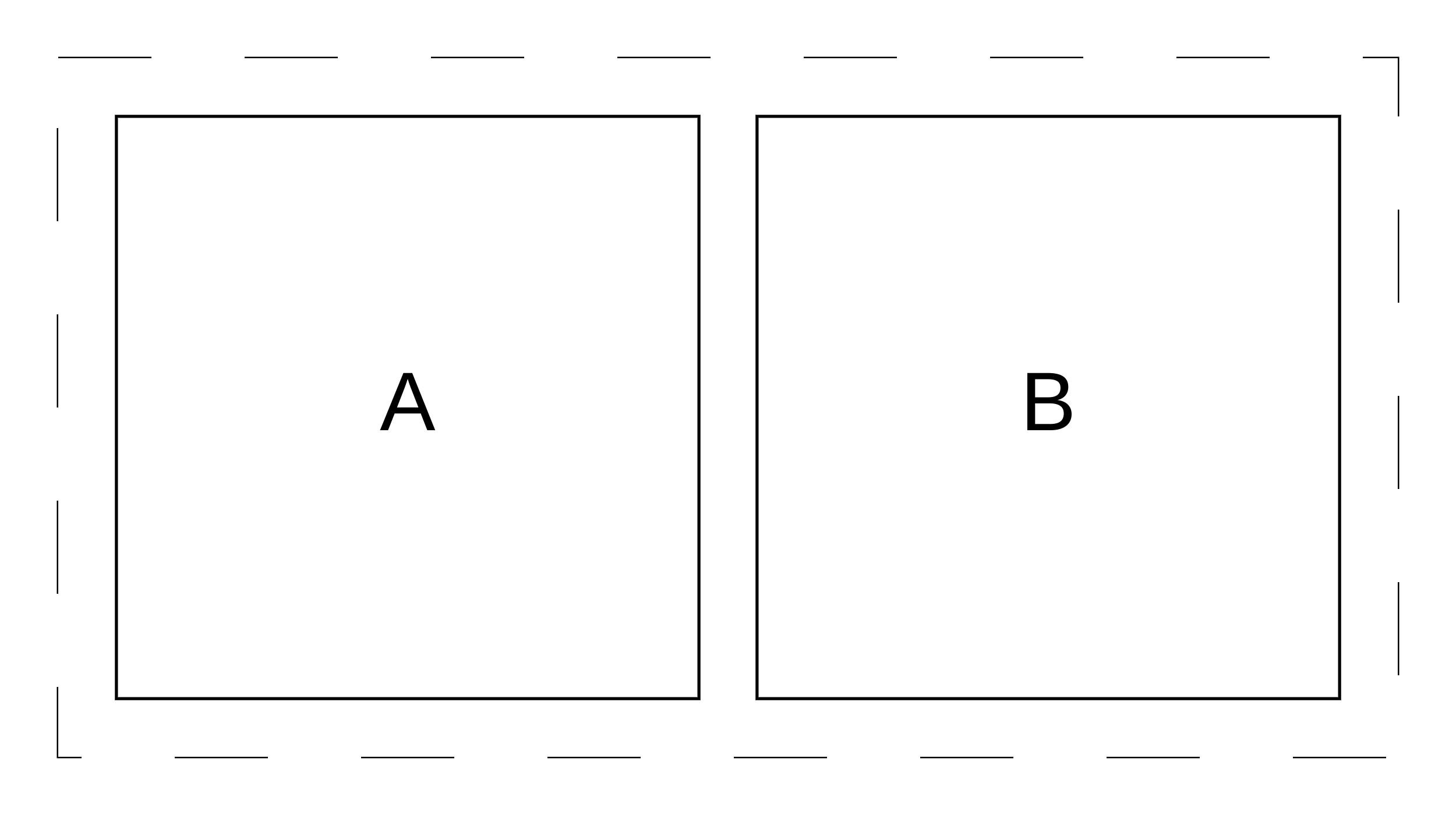}
        \caption{\centering A composed with B.}
        \label{fig:vizx-compose}
    \end{subfigure}
    
    \end{subfigure}    
    \begin{subfigure}[t]{.14\textwidth}
        \centering
        \includegraphics[width=\linewidth]{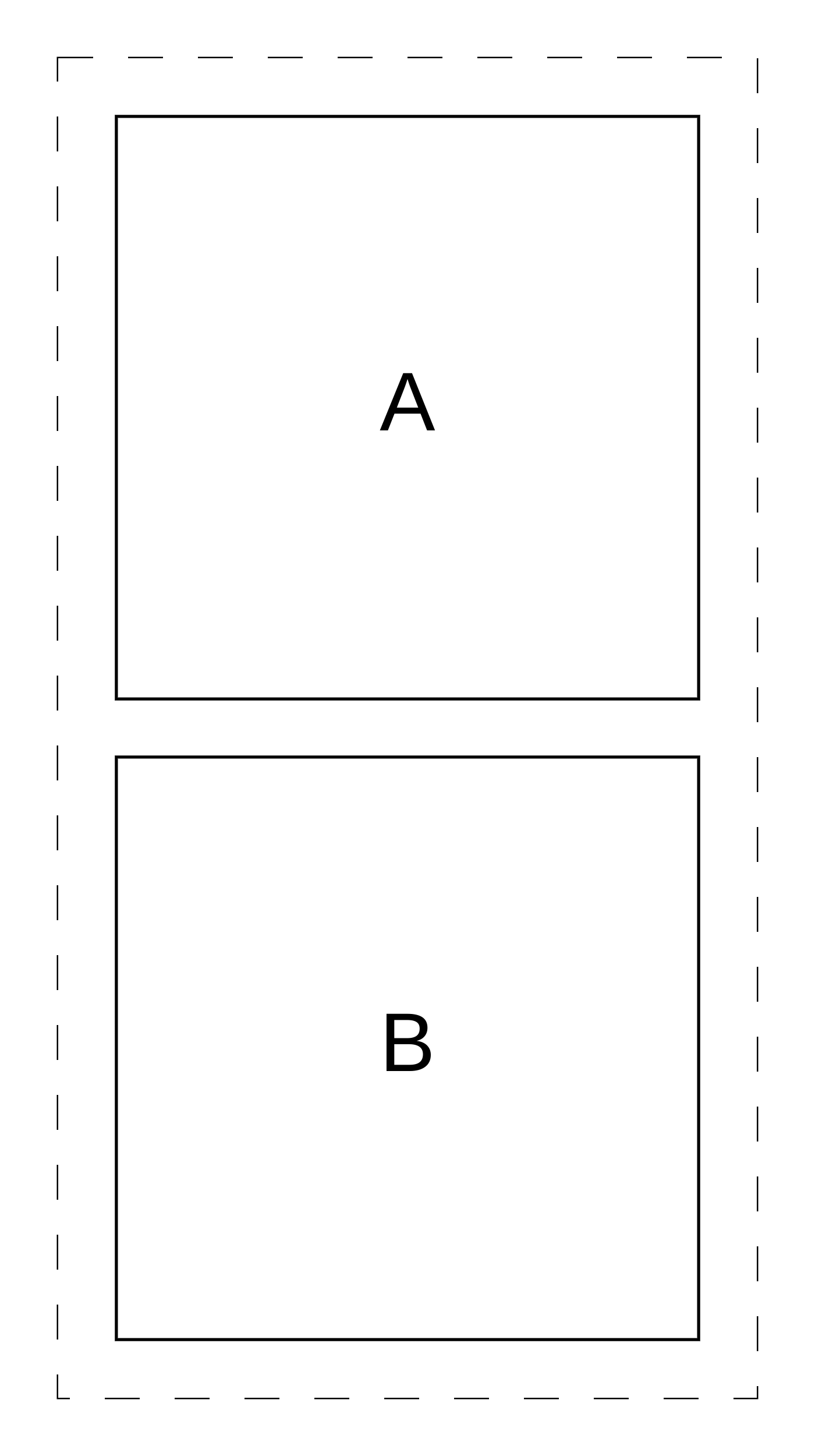}
        \caption{\centering A stacked with B.}
        \label{fig:vizx-stack}
    \end{subfigure}    
    \caption{\vizx visualization of \vyzx constructors and functions, as in \Cref{sec:vyzx-diags}.}
    \label{fig:vizx-constructs}
\end{figure}
The block structure is a key feature that distinguishes \vizx diagrams from standard visualizations of \zxdiags. 
\zxdiags are built around the principle of connectivity, which is useful for pen and paper reasoning but obfuscates details about associativity at the semantic level.
\vyzx proofs manipulate the associativity and subdiagrams of the block form rather than just the connectivity of the graphical form. 
Furthermore, parametric properties such as the number of inputs and outputs for casts must be kept intact in the visualizations, something the standard representation does not make room for.

\Cref{fig:vizx-constructs} shows how we visualize constructs from \Cref{sec:vyzx}.
To avoid ambiguity related to associativity information, we also enclose the stacked, composed, or cast diagrams within a dashed boundary.
The styles for the boundaries are slightly varied by construct to improve readability when multiple dashed lines are side-by-side.

\vyzx diagrams are frequently more complex than just the single elements explored above. 
In \Cref{sec:prf-structural-example}, we show a graphical proof that uses \zxcalc rewriting and \vizx to show that the standard circuit for constructing a Bell pair is equivalent to a cup. 

Visualization makes it much simpler to interpret the structure of terms.
The visualization clarifies how the subterms fit into the larger associativity structure.
We discuss how this aids proof engineering practice in \Cref{sec:vizx-eng}. 
As terms get more complex, the visualizations significantly enhance the user's experience with parsing the proof goal. 
Using color for the \coqe{Z} and \coqe{X} spiders
%
%Color 
gives us another layer of visual communication that text cannot provide.
Associativity can be challenging to understand from the term string, but the visual element makes it easy to parse.
Additionally, it is often hard to tell whether subterms are adjacent (up to associativity) in a term string, particularly when they appear in nested horizontal and vertical compositions. 
In contrast, the visual representation clearly presents the positional relationship between subterms, while also helping guide the user in reassociating the goal to directly expose adjacent terms.

First, the input \vyzx proof state is lexed and parsed.
We explicitly keep all notations and syntactic sugar during this process to stay faithful to the user's textual goal state. 
%Arbitrary \vyzx proof states produced by Coq can be parsed. 
% 
Next, the parsed term and all nested sub-terms are assigned graphical objects, which \vizx places on an HTML canvas according to a user-provided scale.
The rendering happens in a two-pass system. First, each element is assigned a base position and size based on stack and composition information. 
In the event of a mismatch in the sizes of composed elements, the second pass takes note of this discrepancy and increases the size of the smaller element.
An example of this can be seen in \Cref{fig:grow-z}.
In this example, we initially determine the native size of all elements.
All elements except the rightmost spider also keep this size through the second pass, as they are not stacked or composed with any element that has a larger minimum size.
The rightmost spider, however, is composed with the stack of two base-size elements, so the second pass increases its vertical size to make it visually coherent.
We handle text by appropriately scaling and wrapping it when necessary.
The term is then rendered using the determined layout, with additional visual components including the formatted text and color. 
Finally, this canvas is displayed as a webview in the VSCode user interface and rendered as a panel alongside the code and goals.

\section{Verifying a Complete Equational Theory}\label{sec:vyzx}

\begin{figure}[b]
    \begin{subfigure}{0.5\textwidth}
        \centering
        \includegraphics[width=\textwidth]{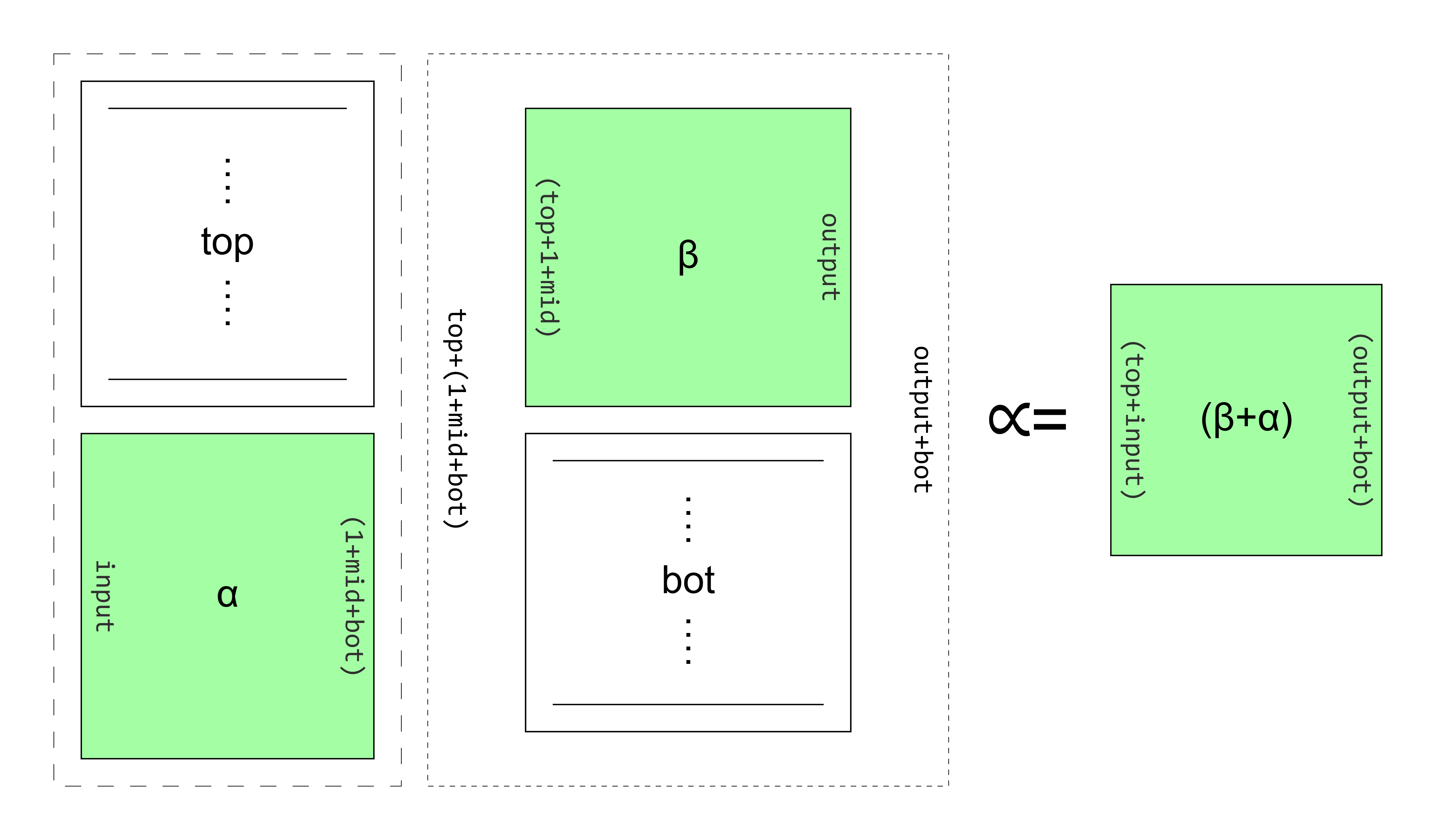}
        \caption{Spider fusion in \vyzx}\label{fig:vizx-spider-fusion}
    \end{subfigure}\begin{subfigure}{0.5\textwidth}
        \centering
        \includegraphics[width=\textwidth]{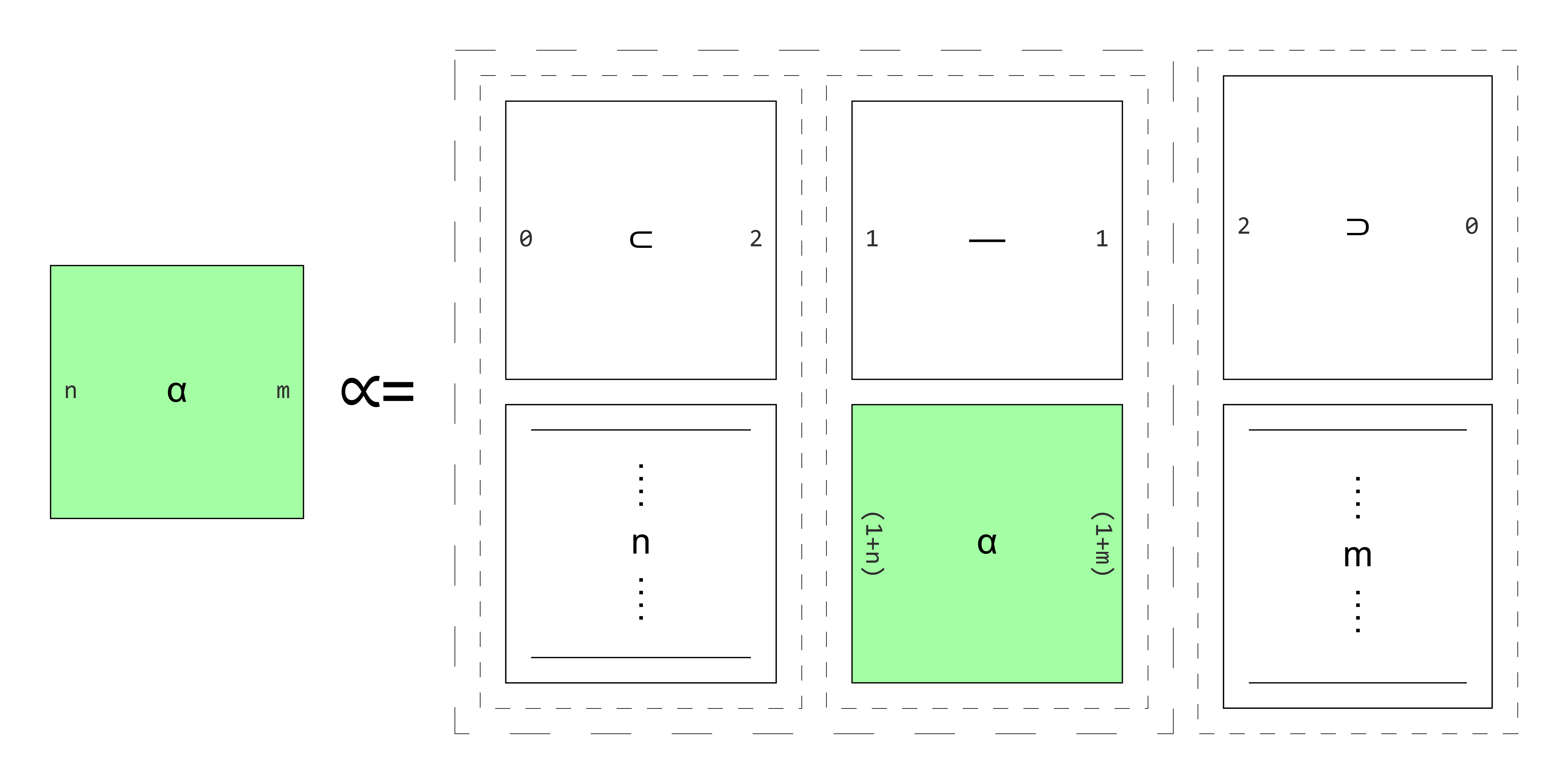}
        \caption{Self-loop removal in \vyzx}
        \label{fig:vizx-self-loop}
    \end{subfigure}
        \caption{Select rules from \vyzx}\label{fig:selec-rules-vizx}
\end{figure}

To encode the \zxcalc, we need both diagrams and rules.
\Cref{sec:vyzx-diags} described how we can encode any \zxdiag in \Coq.
Here, we show how to encode the canonical set of rules from \Cref{sec:zx} using the visual representation generated by \vizx. 

\subsubsection*{Only connectivity matters}\label{sec:vyzx-ocm}
\begin{figure}
    \centering
    \begin{subfigure}[t]{.48\textwidth}
        \centering
        \includegraphics[width=\textwidth]{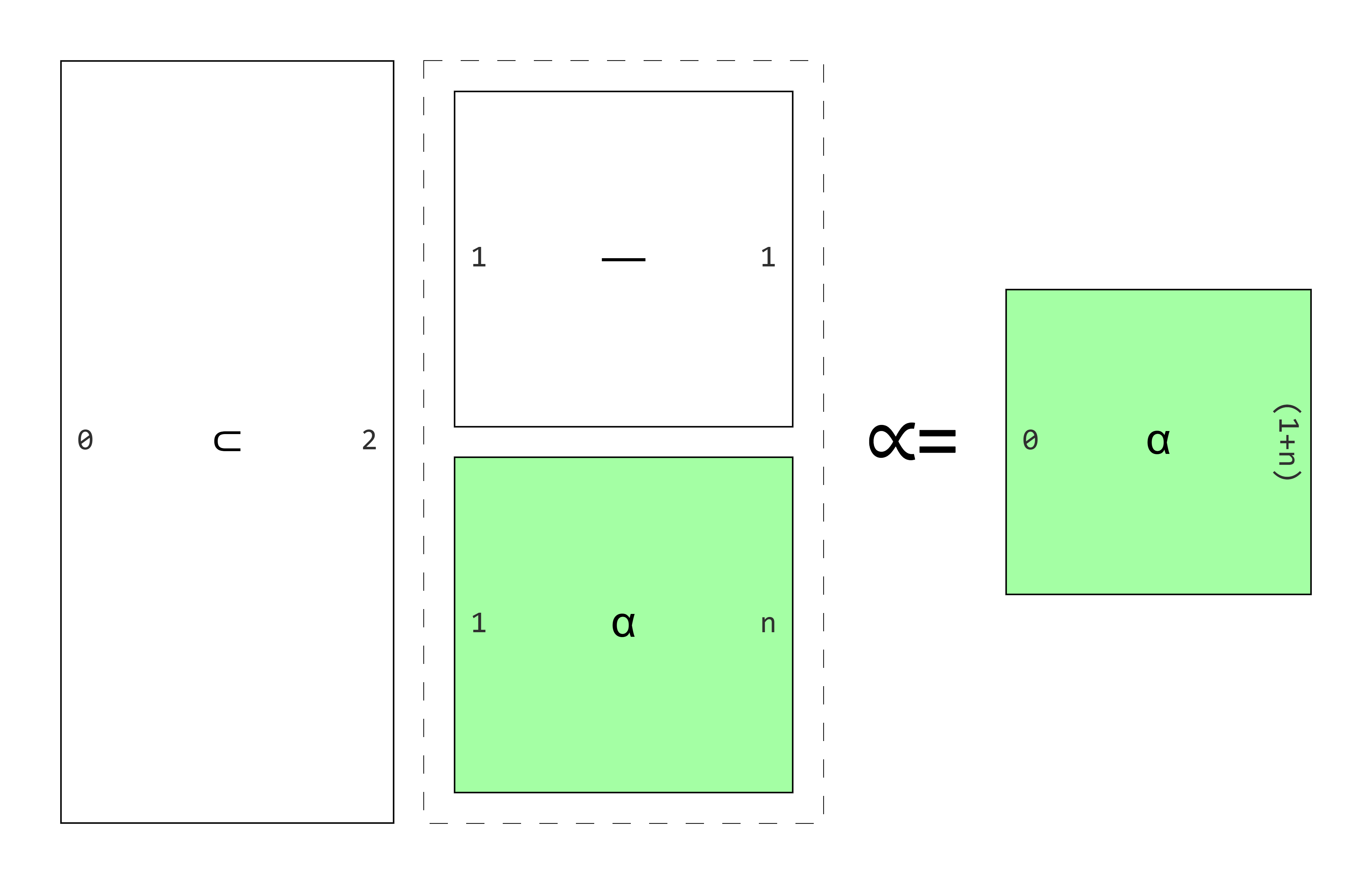}
        \caption{\centering The wrap over top left lemma used to express only connectivity matters}
        \label{fig:vizx-wrap-over}
    \end{subfigure}
    \begin{subfigure}[t]{.48\textwidth}
        \centering
        \includegraphics[width=\textwidth]{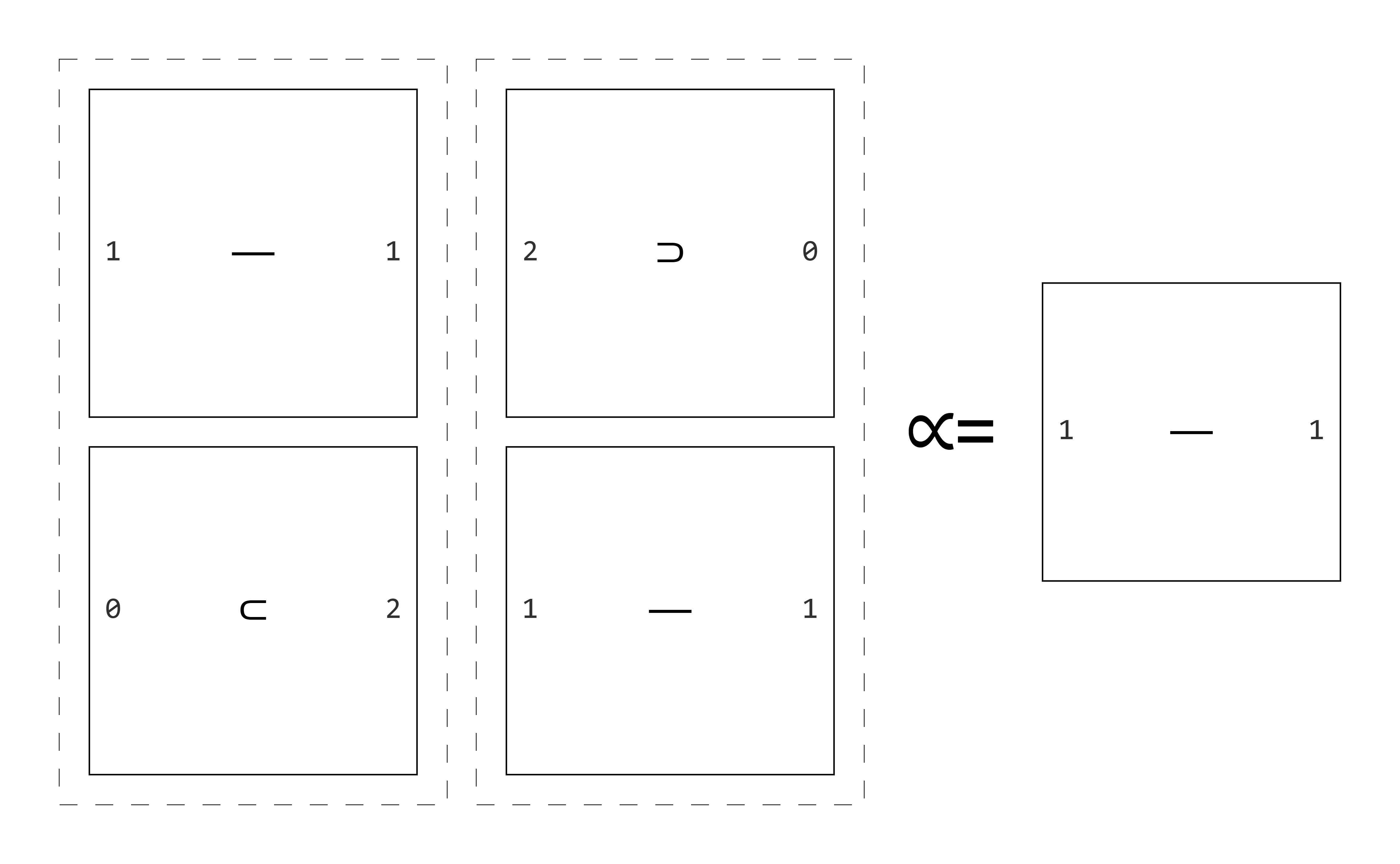}
        \caption{\centering A yanking lemma used to simplify connectivity information.}
        \label{fig:vizx-yank}
    \end{subfigure}
    \caption{\centering Visualizing two ``only connectivity matters'' lemmas in \vyzx}
    \label{fig:vizx-ocm}
\end{figure}

The idea behind ``only connectivity matters'' is that we can %freely bend wires at will.
arbitrarily deform a diagram without changing its semantics.
The ability to deform diagrams is nicely generated by a small set of axioms~\cite[Section 2.2]{jeandel2018complete}.
Of the ``only connectivity matters'' rules given by Jeandel et al., one significant rule allows the free conversion between inputs and outputs.
We provide wrapping lemmas like the one shown in \Cref{fig:vizx-wrap-over} (as a string diagram in \Cref{fig:ocm-rules}) to make this possible within our block structure.
Likewise, the yanking equations (see \Cref{fig:vizx-yank} and \Cref{fig:ocm-rules}) allow us to remove redundant cups and caps from our diagrams. 
By proving the lemmas given by Jeandel et al., we encapsulate ``only connectivity matters'' directly rather than assuming it by using a datatype (such as an adjacency matrix) that only explicitly encodes connection information.

\subsubsection*{Spider fusion}

Spider fusion is perhaps the most interesting and important rule in the \zxcalc.
We implement it in \vyzx by explicitly encoding the number of inputs and outputs of each spider and the connections between the two spiders.
Notice that the dimensions in \Cref{fig:vizx-spider-fusion} across the cast are \coqe{top + (1 + mid) + bot} and \coqe{top + (1 + mid + bot)} respectively. 
To unify this discrepancy, we must cast (see \Cref{sec:vyzx-diags}) one side of the composition.
This rule also highlights that we can express \vyzx rules in multiple ways.
For example, we assume that the left node in the fusion expression is above the wires and the right node is below the wires. 
There are, however, many combinations of different locations for the rules.
Further, we could cast either the left or right side. 
For the interested reader, in \Cref{sec:prf-structural-example}, we show the proof engineering implications to deal with such structural challenges. 

\subsubsection*{Self-loop removal}

\Cref{fig:vizx-self-loop} shows self-loop removal in \vyzx. 
We express self-loop removal with the restriction that the self-loop has come from a given spider's top inputs/outputs.
However, using swaps constructors, we can swap any output of a spider to become the uppermost.
We can then remove swaps fully connected to a spider using other lemmas.

\subsubsection*{Other rules}
The \emph{Hopf rule}, \emph{Bi-algebra rule} and \emph{state-copy-rule} can be directly stated in \vyzx cast-free, as shown in \Cref{fig:vizx-hopf,fig:vizx-state-copy}

\subsubsection*{Transpose \& Adjoint}

As we can see, \vyzx rules are often restricted because they require certain relative positioning.
However, using \vyzx, we can rearrange the structure if necessary or prove lemmas with different positioning of components (e.g., for \Cref{fig:vizx-spider-fusion} we also have a version where we have the left green spider on the bottom and the right green spider on the top).
We can automate these changes using the transpose, adjoint, and color swap operations.

All told, VyZX implements Jeandel et al.'s \cite{jeandel2018diagrammatic} complete set of rewrite rules for the \zxcalc.
This allows us to prove two diagrams equivalent without having to rely on the matrix semantics or any matrix-level manipulations. 
The full set of rewrite rules is given in \vizx form in \Cref{sec:equational-theory}.

\section{Proving Universality}
\label{sec:universality}

It is well known that \zxdiags are universal for linear maps: We can construct a \zxdiag with $n$ inputs and $m$ outputs corresponding to any complex-valued $2^m$-by-$2^n$ matrix.
Universality means \zxdiags can encode any quantum process and, by completeness (see \Cref{sec:vyzx}), that the \zxcalc can prove the equality between of any two equivalent processes.
The standard proof of universality begins by showing that \zxdiags are universal for quantum computation, i.e. unitaries, by creating an arbitrary state, and using process-state duality to get any arbitrary matrix \cite{coecke2017picturing} \cite{vandewetering2020zxcalculus}. 
We take a different approach to universality by first showing that \zxdiags can represent any scalar, and then using an inductively defined addition of diagrams (following \cite{Jeandel2024AdditionZX}) to form linear combinations of basis elements. 
This strategy allows us to prove only facts about the \zxcalc, instead of quantum programs, while also introducing the separately useful constructs of addition and scaling of diagrams.
% Although universality can be used to indirectly define these (or any other) operations, inductive definitions benefit by reflecting the structure of the input diagrams in the output diagram \cite{Jeandel2024AdditionZX}.
In practice, proofs of universality can produce exponentially-large diagrams which are not particularly useful. 
Our proof is the same, as it forms a linear combination of the exponential-size basis. 

If all the entries of a matrix have preferred, explicit representations, our procedure could be very easily adapted to use those representations to scale the basis vectors instead of using the generic, often very messy gadgets generated by scalar universality.
Explicit representations can be given for many common subrings of $\mathbb{C}$, such as $\mathbb{Z}[\frac{1}{\sqrt2},i]$, which corresponds to the universal Clifford+T gate set for quantum computing~\cite{Giles_2013}.
% However, if all the entries of a matrix can be constructively represented by gadgets, our proof can be adapted trivially to give an entirely explicit diagram, though still exponential in size.
% This applies to matrices over many common subrings of $\mathbb{C}$, such as $\mathbb{Z}[\frac{1}{\sqrt2},i]$, which corresponds to the universal Clifford+T gate set for quantum computing~\cite{Giles_2013}.

% The gadget scale factors of that linear combination are in general very messy as diagrams because they come from a fully-generic function converting any complex number to a gadget. 
% If we restrict to matrices with entries in a subring of $\mathbb{C}$ whose elements (or simply generators) have clean expressions as diagrams, these scalars can instead be given explicit gadgets which may be simpler.
% The diagrams produced this way would still be exponential, but would not rely on operations on the real numbers which do not evaluate automatically.

\subsection{Scalar Universality}

The simplest case of universality is for matrices $A : \mathbb{C}^{2^0}\to\mathbb{C}^{2^0}$, which correspond directly to scalars in $\mathbb{C}$. 
In this subsection, we will refer to a 1-by-1 matrix and its sole entry interchangeably. 
We want to represent any complex number $z$ by a gadget \coqe{g : ZX 0 0} such that $\sem{\texttt{g}} = z$.
This is the algorithm that is used for the function \coqe{zx_of_const} mentioned in \Cref{sec:vyzx-diags}.
Note that gadgets can be easily multiplied by composing them together (either horizontally or vertically), so we can construct gadgets factor-by-factor.

\begin{figure}[t]
    \centering
    \begin{subfigure}[b]{.30\textwidth}
        \centering
        \begin{tikzpicture}
	\begin{pgfonlayer}{nodelayer}
		\node [style=gray phase dot] (0) at (-1, 0) {};
		\node [style=white phase dot] (1) at (1, 0) {$\pi$};
		\node [style=none] (2) at (-1, 1) {};
		\node [style=none] (3) at (1, -0.5) {};
	\end{pgfonlayer}
	\begin{pgfonlayer}{edgelayer}
		\draw (0) to (1);
	\end{pgfonlayer}
\end{tikzpicture}
        \caption{\centering A gadget representing $\sqrt{2}$}
        \label{fig:zx-sqrt2}
    \end{subfigure}
    \begin{subfigure}[b]{.30\textwidth}
        \centering
        \begin{tikzpicture}
	\begin{pgfonlayer}{nodelayer}
		\node [style=gray phase dot] (0) at (0, 0) {};
		\node [style=gray phase dot] (1) at (0, 0) {$\alpha$};

		\node [style=none] (2) at (-1, 1) {};
		\node [style=none] (3) at (1, -0.5) {};
	\end{pgfonlayer}
\end{tikzpicture}
        \caption{\centering A gadget representing $1+e^{i \alpha}$}
        \label{fig:zx-1-exp}
    \end{subfigure}
    \begin{subfigure}[b]{.30\textwidth}
        \centering
        \begin{tikzpicture}
	\begin{pgfonlayer}{nodelayer}
		\node [style=white phase dot] (0) at (-1, 1) {$\alpha$};
		\node [style=white phase dot] (1) at (-1, -0.5) {};
		\node [style=gray phase dot] (2) at (1, 1) {$\pi$};
		\node [style=gray phase dot] (3) at (1, -0.5) {};
	\end{pgfonlayer}
	\begin{pgfonlayer}{edgelayer}
		\draw (0) to (2);
		\draw (1) to (3);
		\draw [bend right, looseness=1.25] (1) to (3);
		\draw [bend left, looseness=1.25] (1) to (3);
	\end{pgfonlayer}
\end{tikzpicture}
        \caption{\centering A gadget representing $e^{i \alpha}$}
        \label{fig:zx-exp}
    \end{subfigure}
    \caption{\centering Gadgets representing basic scalars}
    \label{fig:zx-gadgets}
\end{figure}   

% Proving scalar universality is relatively direct. 
First, we construct a gadget representing $\sqrt{2}$ (\Cref{fig:zx-sqrt2}).
By combining copies of this gadget, we then have gadgets representing $\sqrt{2}^n$ for any natural $n$. 
Moreover, the empty Z-spider with phase $\alpha$ represents $1 + e^{i \alpha}$, whose magnitude ranges between $0$ and $2$, inclusive (\Cref{fig:zx-1-exp}).
Together, these give gadgets with any given magnitude. 
We then only need gadgets for each phase, for which we use a gadget representing $e^{i \beta}$ (\Cref{fig:zx-exp}). 
All together, given a scalar $z$, we pick a large enough $n$ that $\|z / \sqrt{2}^{\ n}\| \le 2$ (for instance, $n = \lfloor\log_{\sqrt{2}}\frac{z}{2}\rfloor+1$ suffices). 
Denoting $z / \sqrt{2}^{\ n}$ by $w$, we find $\alpha$ with $\|1+e^{i \alpha}\| = \|w\|$ ($\alpha = \arccos\left(\frac{\|w\|^2}{2}-1\right)$ works). Then $w = (1 + e^{i \alpha}) \cdot e^{\left(i \arg \frac{w}{1+e^{i \alpha}} \right)}$, so 
\[ z = \sqrt{2}^{\ n} \cdot (1 + e^{i \alpha}) \cdot e^{\left(i \arg \frac{w}{1+e^{i \alpha}} \right)}.\]
Composing the gadgets representing each of these factors gives a gadget representing $z$, as desired.

% (wjbs) I don't think this paragraph adds very much, but I'll leave it here in case anyone disagrees

% Gadgets are particularly easy to reason about because they commute with horizontal and vertical composition of diagrams, so all gadgets can be pulled to the outermost part of a block diagram.
% In light of Only Connectivity Matters, this represents the diagrammatic fact that a gadget is disconnected from the rest of the diagram, so can be pulled to the outside.
% Universality for scalars amounts to a function \coqe{zx_of_const : C -> ZX 0 0} assigning to each constant $z$ a gadget with semantic $z$. 
% Then, we can define a function \coqe{zx_scale} which scales a diagram \coqe{zx} by \coqe{z} as \coqe{zx_of_const z ↕ zx}, which we denote \coqe{z .* zx}.
% These scalar factors pass through horizontal and vertical composition, so can be easily moved to the outermost position using proof automation we provide (\coqe{distribute_zxscale}). 
% When working up to proportionality, rather than semantic equality, non-zero scale factors can be absorbed, meaning \coqe{zx0 \propto z .* zx1 <-> zx0 \propto zx1} if \coqe{z <> 0}.

\subsection{Summing Diagrams}

The key step to proving universality is the following theorem.

\begin{theorem}[Summation]
    Given any two \zxdiags ${ZX}_1$ and ${ZX}_2$ (of the same dimensions), we can construct a diagram ${ZX}$ such that
     \[ \sem{ZX} = \sem{ZX_1} + \sem{ZX_2} \]
\end{theorem}
This is Theorem 4.8 in Jeandel et al.~\cite{Jeandel2024AdditionZX}, and our formalization directly follows their proof.
We point the reader there for the details, giving only a sketch here. 
The key concept of their inductive definition of addition is the notion of a \textit{controlizer} of a diagram.
Controlizers generalize the notion of a controlled state by converting diagrams to states using the \choijam isomorphism, also known as process-state duality (or simply, `wrapping around the input/output wires').
Up to a global scalar, a controlizer of a diagram is just a controlled state representing the diagram, under process-state duality. 
Crucially, these controlizers can be added diagrammatically. 
So, if we can construct controlizers for two diagrams, we can add them together. 
Again following \cite{Jeandel2024AdditionZX}, we define a controlizer for every \zxdiag inductively in terms of the block structure, so that we can add any two diagrams together. 
This allows us to inductively construct the sum of any two diagrams, and then arbitrary finite sums.
% Doing so, we obtain inductive sums of \zxdiags which are constructive (and effective\footnote{Though in practice we never evaluate the addition of diagrams as they are very large, the process is entirely effective, even with non-constructive real number phases}).

\subsection{Full Universality}

Having proven universality for scalars and defined the addition of diagrams, it is simple to prove universality for all matrices.
For any scalar $c$ and \zxdiag $zx$, we can form the scaled diagram $c \cdot zx$ by stacking onto $zx$ a gadget representing $c$.
By the \choijam isomorphism, if we can create any state, we can create any process. 
A state is a vector in the space $\mathbb{C}^{2^n}$, which has a basis given by $\left\{ |x\rangle : x\in{\{0,1\}}^n \right\}$, where
\[|(x_1,\dots,x_n)\rangle=|x_1\rangle\otimes\cdots\otimes|x_n\rangle.\] 
The states $|0\rangle$ and $|1\rangle$ are represented by X-spiders with phases $0$ and $\pi$, respectively, each scaled by $2^{-1/2}$, so by stacking these scaled spiders we obtain a basis for $\mathbb{C}^{2^n}$ represented by \zxdiags; we denote these diagrams by $|x\rangle$ as well.
Then, a state $(v_1,\dots,v_{2^n})\in\mathbb{C}^{2^n}$ is represented by the diagram $\sum_{x\in{\{0,1\}}^n} v_{x}\cdot|x\rangle$ (identifying a bitstring $x$ with its numerical value).
We can also provide a translation that directly represents a matrix $A\in\mathbb{C}^{2^n\times 2^m}$ as $\sum_{i\in{\{0,1\}}^n,\ j\in{\{0,1\}}^m} {A_{i,j} \cdot |i\rangle\leftrightarrow\langle j|}$, defining $\langle j|$ to be the diagrammatic transpose of $|j\rangle$, essentially skipping the previous step. 
% \to do{Add a figure for a diagram of a simple matrix. Maybe full diagram for hadamard. Maybe Y?}

% There are several approaches to proving universality, such as by proving universality of a gate set. 
% We chose instead to prove it by constructing sums of \zxdiags, following \cite{Jeandel2024AdditionZX}, as this is more closely related to the \zxcalc. 
% To prove universality in this way, we first show we can construct sums of arbitrary \zxdiags, meaning diagrams whose semantics are the entry-wise sum of the semantics of the summand diagrams. 
% Then, we construct the standard basis for $2^m$-by-$2^n$ matrices as \zxdiags and sum scaled copies of these basis elements to get the final diagram.
% Writing $M_{i,j}$ for the matrix with $1$ at entry $i,j$ and $0$ elsewhere, we essentially use the identity $A = \sum_{i,j} A_{i,j} \cdot M_{i,j}$, so that we need only construct the $M_{i,j}$ as \zxdiags, which straightforwardly reduces to constructing the standard basis vectors $e_i$, which reduces to constructing the states $|0\rangle$ and $|1\rangle$, each of which is a scaled Z-spider. 

\section{Reasoning about Circuits with \vyzx}
\label{sec:circuits}

\subsubsection*{Ingesting Quantum Circuits}\label{sec:ingest}

Quantum programs are usually represented using the circuit model~\cite{deutsch1989circuits,nielsen2010}.
\sqir~\cite{hietala2021sqir} embeds quantum circuits in \Coq, with \qlib matrix semantics~\cite{QuantumLib}.
Fundamentally, its representation of circuits is an inductive structure that allows circuit components with \(n\) qubits (i.e., inputs and outputs) to be created.
Quantum circuits in general work on a qubit-based model, circuits and all sub-circuits that are horizontally composed are constant size and induce square matrices.
Circuits comprise a composition of components of size \(n\).
Each component either has no gate, a one-qubit gate operating on qubit \(q < n\), or a two-qubit gate operating on qubits \(p, q < n\) where \(p \neq q\).
We immediately see that this model differs substantially from \vyzx's model: diagrams have \(n\) inputs and \(m\) outputs for any $n$ or $m$.
Moreover, \sqir circuit components explicitly specify gate locations instead of stacking diagrams.

\begin{figure}[h]
    \begin{subfigure}[t]{.48\textwidth}
        \begin{subfigure}[b]{\textwidth}
            \centering
            ~\\[15pt]
            \begin{tikzpicture}
	\begin{pgfonlayer}{nodelayer}
		\node [style=none] (5) at (-2, 0) {};
		\node [style=none] (6) at (2, 0) {};
		\node [style=none] (7) at (-2, 1) {};
		\node [style=none] (8) at (2, 1) {};
		\node [style=none] (9) at (-2, 3) {};
		\node [style=none] (10) at (2, 3) {};
		\node [style=none] (11) at (-2, -1) {};
		\node [style=none] (12) at (2, -1) {};
		\node [style=none] (13) at (-2, -3) {};
		\node [style=none] (14) at (2, -3) {};
		\node [style=none] (15) at (0, 2.25) {$\vdots$};
		\node [style=none] (16) at (0, -1.75) {$\vdots$};
		\node [style=hadamard] (18) at (0, 0) {Gate};
		\node [style=none] (19) at (-3, 3) {1};
		\node [style=none] (20) at (-3, 1) {$q-1$};
		\node [style=none] (21) at (-3, 0) {$q$};
		\node [style=none] (22) at (-3, -1) {$q+1$};
		\node [style=none] (23) at (-3, -3) {$n$};
	\end{pgfonlayer}
	\begin{pgfonlayer}{edgelayer}
		\draw (7.center) to (8.center);
		\draw (10.center) to (9.center);
		\draw (11.center) to (12.center);
		\draw (13.center) to (14.center);
		\draw (5.center) to (18);
		\draw (18) to (6.center);
	\end{pgfonlayer}
\end{tikzpicture}
            \caption{\centering Ingestion of a one qubit gate acting on qubit \(q\) into a \zxdiag}\label{fig:ingest-1-q}
        \end{subfigure}
    \end{subfigure}
    \begin{subfigure}[t]{.48\textwidth}
        \begin{subfigure}[b]{\textwidth}
            \centering
            \scalebox{0.75}{\begin{tikzpicture}
	\begin{pgfonlayer}{nodelayer}
		\node [style=none] (1) at (-5, 1) {};
		\node [style=none] (2) at (-5, 0) {};
		\node [style=none] (4) at (-5, -4) {};
		\node [style=none] (5) at (-3, -4) {};
		\node [style=none] (7) at (-3, 0) {};
		\node [style=none] (8) at (-3, 1) {};
		\node [style=none] (10) at (-3, 3) {};
		\node [style=none] (11) at (-5, 3) {};
		\node [style=none] (12) at (-1, 0) {};
		\node [style=none] (13) at (-1, 1) {};
		\node [style=none] (15) at (-1, 3) {};
		\node [style=none] (16) at (-1, -4) {};
		\node [style=none] (17) at (1, -4) {};
		\node [style=none] (18) at (1, 1) {};
		\node [style=none] (19) at (1, 0) {};
		\node [style=none] (21) at (1, 3) {};
		\node [style=none] (25) at (1, 2) {};
		\node [style=none] (26) at (1, 2) {};
		\node [style=none] (27) at (1, 2) {.};
		\node [style=none] (28) at (1, 1.75) {};
		\node [style=none] (29) at (1, 1.75) {.};
		\node [style=none] (30) at (1, 2.25) {};
		\node [style=none] (31) at (1, 2.25) {};
		\node [style=none] (32) at (1, 2.25) {.};
		\node [style=none] (33) at (-5, 2) {};
		\node [style=none] (34) at (-5, 2) {};
		\node [style=none] (35) at (-5, 2) {.};
		\node [style=none] (36) at (-5, 1.75) {};
		\node [style=none] (37) at (-5, 1.75) {.};
		\node [style=none] (38) at (-5, 2.25) {};
		\node [style=none] (39) at (-5, 2.25) {};
		\node [style=none] (40) at (-5, 2.25) {.};
		\node [style=none] (41) at (-5, -1) {};
		\node [style=none] (42) at (-5, -3) {};
		\node [style=none] (43) at (-3, -3) {};
		\node [style=none] (44) at (-1, -3) {};
		\node [style=none] (45) at (1, -3) {};
		\node [style=none] (46) at (1, -1) {};
		\node [style=none] (47) at (-1, -1) {};
		\node [style=none] (48) at (-3, -1) {};
		\node [style=none] (49) at (-5, -7) {};
		\node [style=none] (50) at (-3, -7) {};
		\node [style=none] (51) at (-3, -5) {};
		\node [style=none] (52) at (-5, -5) {};
		\node [style=none] (53) at (-1, -7) {};
		\node [style=none] (54) at (-1, -5) {};
		\node [style=none] (55) at (1, -7) {};
		\node [style=none] (56) at (1, -5) {};
		\node [style=none] (57) at (1, -6) {};
		\node [style=none] (58) at (1, -6) {};
		\node [style=none] (59) at (1, -6) {.};
		\node [style=none] (60) at (1, -6.25) {};
		\node [style=none] (61) at (1, -6.25) {.};
		\node [style=none] (62) at (1, -5.75) {};
		\node [style=none] (63) at (1, -5.75) {};
		\node [style=none] (64) at (1, -5.75) {.};
		\node [style=none] (65) at (-5, -6) {};
		\node [style=none] (66) at (-5, -6) {};
		\node [style=none] (67) at (-5, -6) {.};
		\node [style=none] (68) at (-5, -6.25) {};
		\node [style=none] (69) at (-5, -6.25) {.};
		\node [style=none] (70) at (-5, -5.75) {};
		\node [style=none] (71) at (-5, -5.75) {};
		\node [style=none] (72) at (-5, -5.75) {.};
		\node [style=none] (73) at (1, -2) {};
		\node [style=none] (74) at (1, -2) {};
		\node [style=none] (75) at (1, -2) {.};
		\node [style=none] (76) at (1, -2.25) {};
		\node [style=none] (77) at (1, -2.25) {.};
		\node [style=none] (78) at (1, -1.75) {};
		\node [style=none] (79) at (1, -1.75) {};
		\node [style=none] (80) at (1, -1.75) {.};
		\node [style=none] (81) at (-5, -2) {};
		\node [style=none] (82) at (-5, -2) {};
		\node [style=none] (83) at (-5, -2) {.};
		\node [style=none] (84) at (-5, -2.25) {};
		\node [style=none] (85) at (-5, -2.25) {.};
		\node [style=none] (86) at (-5, -1.75) {};
		\node [style=none] (87) at (-5, -1.75) {};
		\node [style=none] (88) at (-5, -1.75) {.};
		\node [style=gray dot] (89) at (-2, 1) {};
		\node [style=white dot] (90) at (-2, 0) {};
		\node [style=none] (91) at (2, 3) {};
		\node [style=none] (92) at (2, 1) {};
		\node [style=none] (93) at (2, 0) {};
		\node [style=none] (94) at (2, -1) {};
		\node [style=none] (95) at (2, -3) {};
		\node [style=none] (96) at (2, -4) {};
		\node [style=none] (97) at (2, -5) {};
		\node [style=none] (98) at (2, -7) {};
		\node [style=none] (99) at (-6, -7) {};
		\node [style=none] (100) at (-6, -5) {};
		\node [style=none] (101) at (-6, -4) {};
		\node [style=none] (102) at (-6, -3) {};
		\node [style=none] (103) at (-6, -1) {};
		\node [style=none] (104) at (-6, 0) {};
		\node [style=none] (105) at (-6, 1) {};
		\node [style=none] (106) at (-6, 3) {};
		\node [style=none] (107) at (-7, 1) {$p$};
		\node [style=none] (108) at (-7, 0) {$p+1$};
		\node [style=none] (109) at (-7, 3) {$1$};
		\node [style=none] (110) at (-7, -4) {$q$};
		\node [style=none] (111) at (-7, -3) {$q-1$};
		\node [style=none] (112) at (-7, -1) {$p+2$};
		\node [style=none] (113) at (-7, -5) {$q+1$};
		\node [style=none] (114) at (-7, -7) {$n$};
	\end{pgfonlayer}
	\begin{pgfonlayer}{edgelayer}
		\draw [in=-180, out=0, looseness=0.75] (4.center) to (7.center);
		\draw [in=180, out=0, looseness=0.75] (2.center) to (5.center);
		\draw (1.center) to (8.center);
		\draw (10.center) to (11.center);
		\draw (15.center) to (10.center);
		\draw (13.center) to (18.center);
		\draw (21.center) to (15.center);
		\draw (41.center) to (48.center);
		\draw (48.center) to (47.center);
		\draw (47.center) to (46.center);
		\draw (45.center) to (42.center);
		\draw [in=180, out=0, looseness=0.75] (16.center) to (19.center);
		\draw [in=180, out=0, looseness=0.75] (12.center) to (17.center);
		\draw (49.center) to (50.center);
		\draw (51.center) to (52.center);
		\draw (54.center) to (51.center);
		\draw (50.center) to (53.center);
		\draw (53.center) to (55.center);
		\draw (56.center) to (54.center);
		\draw (8.center) to (89);
		\draw (89) to (13.center);
		\draw (12.center) to (90);
		\draw (90) to (7.center);
		\draw (90) to (89);
		\draw (5.center) to (16.center);
		\draw (11.center) to (106.center);
		\draw (21.center) to (91.center);
		\draw (18.center) to (92.center);
		\draw (19.center) to (93.center);
		\draw (46.center) to (94.center);
		\draw (45.center) to (95.center);
		\draw (17.center) to (96.center);
		\draw (97.center) to (56.center);
		\draw (98.center) to (55.center);
		\draw (49.center) to (99.center);
		\draw (52.center) to (100.center);
		\draw (42.center) to (102.center);
		\draw (101.center) to (4.center);
		\draw (41.center) to (103.center);
		\draw (104.center) to (2.center);
		\draw (105.center) to (1.center);
	\end{pgfonlayer}
\end{tikzpicture}}
            \caption{\centering Constructing CNOT gate on qubits \(p\) and \(q\) (where $p<q$) for a circuit with \(n\) qubits)}\label{fig:ingest-cnot}
        \end{subfigure}
    \end{subfigure}
    \caption{\centering Construction of \sqir operations in \vyzx}
\end{figure}

Ingesting circuits is the process of converting \sqir circuits into \vyzx diagrams.
In the following, we describe our \Coq 
formalization of \sqir circuit ingestion in \vyzx.
Performing this transformation for a circuit with \(n\) qubits, we build components of type \zx{n}{n} for each gate application and then compose them appropriately.

WLOG, we can assume that 1-qubit gates are Hadamard, Pauli X, or Rz gates with rotation \(\alpha\), and 2-qubit gates are CNOT gates between two qubits\footnote{We can make this assumption as \voqc can verifiably translate any \sqir circuit into the gate set that only contains these terms. This gate set is commonly referred to as the RzQ gate set~\cite{hietala2021voqc,nam2018automated}.}. 
For the case of 1-qubit gates, we define constructions \coqe{ZX_H : ZX 1 1}, \coqe{ZX_X : ZX 1 1}, and \coqe{ZX_Rz : \RR -> ZX 1 1} that translate gates to the corresponding \zxdiags as shown in \Cref{sec:gates}.
To include explicit circuit size, we define padding functions, to add an arbitrary number of wires to the bottom or top of a diagram.
We then use these functions to define a function that produces diagrams of the specified height.
For a gate operating on qubit \(q\), we pad our \zx{1}{1} circuit with \(q - 1\) wires above and \(n - q\) wires below.
\Cref{fig:ingest-1-q} shows the result of said transformation. 
To translate CNOT gates, we need to be more thoughtful. 
Since we can't easily connect two arbitrary qubits in \vyzx, we first consider CNOTs on adjacent qubits.
We see that in such situations, we can use the approach for single-qubit gates to correctly place the CNOT using a slight modification of the construction shown in \Cref{fig:ingest-1-q}.
To allow for CNOT gates that do not operate on adjacent qubits, we swap the qubit with the higher index to be next to the qubit with the lower index.
Notice, this is not trivially accomplished with the basic swap operations, as they merely swap adjacent qubits.
Unlike the adjacent swaps, these swaps are not basic constructs of our representation; we need to define them.
An arbitrary swap swaps the first and \(n\textsuperscript{th}\) qubits, requires chaining of \(2n - 1\) basic swaps\footnote{The construction works by moving the qubit 1 down to \(n\) and then moving qubit \(n-1\) up to qubit \(1\).}.
Using arbitrary swaps and shifts, we can now interpret any wire crossing.
Hence, in block representation, we can construct a CNOT acting on two qubits \(p,q\) by swapping one qubit next to the other qubit, applying the CNOT, and swapping back, as shown in \Cref{fig:ingest-cnot}.
We can also construct single qubit gates as previously described.
Composition of \sqir terms is represented by composition in our block representation.
We use proof automation to convert any proposition about circuit equality into an equivalent proposition about \zxdiags.

Showing this translation highlights that our syntactic representation of graphical diagrams can interoperate with different notions of semantics.
This is a common challenge, especially in systems relying on axiomatic semantics, since two systems will likely not share the same ground truth.
By using \qlib, \vyzx can be used for proof along with other libraries using \qlib, bridging the common pitfall of having different verification models that only exist in isolation.

\subsubsection*{Peephole optimizations}
Using circuit ingestion, we demonstrate the capabilities of \vyzx by verifying peephole optimizations from Nam et al's optimizer~\cite{nam2018automated} which are previously verified in the \voqc optimizer~\cite{hietala2021voqc}. 

\begin{figure}[h]
    \centering
    \hmaxtikzfig{peephole}
    \caption{Quantum circuit peephole optimizations from figures 4 and 5 in Nam et al.~\cite{nam2018automated}}
    \label{fig:peephole}
\end{figure}

We use automation to convert each circuit into a \zxdiag and then proceed to prove equivalences using \vyzx's rules. 
%
% Across all these peephole optimization proofs, we found that \vyzx was able to produce shorter proofs of the same statement. 
% %
% While there are significant differences in how \voqc and \vyzx approach these problems, that would be true of any two systems we chose to compare. 
%
In \vyzx, we are often able to clearly see the equivalences based on the few ZX rules that exist and have an immediate proof path we can follow. % during proof using \vizx.
Due to \vyzx's approach, we are able to use the fact that some of these rules are corollaries of each other, as they are equivalent save for the fact that they are transposes, complex conjugates, or color swapped versions of each other. 
This is something that is easy to see diagrammatically (using \vizx), but less obvious in circuit representation or textual representation of \zxdiags.

By implementing these optimizations, we show the flexibility \vyzx for reasoning about quantum systems and set the foundations for building a verified optimizer in \vyzx.

\section{Graphical proof in \vyzx}\label{sec:proof-eng}

Every formal verification project must eventually confront proof engineering challenges. 
In \vyzx, it is important to concisely express proofs about the \zxcalc. 
To this end, we prove basic facts about the graphical language through the underlying linear algebraic semantics. 
These basic facts allow us to derive additional lemmas without appealing to linear algebra. 
Proving facts using only graphical reasoning allows for more understandable proofs.
This approach is integral to the school of \emph{Quantum Picturalism}~\cite{coecke2010picturalism,coecke2017picturing}, the philosophy that underlies the \zxcalc and quantum process calculi in general.

The constructors within \vyzx represent graphical objects textually.
This textual representation can be unintuitive.
We must consider how to make it approachable.
Using \vizx, we visualize the proof goal's structure to comprehend the proof state and effectively identify rewrite strategies.

\subsection{Induction}\label{sec:prf-induction}
A central proof tactic we want to bring to graphical calculi is induction, allowing us to prove statements for parametric numbers of inputs or outputs of a given node.
To accomplish this for the most basic case, a single Z spider, we must reduce the number of input or outputs from that spider by 1.
This technique relies on our splitting lemmas, which are themselves proved inductively.
We discuss how we built induction capabilities throughout \vyzx to apply inductive reasoning in practice.

\begin{figure}[h]
   \centering
    \begin{subfigure}{.48\textwidth}
        \centering
        \includegraphics[width=0.85\textwidth]{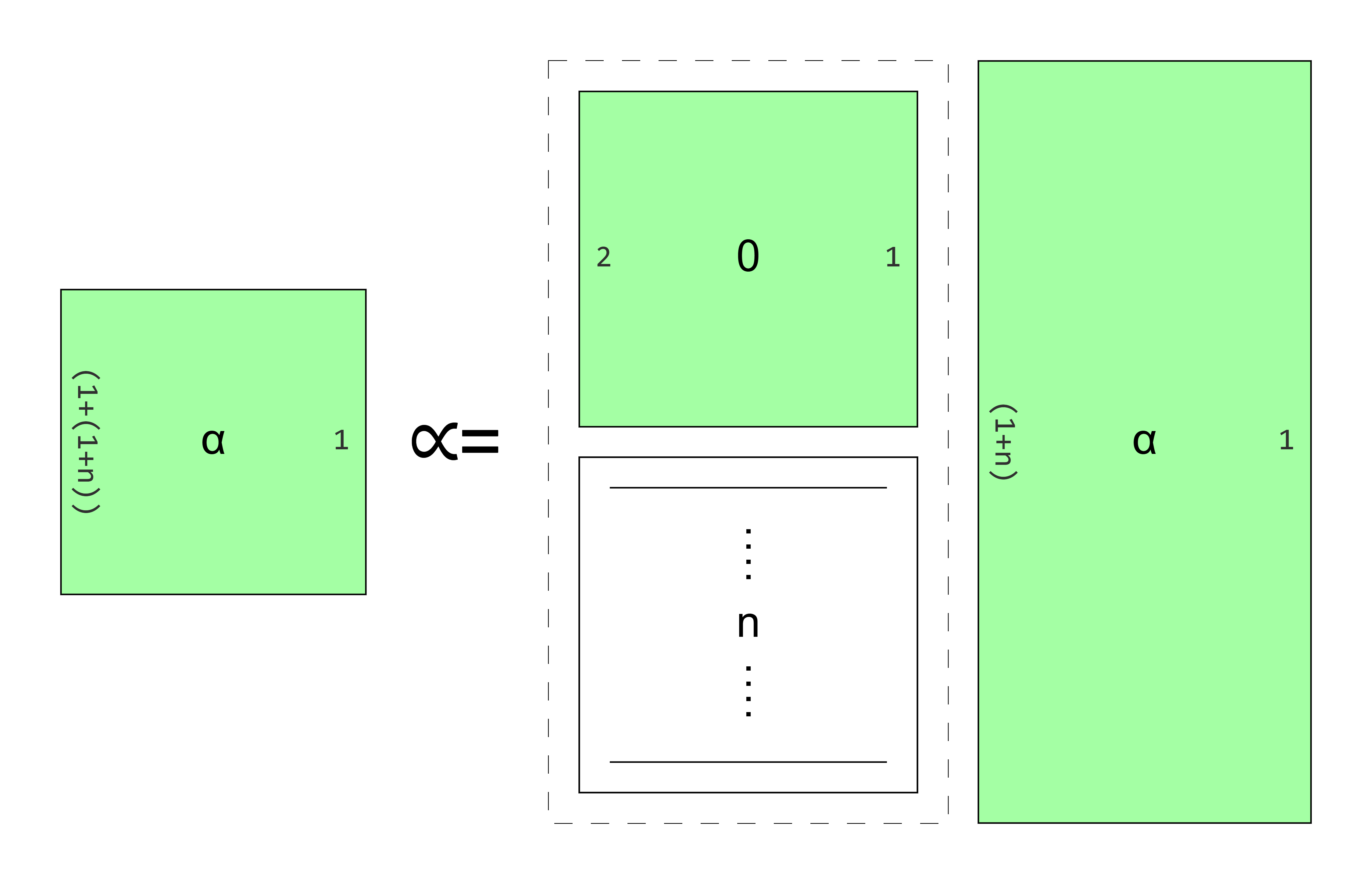}
        \caption{\centering \coqe{Lemma grow_Z_left_2_1}}
        \label{fig:grow-z-1-2}
    \end{subfigure}
    \begin{subfigure}{.48\textwidth} 
        \centering
        \includegraphics[width=0.85\textwidth]{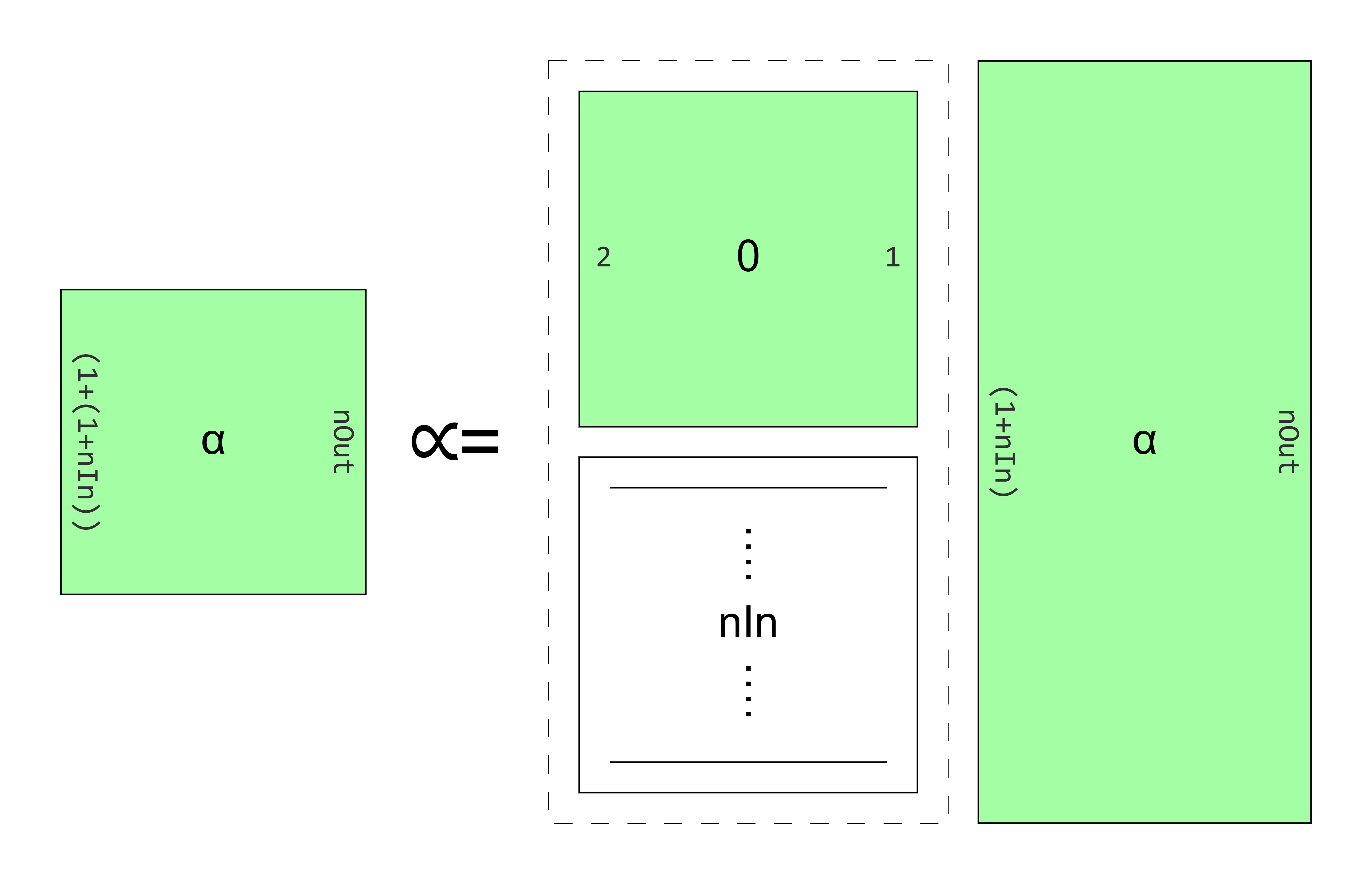}
        \caption{\centering \coqe{Lemma grow_Z_top_left}}
        \label{fig:grow-z}
    \end{subfigure}
    \caption{\centering \coqe{grow_Z_top_left} is built by induction using the base lemma \coqe{grow_Z_left_2_1} which is proven directly through matrix semantics.}
    \label{fig:grow-z-ind}
\end{figure}

Induction on the size of a spider relies on splitting out a two-connection spider.
If we only split a single connection off, we would not reduce the size of the original spider. By splitting two off, we reduce the size of the output, making it something we can apply our inductive hypothesis to.
This splitting is shown by the lemma in \Cref{fig:grow-z-1-2}.
We prove this lemma directly via the underlying semantics by using mathematical properties of composition and stacking.
Using this lemma, we then prove the lemma shown in \Cref{fig:grow-z}.
This lemma allows us to reason inductively over spider sizes and hence allows us to prove parametric rules, such as the wrapping rules shown in \Cref{sec:vyzx-ocm}.
In \Cref{sec:prf-ind-ex} we show an example of a practical application of inductive proof in \vyzx.

\subsection{Proof automation}\label{sec:proof-auto}
Previously, we primarily talked about lemmas that act on Z spiders; however, from the \zxcalc, we know that every valid statement has a valid dual where all spider colors are swapped. 
We encapsulate this through a function that creates the dual diagrams, denoted by $\odot$, and we prove that \coqe{zx0 $\propto$ zx1} implies \coqe{$\odot$ zx0 $\propto$ $\odot$ zx1}.
Similarly, we create a function to transpose diagrams and prove a corresponding lemma. 

\begin{wrapfigure}{r}{.5\textwidth}
    \vspace*{-10pt}
    \centering
    \includegraphics[width=\linewidth]{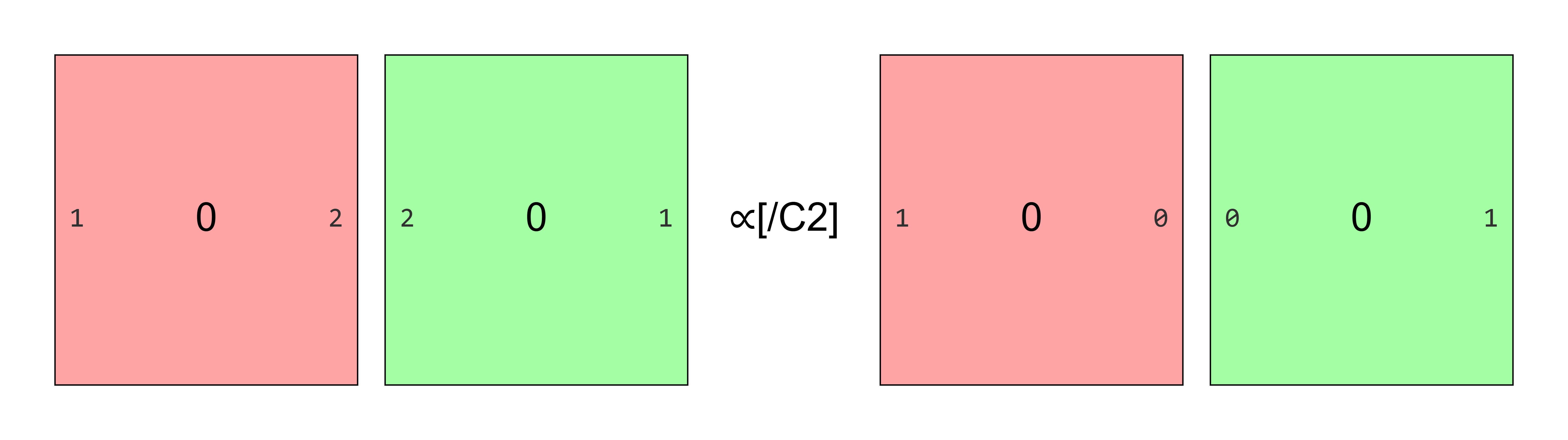}
    \vspace*{-25pt}
    \caption{\centering Color-swapping the hopf rule, as defined in \Cref{fig:hopf},  using automation}
    \label{fig:hopf-dual}
    \vspace*{-15pt}
\end{wrapfigure}

Using \Coq's proof automation features, specifically \coqe{autorewrite} and \coqe{Ltac}, we provide tactics that automatically prove the color-swapped or transposed versions of a lemma. For example, we prove the dual of \coqe{Hopf_rule_Z_X} in \Cref{lst:hopf-dual}, and we show the diagram in \Cref{fig:hopf-dual}. 
The \coqe{colorswap_of} tactic takes the color-swap of both sides of the provided statement, simplifies trivial components, and uses other proven identities of color-swap to simplify further.
We automate proofs about the transposed versions of diagrams using a similar tactic.
We can drastically reduce proof overhead and increase proof maintainability using these two automation tactics.
\begin{lstlisting}[caption={Example of proving the dual of a lemma. See \Cref{fig:hopf-dual} for visualization. In file \texttt{src/DiagramRules/Bialgebra.v}}, label={lst:hopf-dual}]
Theorem Hopf_rule_X_Z : X 1 2 0 ⟷ Z 2 1 0 ∝[/C2] X 1 0 0 ⟷ Z 0 1 0.
Proof. colorswap_of Hopf_rule_Z_X. Qed.
\end{lstlisting}

Along with these proof automation options, we also have a variety of tactics to automatically deal with structural trivialities, such as superfluous stacks of empty diagrams or compositions with \texttt{n\_wire}s.
With this automation, we can deal with most such structures arising from rewrites in practice.
Without the strategies mentioned above, the size of proofs would grow significantly and make them harder to comprehend.
These observations confirm previous findings of domain-specific optimizations being a valuable aide to proof engineering~\cite{ringer2020qedatlarge}.

We also have automation in place to simplify casts.
Besides eliminating trivial casts, the automation attempts to merge multiple casts into a single cast by lifting them to the outermost part of the relevant structure.
With these tools, we have been able to remove most casts within our diagrams automatically.
In cases where our cast automation does not work immediately, we have found that after performing a suitable rewrite within the cast, the automation works successfully.

\subsection{Using visual proof}\label{sec:vizx-eng}

\vyzx aims to prove graphical statements. 
We represent such statements through a syntax tree, as described in \Cref{sec:vyzx}, along with a provided visualization, as described in \Cref{sec:vizx}.
Our visualization aims to convey the core information that \vyzx diagrams capture: its inductive elements and the overall structure.
Specifically, we choose not to use the conventional visualization form of \zxdiags, which focuses on connectivity without conveying the integral structural information we need. 
The integral structural information comes in the form of swaps, caps, and cups, which tell us how wires must bend and cross to make the diagram connect.
This explicit information is not present when you only focus on connectivity.

\vizx is integrated with the rocq-lsp~\cite{lsp} language server to render diagrams of proof goals in the current state automatically. 
When the proof state in focus changes, the rocq-lsp extension sends updated goals to \vizx. 
\vizx then sends the updated diagrams to \vscode, which will display them to the proof engineer.
This simple workflow augments the proof engineer's preexisting workflow by providing a digestible representation of \vyzx diagrams.

\begin{figure}[h]
    \pngfig{ide\_new}{\textwidth}
    \caption{\centering \vizx integrated with the user's proof writing environment.}\label{fig:ide}
\end{figure}

While simple terms are easy to read using just the textual rendering, more complex diagrams benefit from the visualization. 
In our work on developing completeness, we found the (C) rule for completeness~\cite{jeandel2019completeness} (\Cref{fig:completeness-c-zxdiag}) to be difficult to work with textually.
It was hard to understand the positional relationship between subdiagrams and how to make progress simplifying them.
As this is a semantic proof, it was important to control the computation of matrices carefully to avoid large terms that make the proof very slow.
The visualization allowed us to identify repeated structures and easily isolate elements whose linear algebraic representations could be computed explicitly. 
For example, the Z-spider above a wire composed with an X-spider with phase $0$ or $\pi$ appears four times across the two sides, so replacing these structures with their corresponding linear maps simplified the goal substantially.
By repeatedly simplifying subdiagrams, which we could identify using the visualizer, we were able to reduce both structures to a reasonable algebraic form that could be automatically solved.

\begin{lstlisting}[caption={The textual form of the (C) rule for completeness as it appears in the proof environment, visualized in \cref{fig:completeness-c-example}. Found in file \texttt{src/DiagramRules/Completeness.v}}, label={lst:completeness-c-example}]
Z 1 2 0 ⟷ (Z 0 1 β ↕ — ⟷ X 2 1 PI ↕ (Z 0 1 α ↕ — ⟷ X 2 1 0))
⟷ (Z 1 2 β ↕ Z 1 2 α ⟷ (— ↕ X 2 1 γ ↕ —))
⟷ (— ↕ Z 1 2 0 ⟷ (X 2 1 (- γ) ↕ —) ↕ —)
∝= Z 1 2 0 ⟷ (Z 0 1 α ↕ — ⟷ X 2 1 0 ↕ (Z 0 1 β ↕ — ⟷ X 2 1 PI))
   ⟷ (Z 1 2 α ↕ Z 1 2 β ⟷ (— ↕ X 2 1 (- γ) ↕ —))
   ⟷ (— ↕ (Z 1 2 0 ↕ — ⟷ (— ↕ X 2 1 γ)))
\end{lstlisting}

%old c-rule formatting
%Z 1 2 0 ⟷ (Z 0 1 β ↕ — ⟷ X 2 1 PI ↕ (Z 0 1 α ↕ — ⟷ X 2 1 0)) ⟷ (Z 1 2 β ↕ Z 1 2 α ⟷ (— ↕ X 2 1 γ ↕ —)) ⟷ (— ↕ Z 1 2 0 ⟷ (X 2 1 (- γ) ↕ —) ↕ —)
%∝= 
%Z 1 2 0 ⟷ (Z 0 1 α ↕ — ⟷ X 2 1 0 ↕ (Z 0 1 β ↕ — ⟷ X 2 1 PI)) ⟷ (Z 1 2 α ↕ Z 1 2 β ⟷ (— ↕ X 2 1 (- γ) ↕ —)) ⟷ (— ↕ (Z 1 2 0 ↕ — ⟷ (— ↕ X 2 1 γ)))

\begin{figure}[h]
    \centering
    \includegraphics[width=0.75\linewidth]{figures/vizx_C_1.png}
    
    \scalebox{2}{$\propto=$}
    
    \includegraphics[width=0.75\linewidth]{figures/vizx_C_2.png}
    \caption{\centering
    The completeness rule (C), seen textually in \cref{lst:completeness-c-example}.}
    \label{fig:completeness-c-example}
\end{figure}

\begin{figure}
    \ctikzfig{completeness-c}
    \caption{The completeness rule (C) as a \zxdiag.}
    \label{fig:completeness-c-zxdiag}
\end{figure}

\section{VyZX in action}

In this section, we cover a few examples of how \vyzx can be used. Each example covers a different part of \vyzx, showing how we can induct over diagrams, work with diagrams that are close to circuits, and how to handle measurement and correction within \vyzx. All of these examples can be found in \texttt{src/Examples} in the VyZX repository.

\subsection{Inductive proof example: Z absolute fusion}\label{sec:prf-ind-ex}

In this section, we walk through a proof of \emph{absolute fusion}, that any two fully connected spiders can be combined. We give both the tactics used and the diagrams generated by \vizx to give the reader a sense of the proof process.

\begin{lstlisting}
Lemma Z\_absolute\_fusion : forall {n m o} α β, Z n (S m) α ⟷ Z (S m) o β ∝= Z n o (α + β).
Proof.
	intros.
\end{lstlisting}
\begin{figure}[h]
    \centering
    \includegraphics[width=0.5\linewidth]{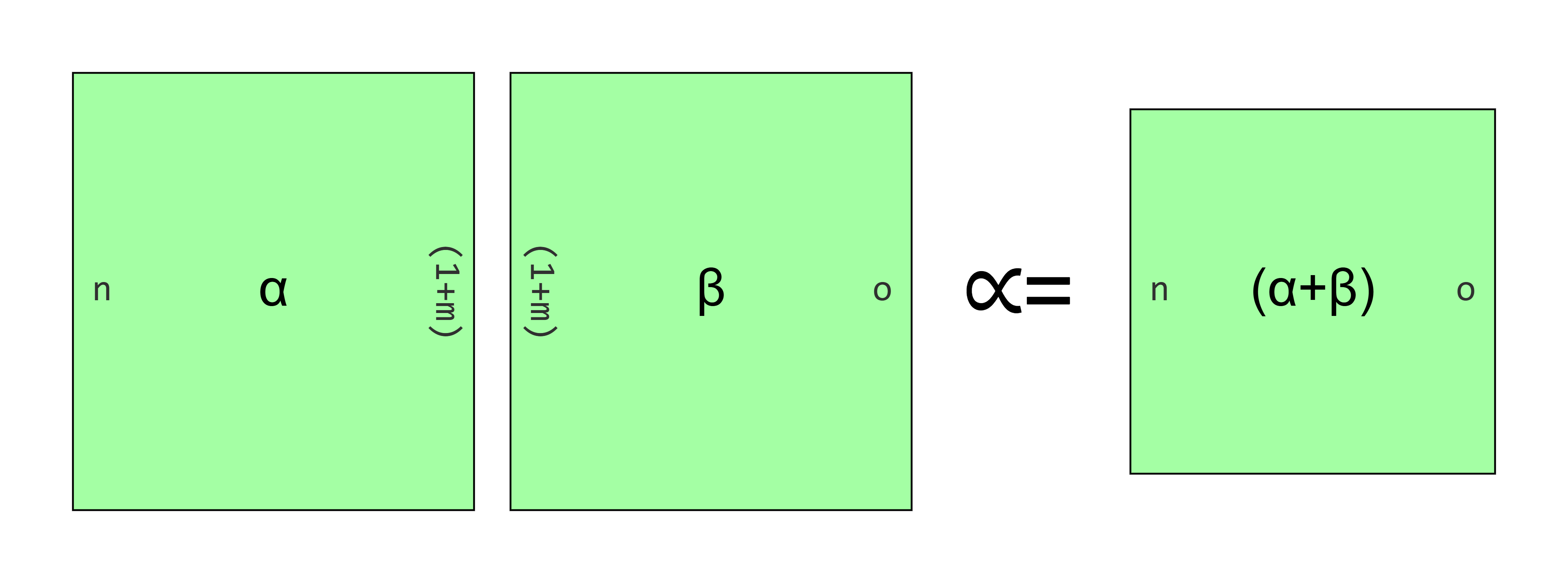}
    \label{fig:z-abs-fusion-lemma}
\end{figure}
This gives us an initial statement for absolute fusion which says that any two Z spiders that are composed together with at least one connection can be fused.
We proceed by induction, giving us a base case where the spiders have exactly one connection and then the case where they have two or more connections.
\begin{lstlisting}
	induction m.
        - (* Base case *)
\end{lstlisting}
\begin{figure}[h]
    \centering
    \includegraphics[width=0.3\linewidth]{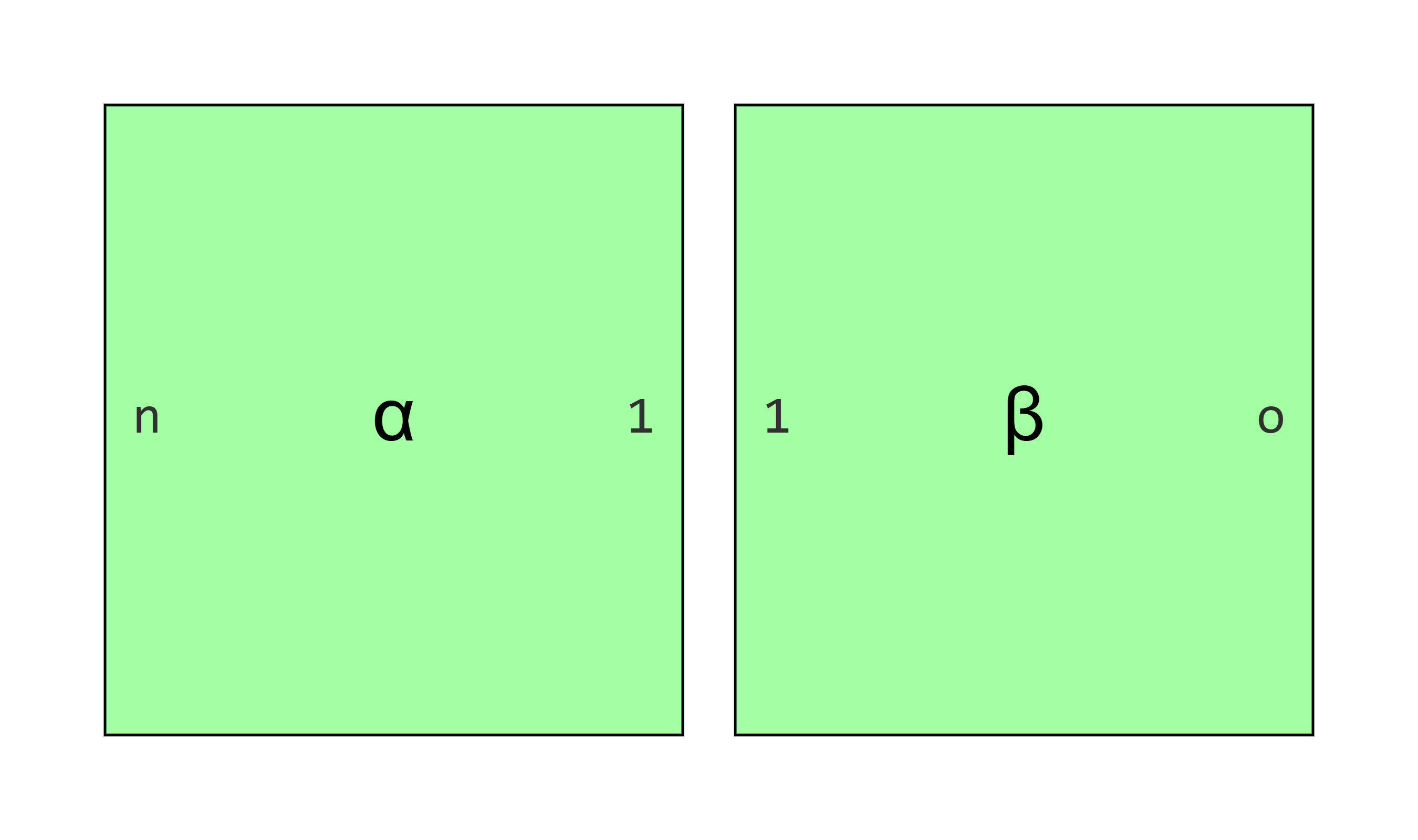}
    \label{fig:placeholder}
\end{figure}
The base case here is handled by a simple computational proof that uses the bra-ket notation for Z spiders.
\begin{lstlisting}
	  apply Z\_spider\_1\_1\_fusion. 
\end{lstlisting}
The inductive case requires that there are at least 2 connections between our spiders.
\begin{lstlisting}
	- (* Inductive case *)
\end{lstlisting}
\begin{figure}[h]
    \centering
    \includegraphics[width=0.3\linewidth]{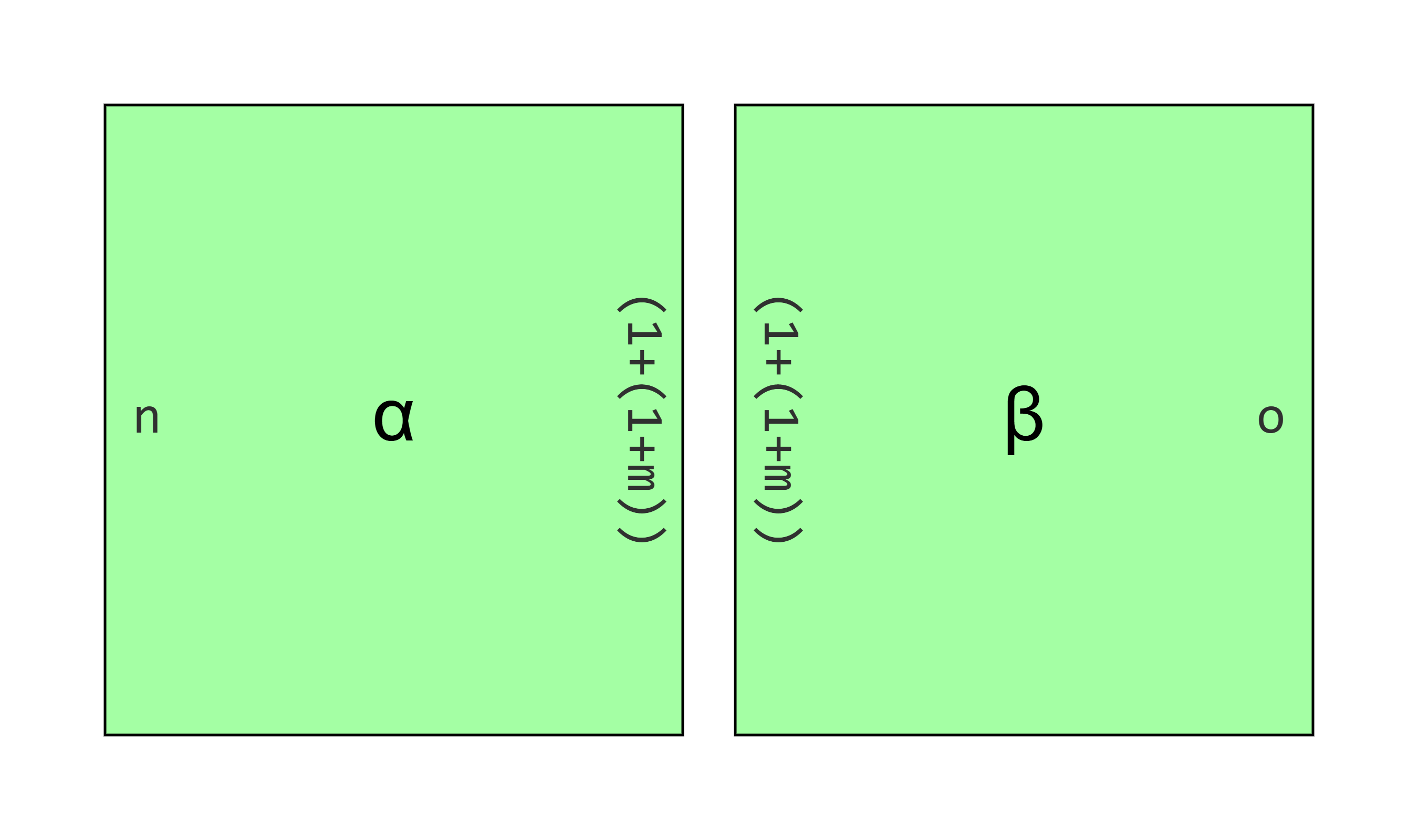}
\end{figure}
In this case, we can split spiders out to the top left and right with finite dimension, allowing us to reason about them more easily.
\begin{lstlisting}
      rewrite grow\_Z\_top\_right, grow\_Z\_top\_left.
\end{lstlisting}
\begin{figure}[h]
    \centering
    \includegraphics[width=0.5\textwidth]{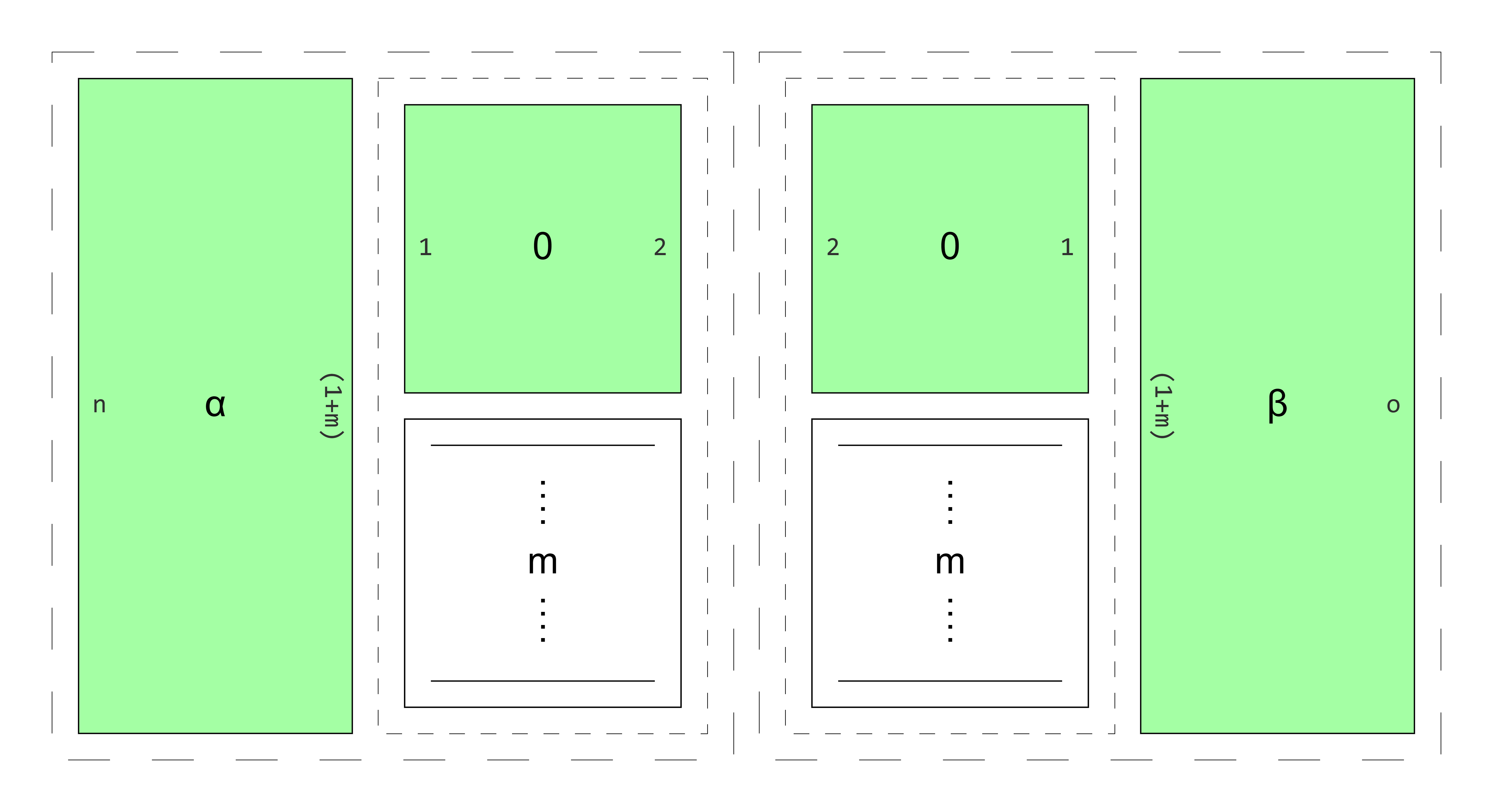}
    \label{fig:placeholder}
\end{figure}
We perform some simple reassociation to juxtapose the top two spiders.
\begin{lstlisting}
      rewrite compose\_assoc.
	   rewrite <- (compose\_assoc ((Z 1 2 0) ↕ (n_wire m)) ((Z 2 1 0) ↕ (n_wire m)) (Z (S m) o β)).
	   rewrite <- stack\_compose\_distr.
\end{lstlisting}
\begin{figure}[h]
    \centering
    \includegraphics[width=0.4\linewidth]{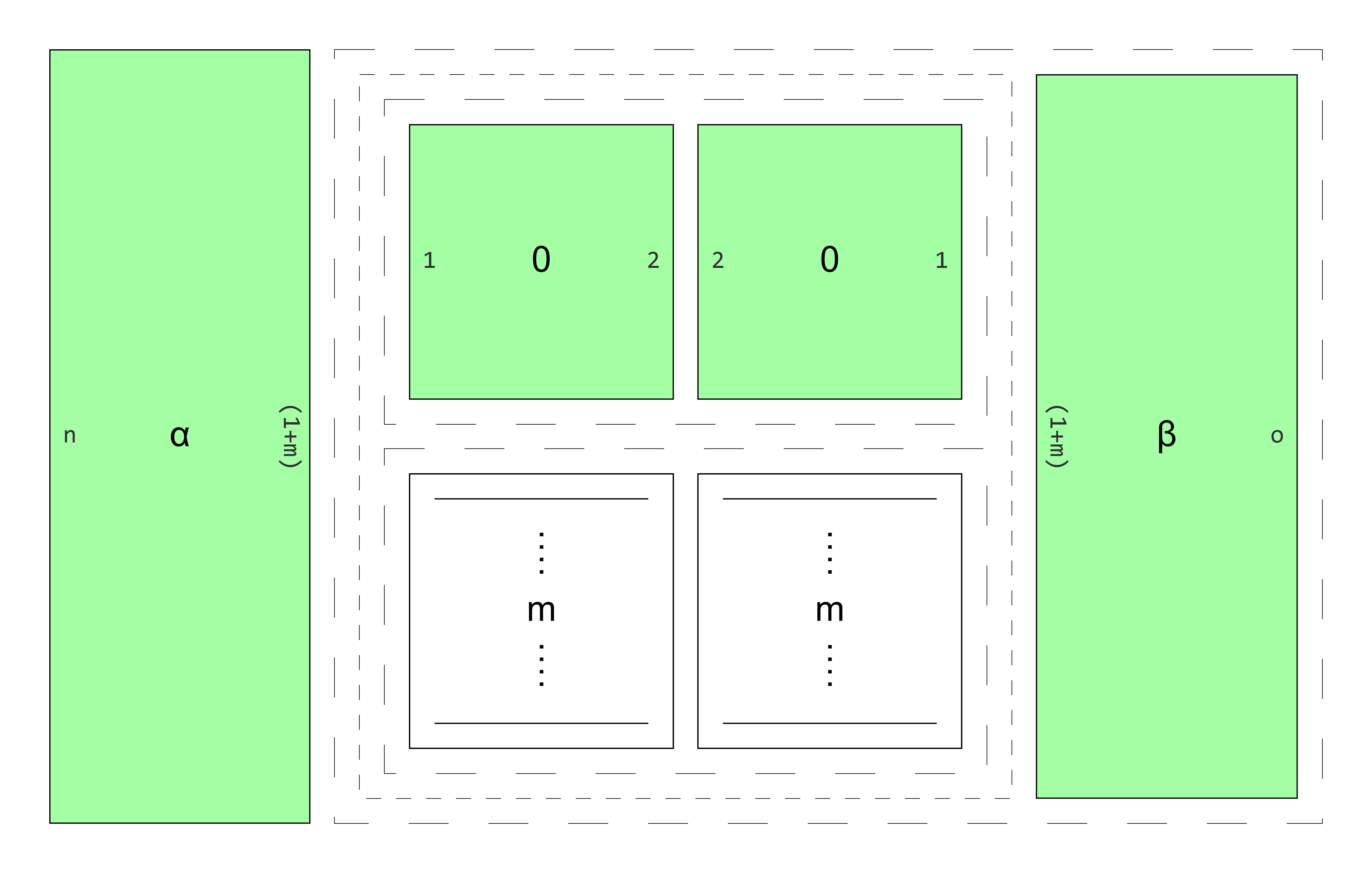}
    \label{fig:placeholder}
\end{figure}
Using a simple lemma, we reduce the diagram further.
\begin{lstlisting}
      rewrite Z\_1\_2\_1\_fusion.
      rewrite Rplus\_0\_l.
\end{lstlisting}
\begin{figure}[h]
    \centering
    \includegraphics[width=0.4\linewidth]{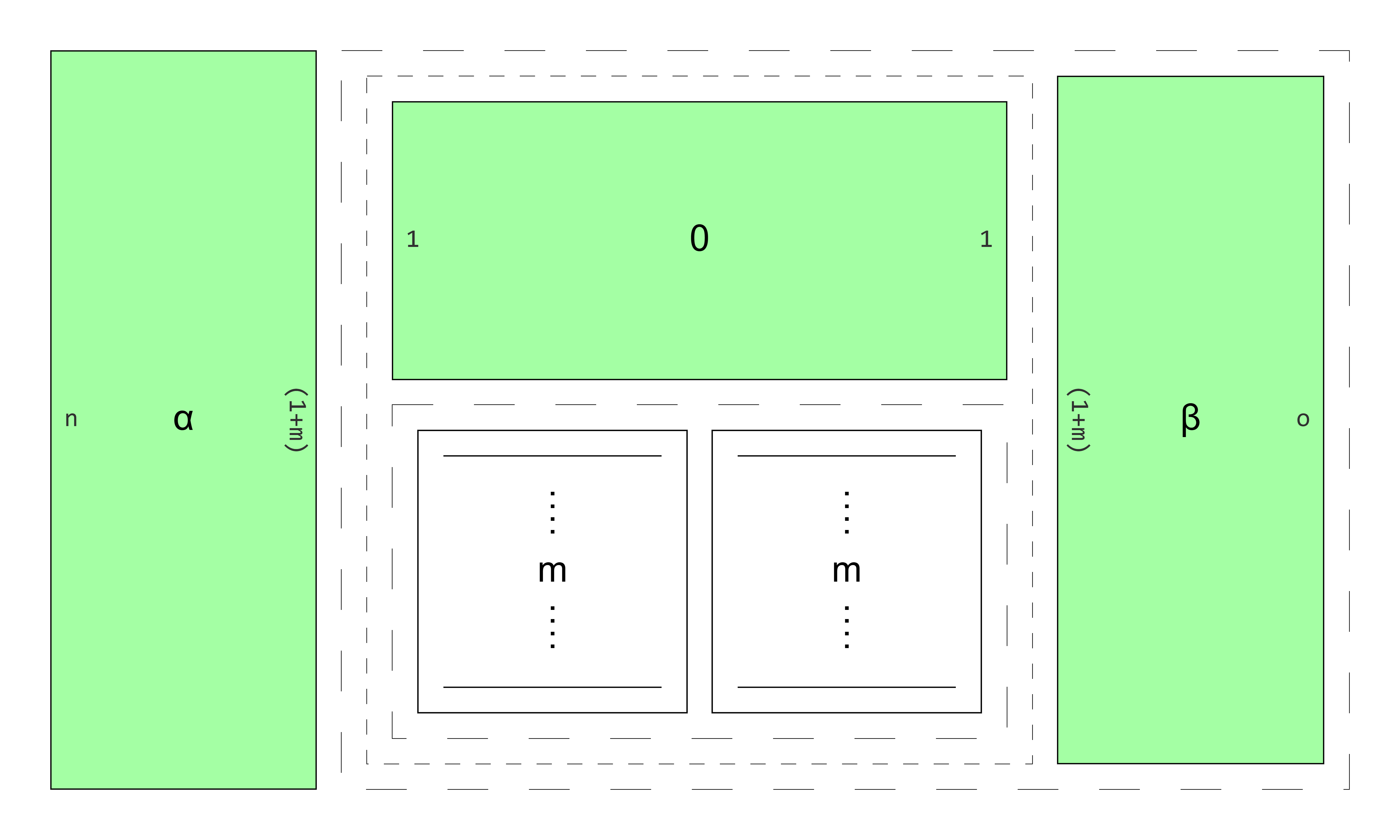}
\end{figure}
Automation can eliminate wires for us, including the spider on top, leaving us with just our inductive hypothesis:
\begin{lstlisting}
	   cleanup\_zx.
\end{lstlisting}
\begin{figure}[h]
    \centering
    \includegraphics[width=0.4\linewidth]{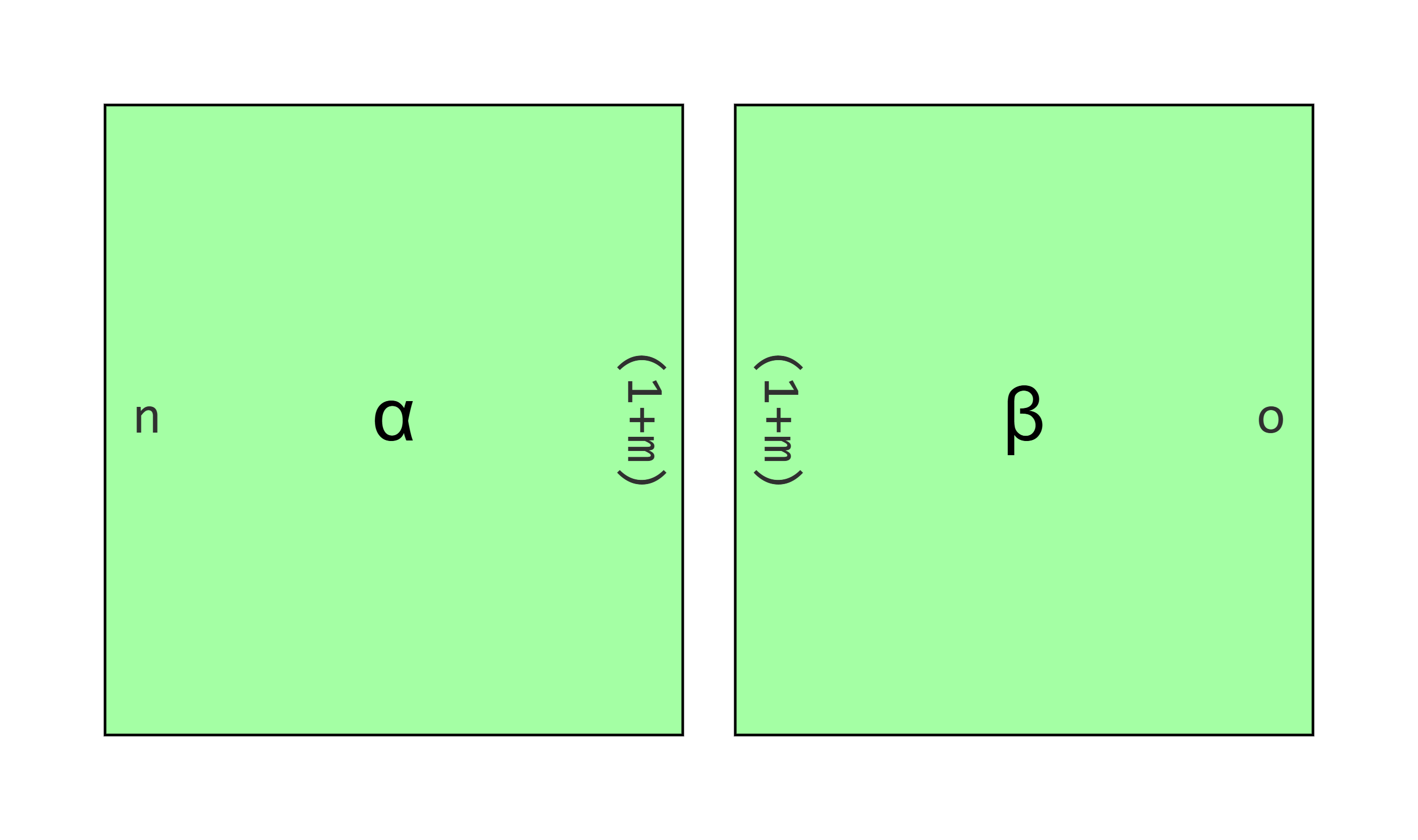}
    \label{fig:placeholder}
\end{figure}

This allows us to complete the proof.
\begin{lstlisting}
      apply IHm.
\end{lstlisting}
\begin{figure}[h]
    \centering
    \includegraphics[width=0.25\linewidth]{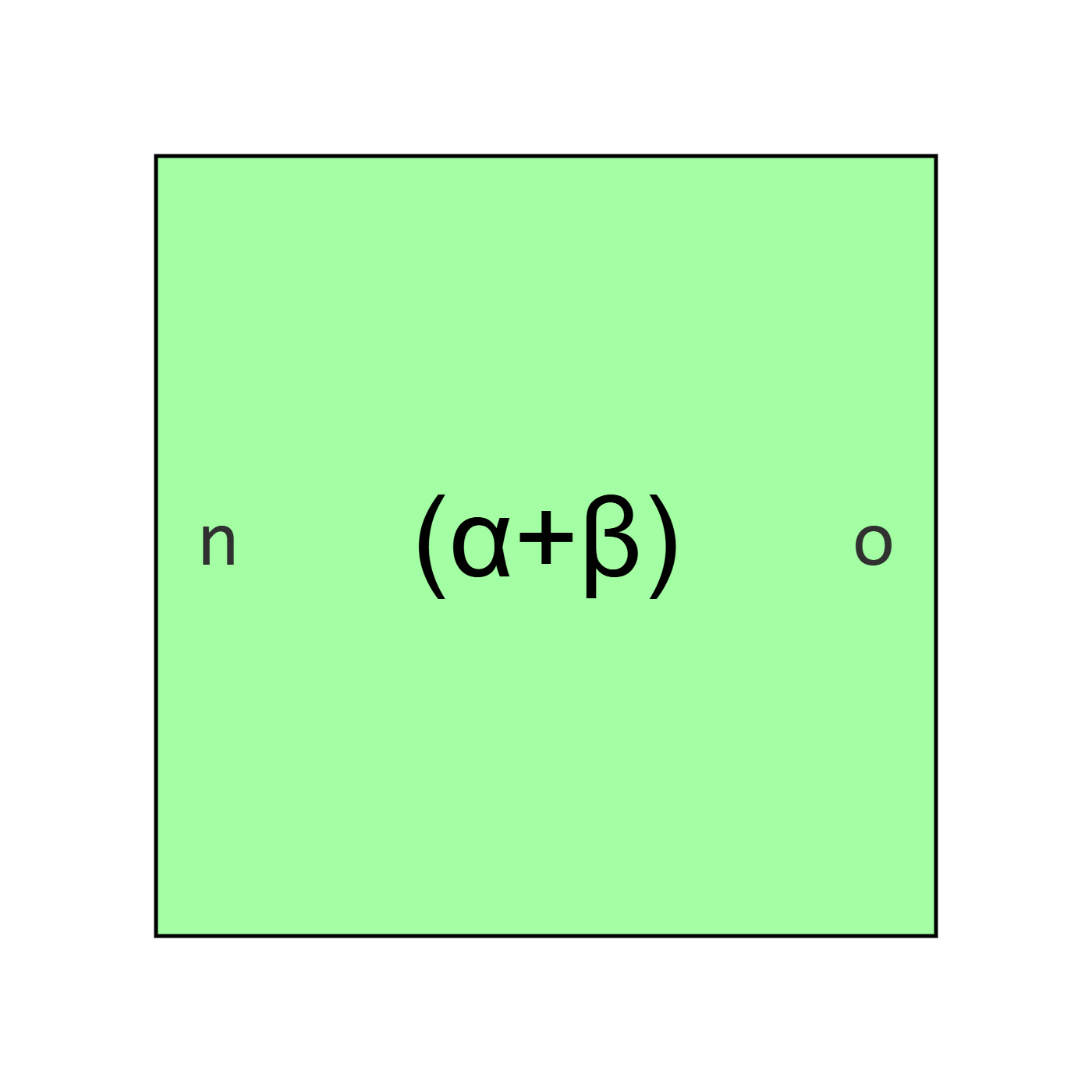}
\end{figure}
\begin{lstlisting}
Qed.
\end{lstlisting}

\begin{figure}[b]
    \centering
    \begin{subfigure}[b]{.53\textwidth}
        \includegraphics[width=\textwidth]{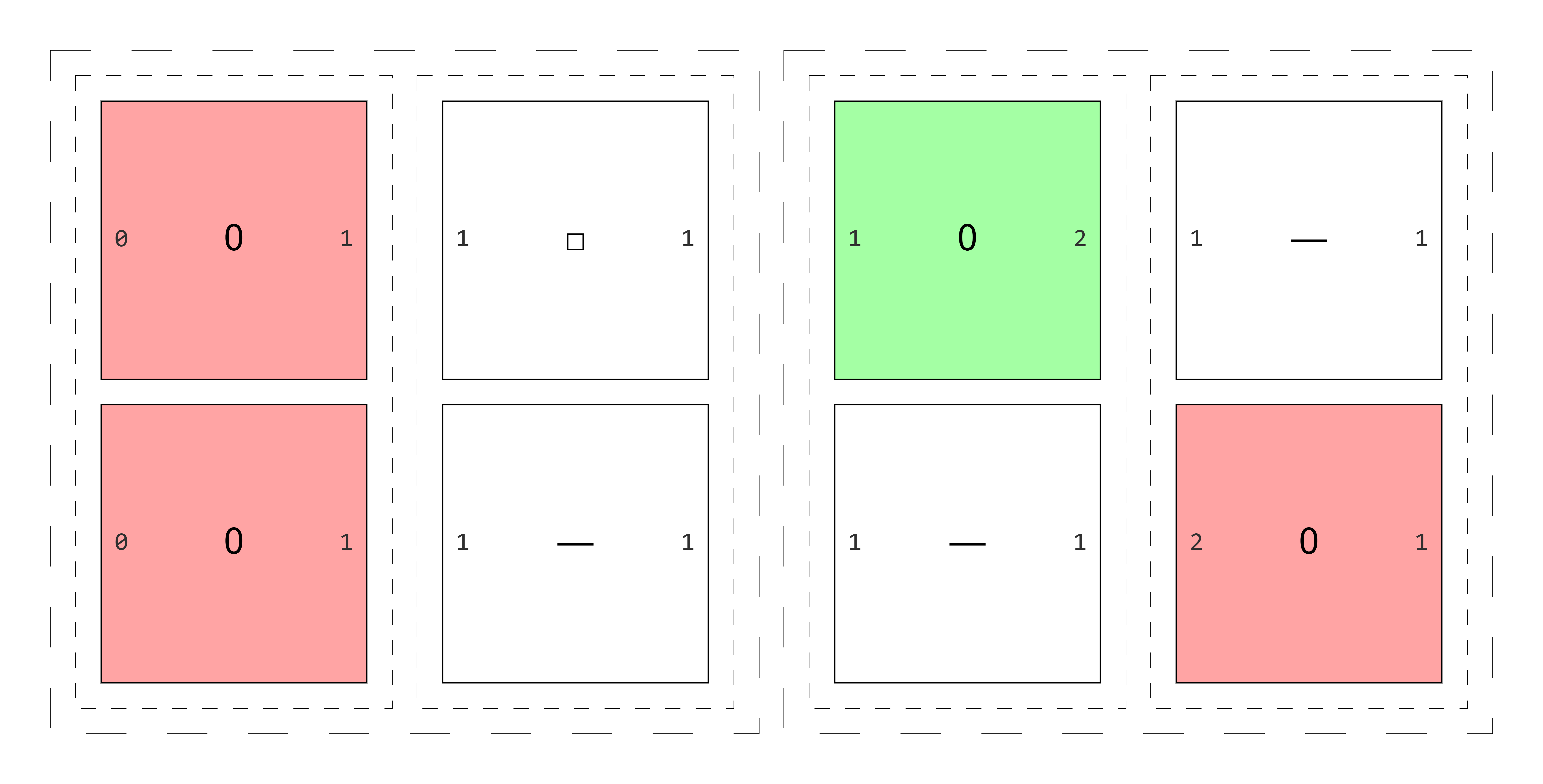}
        \caption{\centering \texttt{bell\_state\_prep\_correct} lemma}
        \label{fig:bell-prep-diag:lemma}
    \end{subfigure}
    \begin{subfigure}[b]{.40\textwidth}
        \includegraphics[width=\textwidth]{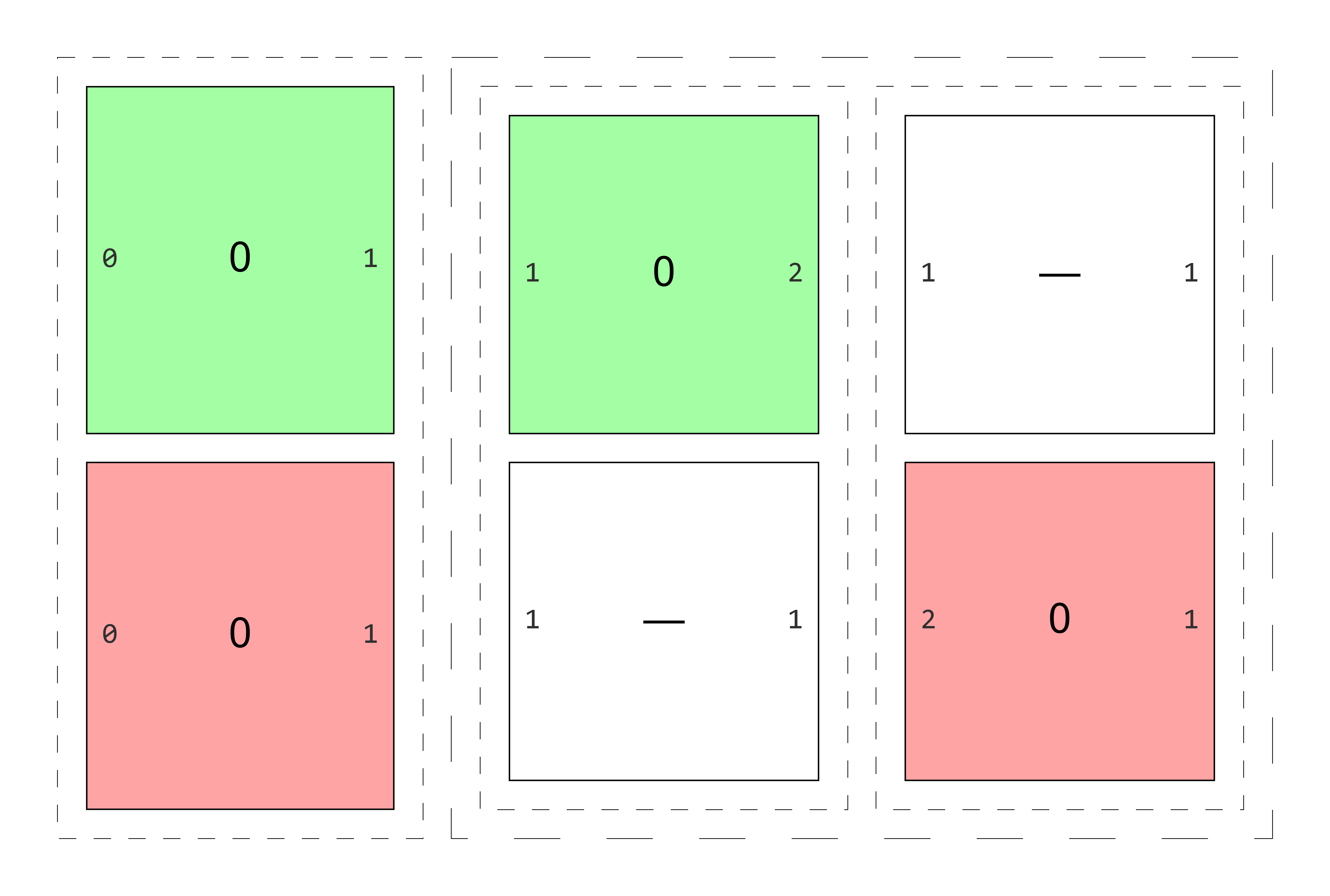}
        \caption{\centering Reassociate and color-swap top-left node}
        \label{fig:bell-prep-diag:1}
    \end{subfigure}
    \begin{subfigure}[b]{.35\textwidth}
        \centering
        \includegraphics[width=\textwidth]{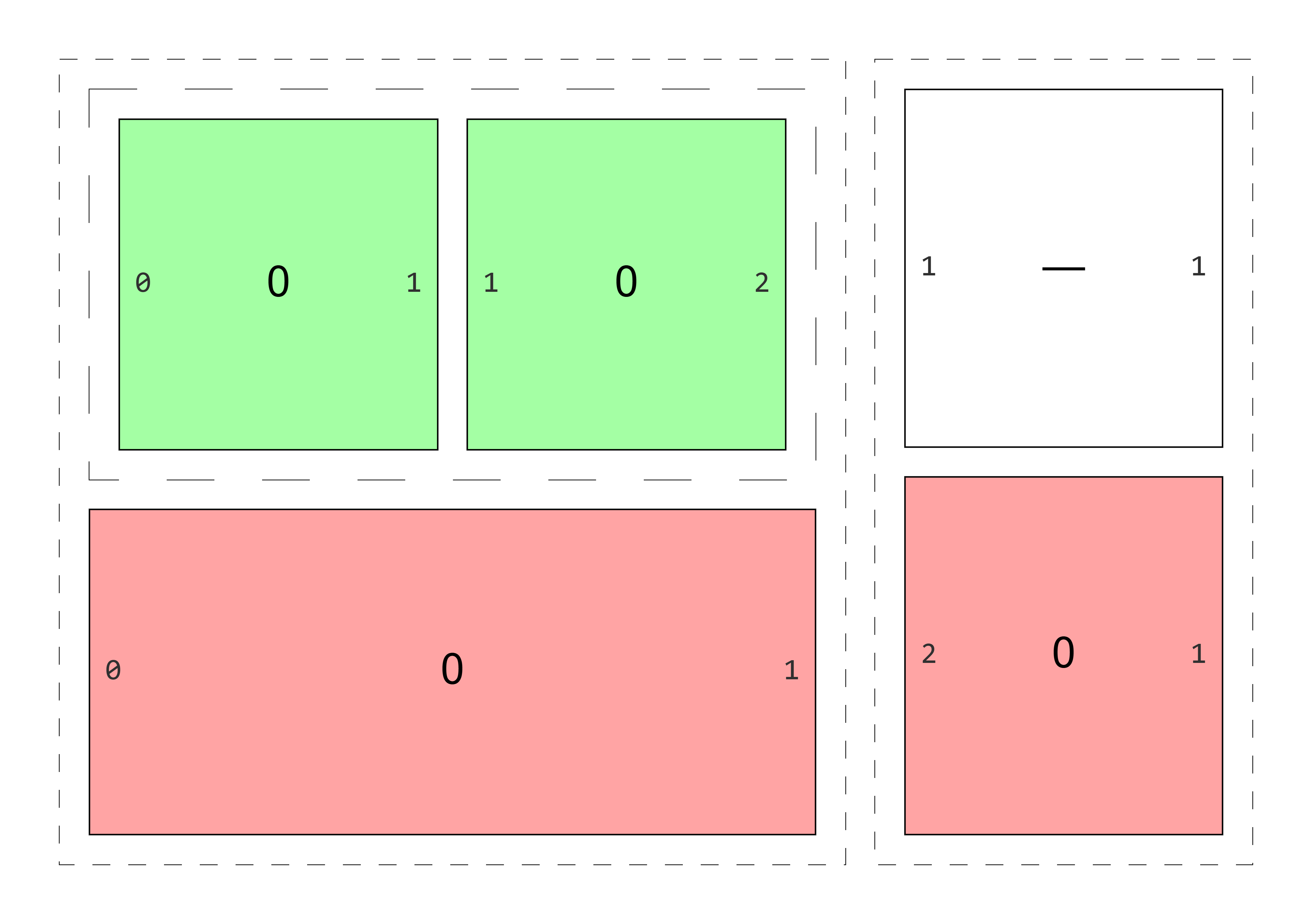}
        \caption{\centering Reassociate for fusion}
        \label{fig:bell-prep-diag:2}
    \end{subfigure}
    \begin{subfigure}[b]{.25\textwidth}
        \centering
        \includegraphics[width=\textwidth]{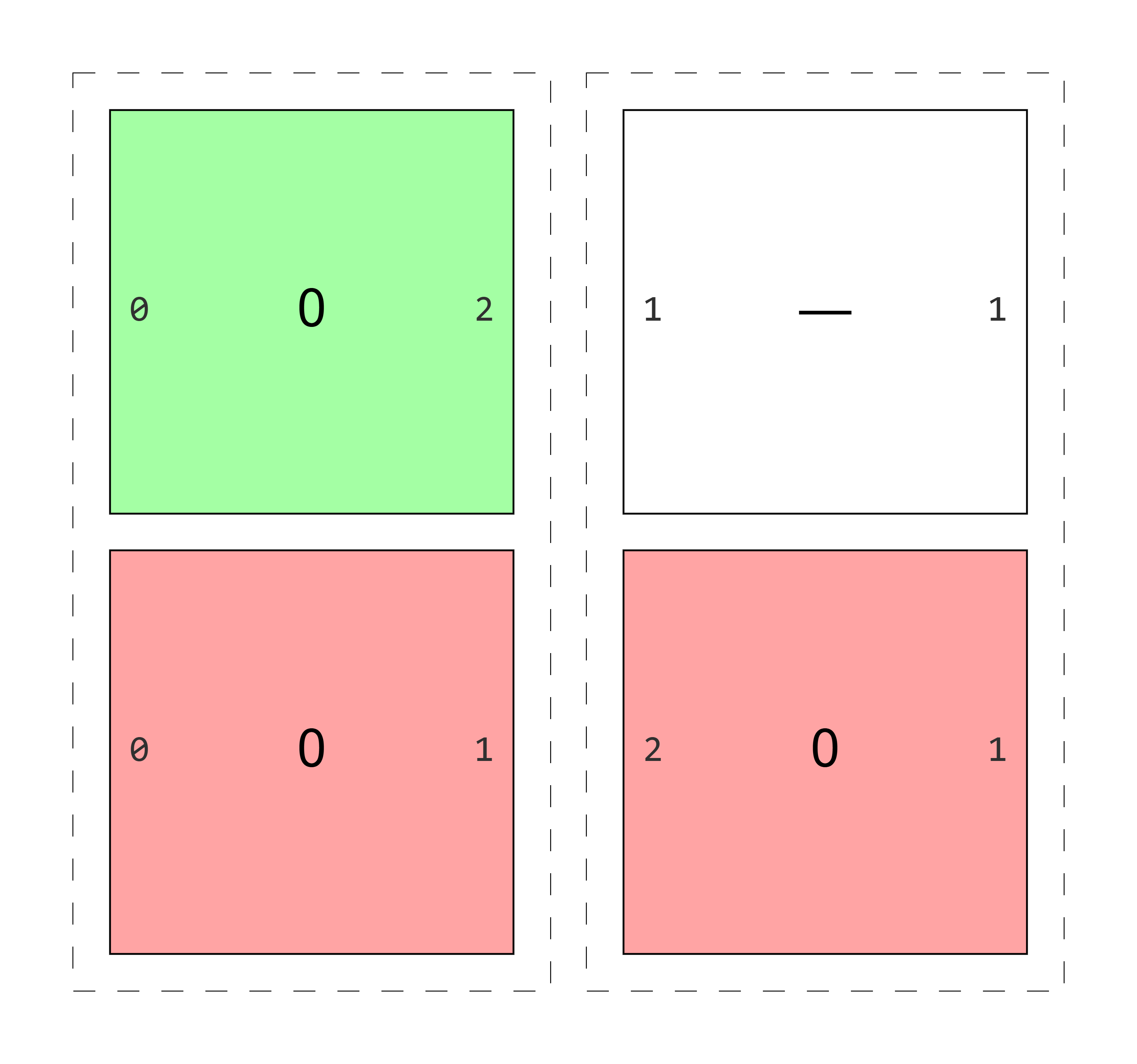}
        \caption{\centering Fuse top-left spiders}
        \label{fig:bell-prep-diag:3}
    \end{subfigure}
    \begin{subfigure}[b]{.30\textwidth}
        \includegraphics[width=\textwidth]{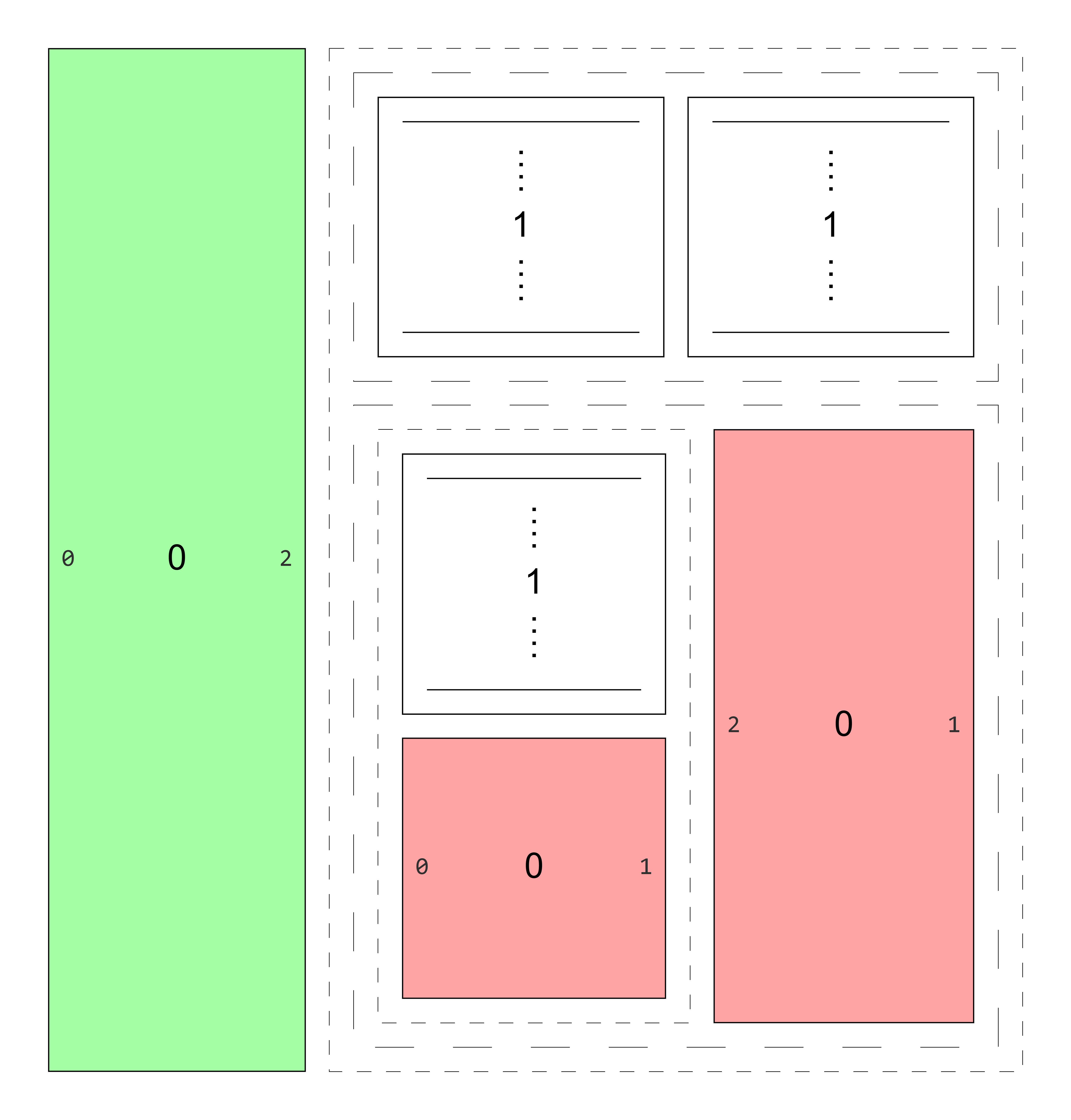}
        \caption{\centering Add auxiliary wires and reassociate}
        \label{fig:bell-prep-diag:4}
    \end{subfigure}
    \begin{subfigure}[b]{.34\textwidth}
        \includegraphics[width=\textwidth]{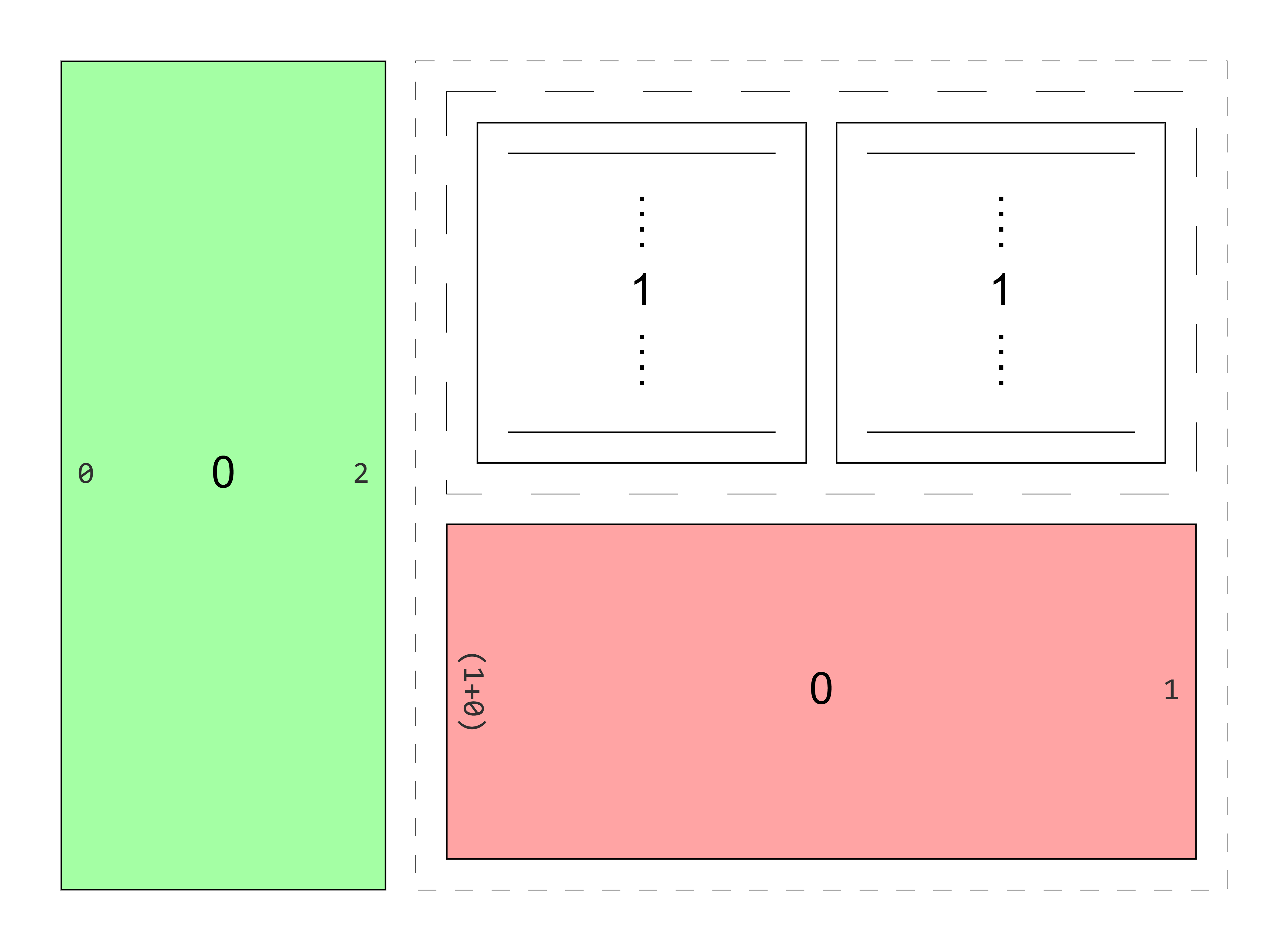}
        \caption{\centering Fuse bottom-right spiders}
        \label{fig:bell-prep-diag:5}
    \end{subfigure}
    \begin{subfigure}[b]{.25\textwidth}
        \includegraphics[width=\textwidth]{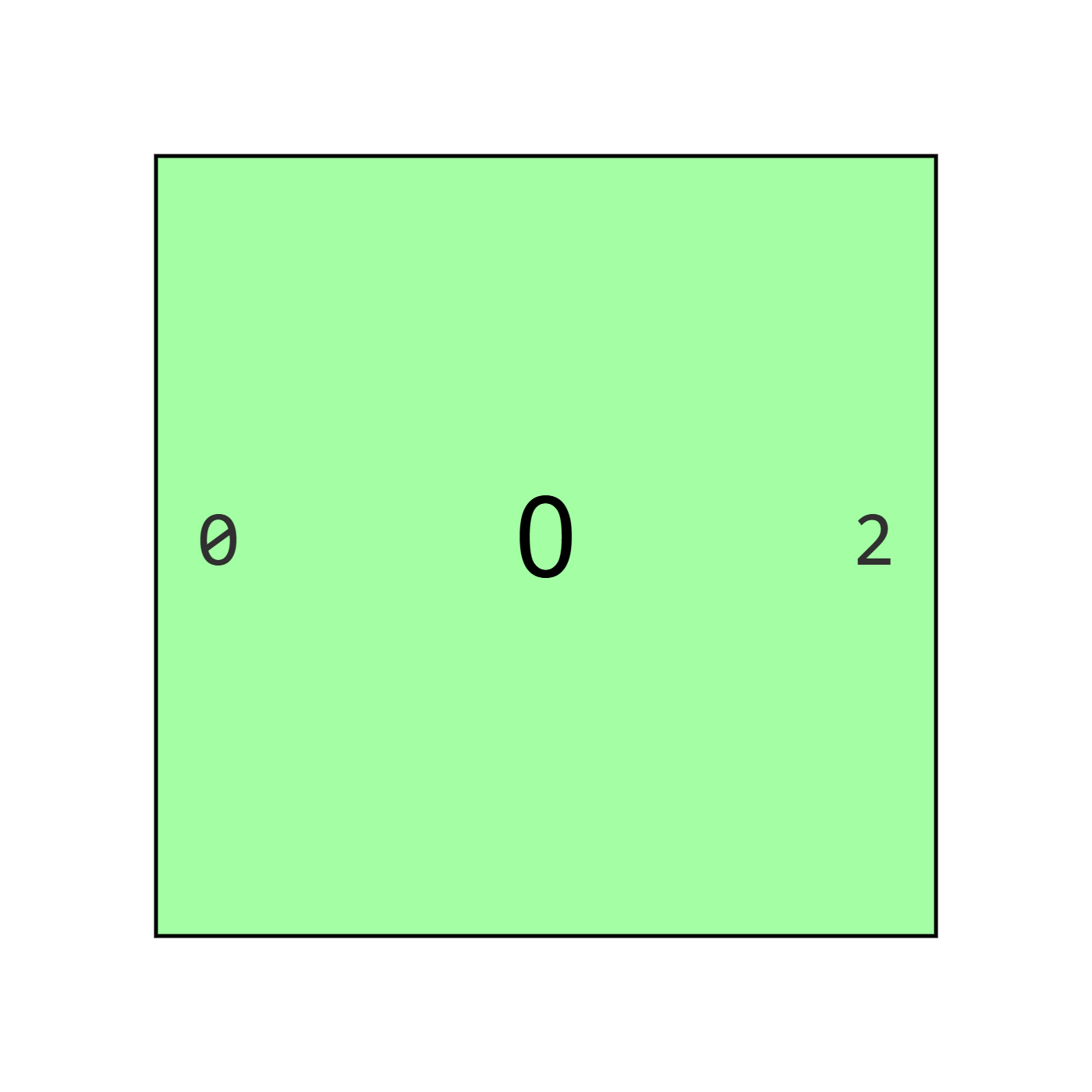}
        \caption{\centering Simplify wires and spider}
        \label{fig:bell-prep-diag:6}
    \end{subfigure}
    \begin{subfigure}[b]{.25\textwidth}
        \includegraphics[width=\textwidth]{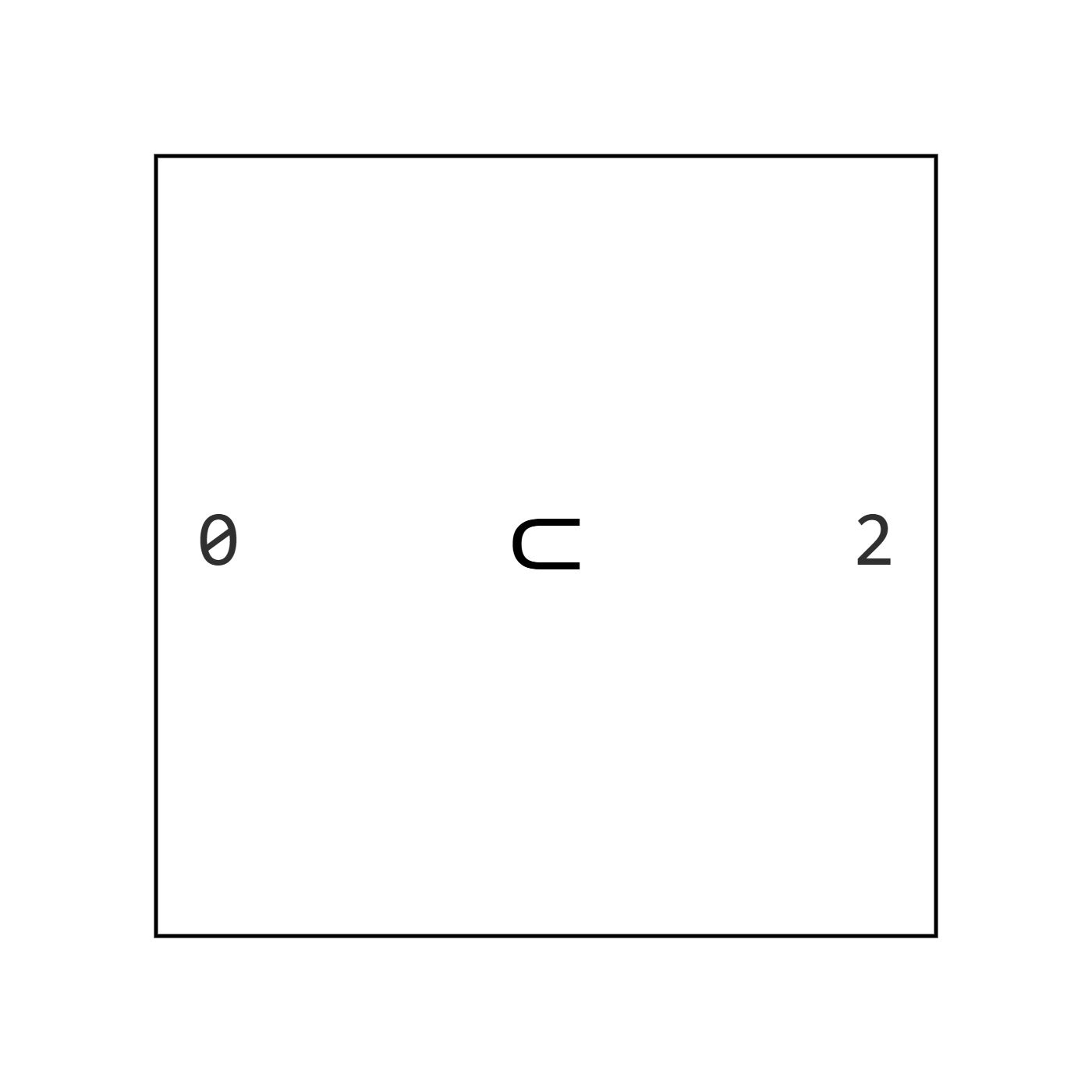}
        \caption{\centering Convert spider into cap}
        \label{fig:bell-prep-diag:7}
    \end{subfigure}
    \caption{\centering Proof of the correctness of Bell state preparation in \vyzx. See \Cref{sec:prf-structural-example}.}
    \label{fig:bell-prep-diag}
\end{figure}

\subsection{Structural Proof Example: Bell state preparation}\label{sec:prf-structural-example}

To show an example of structural editing within \vyzx, we prove the correctness of the Bell state preparation circuit.
First, we transform the standard Bell pair circuit into a \zxdiag by translating gates as shown in \Cref{sec:gates}.
\Cref{fig:bell-prep-diag:lemma} shows the outcome of this translation: The two $X$ spiders on the left prepare qubits in the $\ket{0}$ state, which is followed by a Hadamard and CNOT gate. 
We start the proof by reassociating to use bi-Hadamard color-swapping on the top wire in \Cref{fig:bell-prep-diag:1}.
After that, we reassociate (see \Cref{fig:bell-prep-diag:2}) and fuse the two Z spiders (\Cref{fig:bell-prep-diag:3}).
The next step is to reassociate (\Cref{fig:bell-prep-diag:4}) and fuse the X spiders (\Cref{fig:bell-prep-diag:5}).
Finally, in \Cref{fig:bell-prep-diag:6} we simplify the resulting diagram and then convert the spider into a cap in \Cref{fig:bell-prep-diag:7}.

\begin{figure}[b]
    \centering
    \begin{subfigure}[b]{\textwidth}
        \centering 
        \includegraphics[width=.70\textwidth]{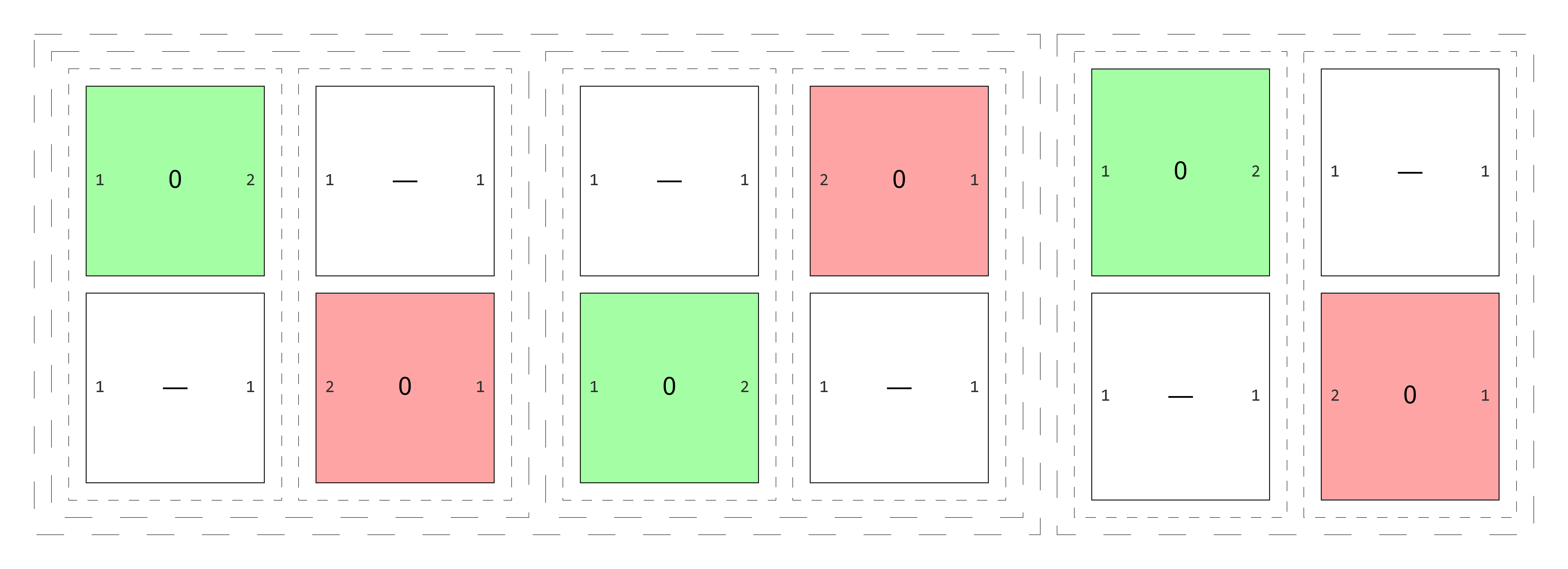}
        \caption{\centering Left-hand side of three CNOTs to SWAP lemma}
        \label{fig:cnot2swap:lemma}
    \end{subfigure}
    % \begin{subfigure}[b]{\textwidth}
    %     \centering 
    %     \includegraphics[width=.70\textwidth]{figures/cnot2swap_step2_lhs.png}
    %     \caption{\centering Flip the rightmost CNOT gate}
    %     \label{fig:cnot2swap:swap-right}
    % \end{subfigure}
    \begin{subfigure}[b]{\textwidth}
        \centering
        \includegraphics[width=.70\textwidth]{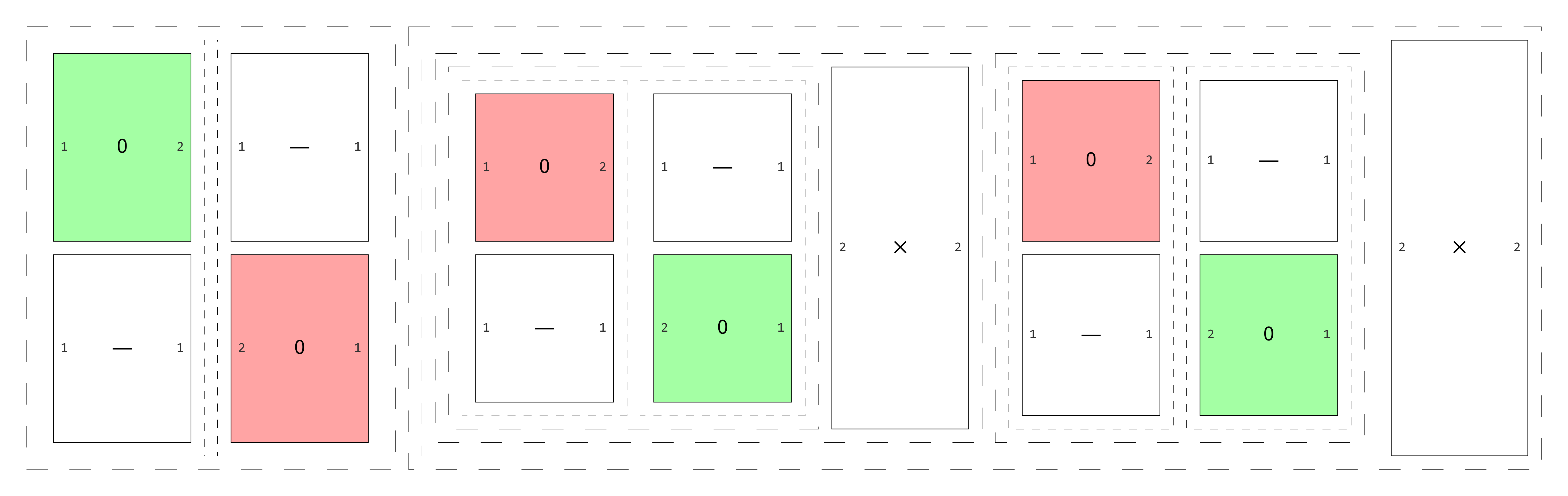}
        \caption{\centering Flip the rightmost CNOT gate and reassociate to apply the Bialgebra rule}
        \label{fig:cnot2swap:reassoc-bialg}
    \end{subfigure}
    
    \begin{subfigure}[b]{.345\textwidth}
        \centering
        \includegraphics[width=\textwidth]{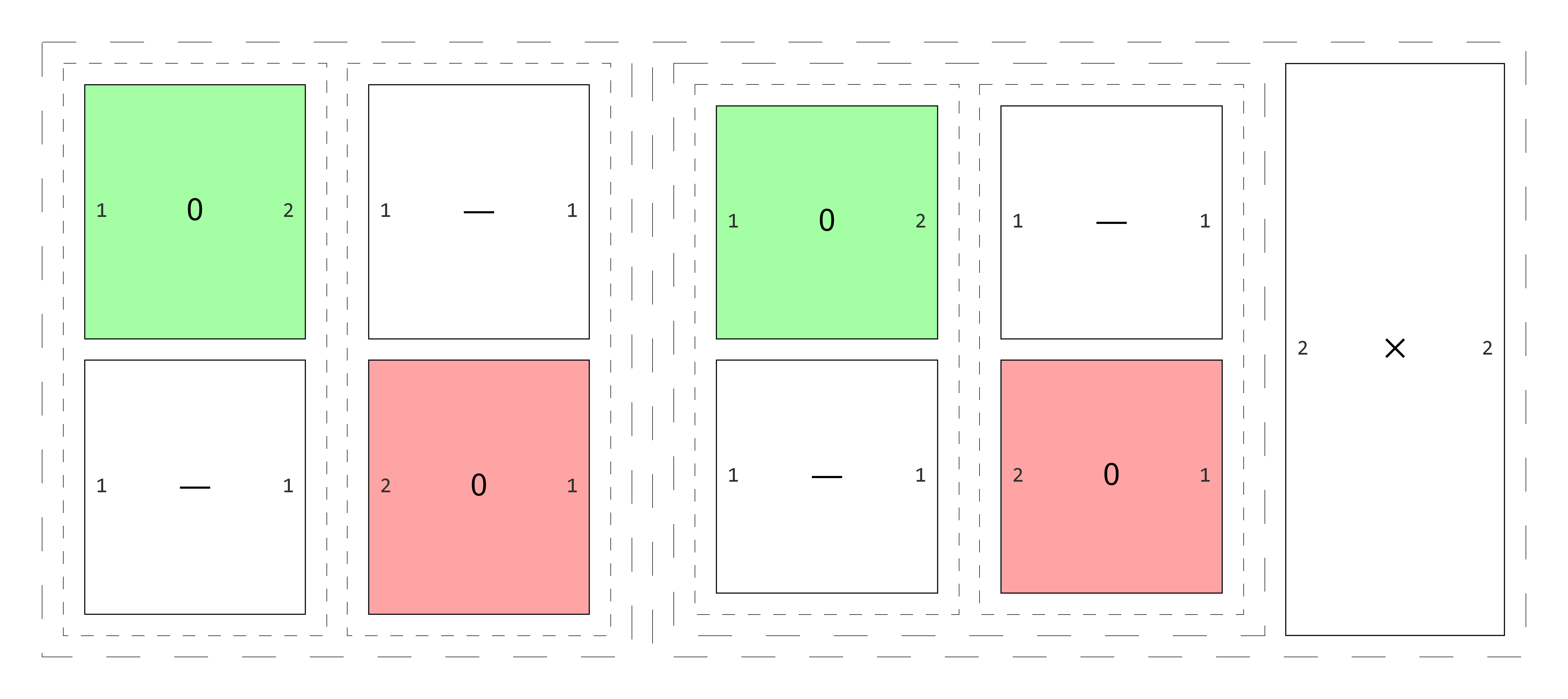}
        \caption{\centering Apply Bialgebra rule}
        \label{fig:cnot2swap:bialg}
    \end{subfigure}
    \begin{subfigure}[b]{.33\textwidth}
        \includegraphics[width=\textwidth]{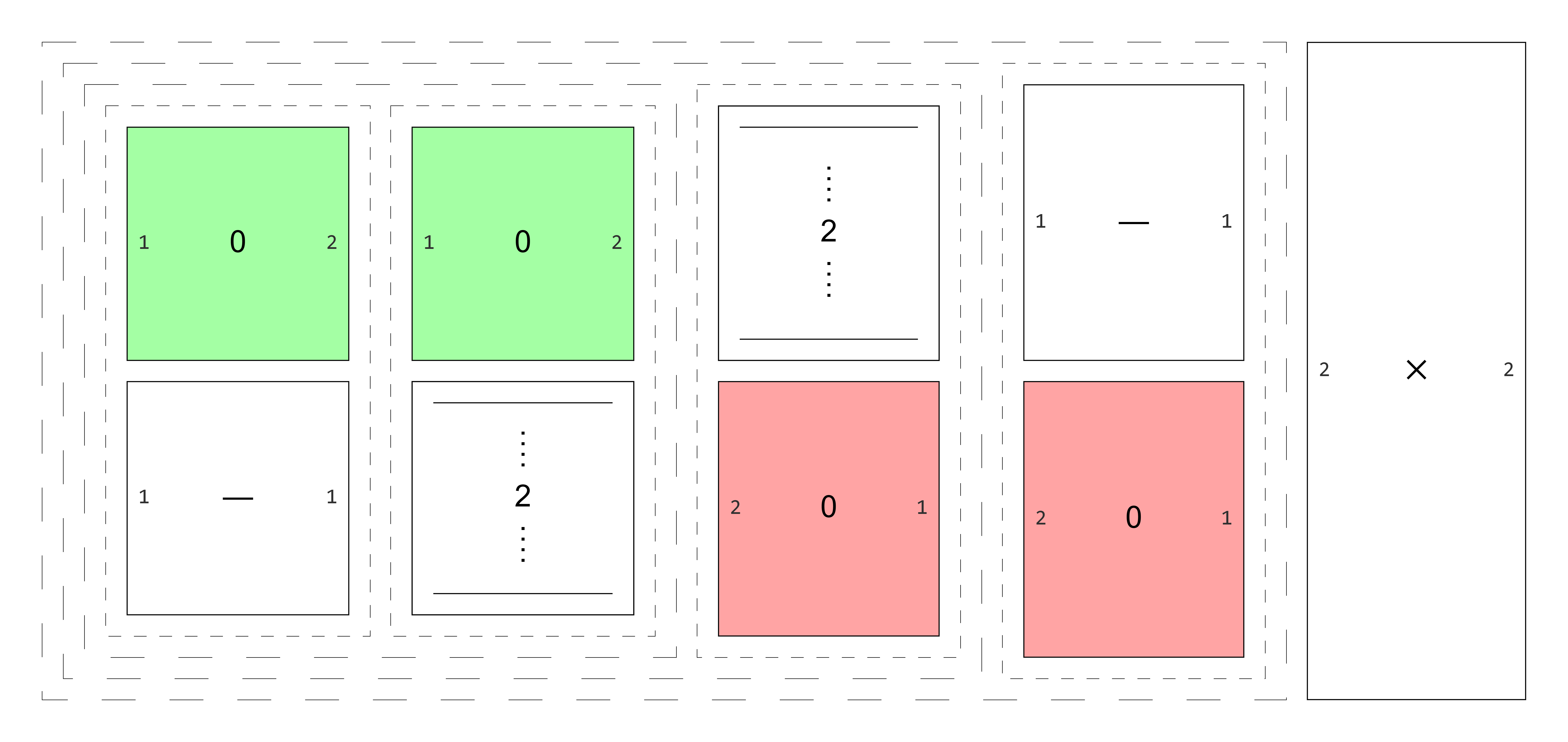}
        \caption{\centering Reassociate for fusion}
        \label{fig:cnot2swap:reassoc-fusion}
    \end{subfigure}
    
    \begin{subfigure}[b]{.275\textwidth}
        \includegraphics[width=\textwidth]{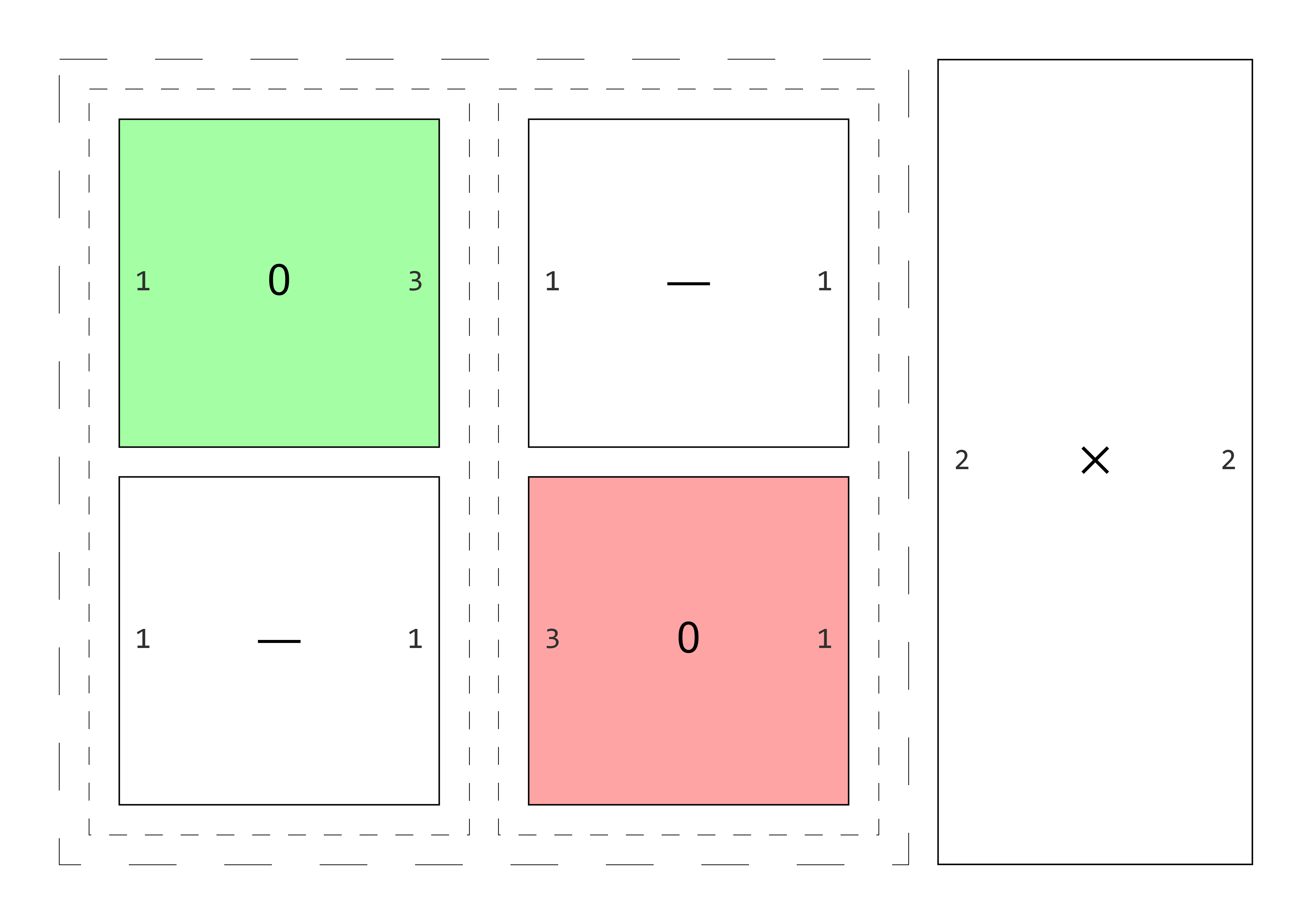}
        \caption{\centering Fuse both pairs of spiders}
        \label{fig:cnot2swap:fusion}
    \end{subfigure}
    \begin{subfigure}[b]{.19\textwidth}
        \includegraphics[width=\textwidth]{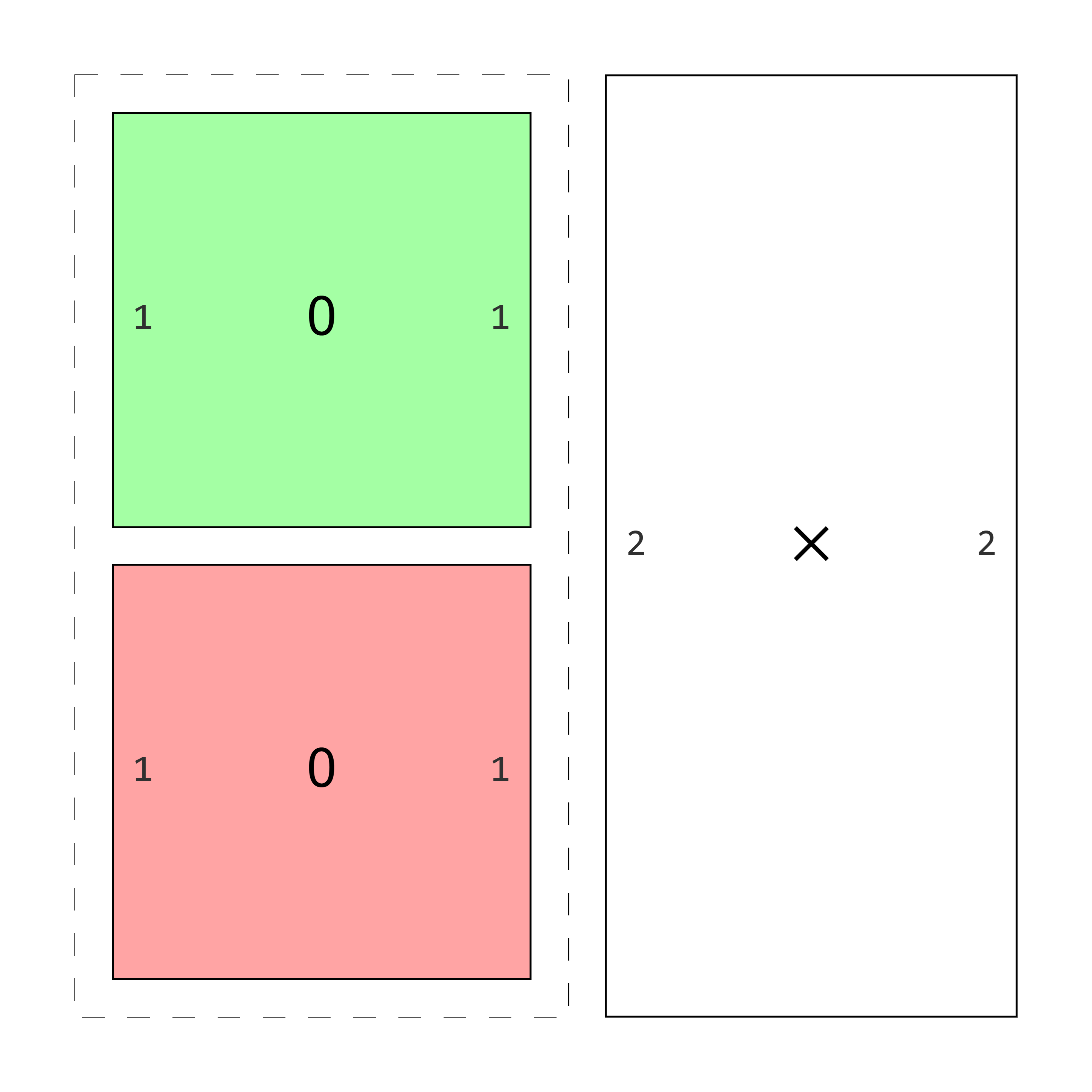}
        \caption{\centering Apply the Hopf rule}
        \label{fig:cnot2swap:hopf}
    \end{subfigure}
    \begin{subfigure}[b]{.16\textwidth}
        \includegraphics[width=\textwidth]{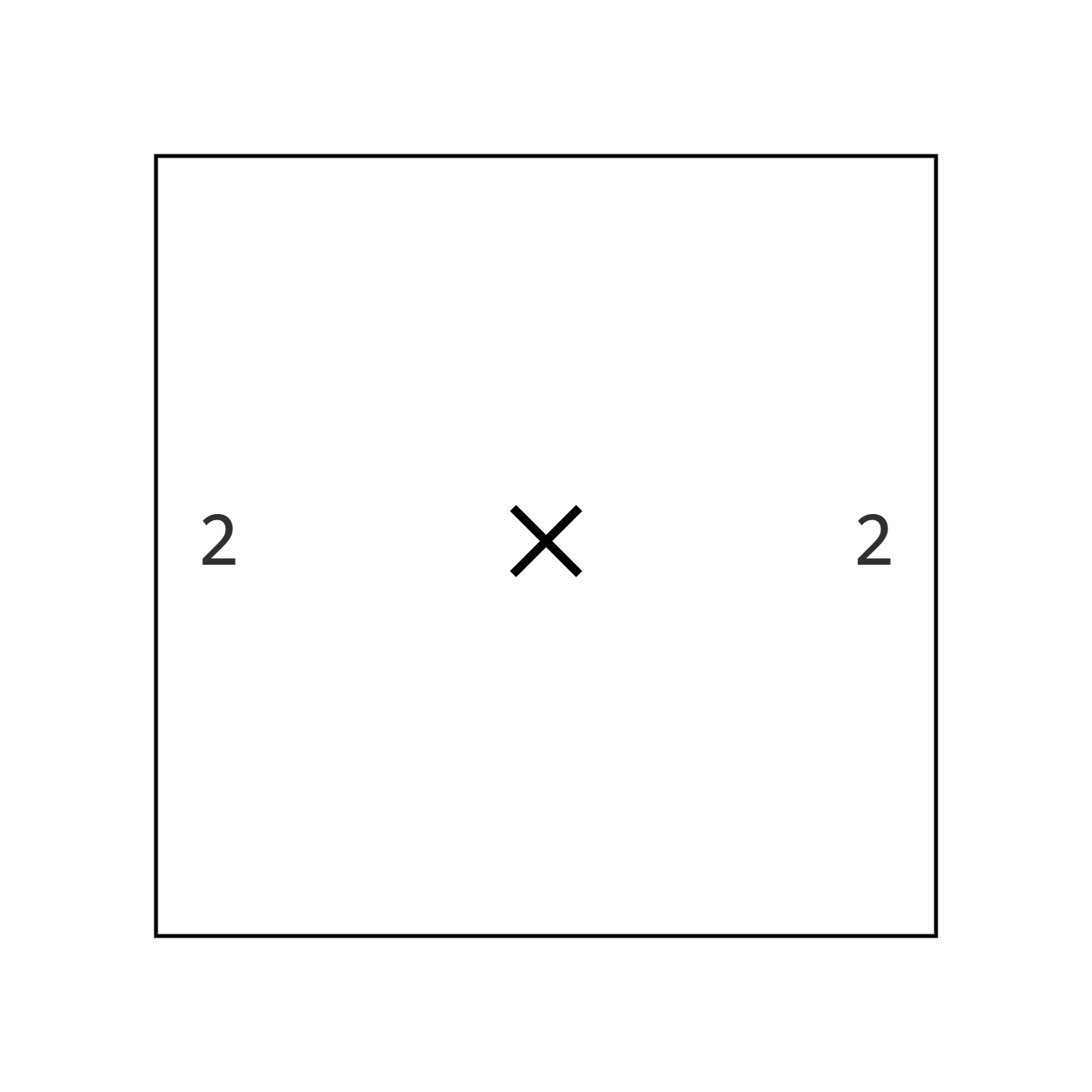}
        \caption{\centering Convert spiders into wires}
        \label{fig:cnot2swap:wire}
    \end{subfigure}
    \caption{\centering Proof that a composition of three CNOT diagrams is a SWAP.}
    \label{fig:cnot2swap}
\end{figure}

\subsection{Circuit Proof Example: 3 CNOTs are SWAP}

\vyzx allows us to use \zxcalc reasoning techniques to prove circuit identities. 
A simple circuit identity is that the composition of three CNOT gates (with alternating target and control qubits) is %, up to a global scalar, 
equivalent to
a SWAP gate. 
Recall the ZX definition of CNOT given in \Cref{fig:cnot}.
% This result can be understood intuitively by considering its action on classical states.
% The CNOT gate sends the state $\ket{b,c}$ to the state $\ket{b,c\oplus b}$, assuming the first register is the control and the second the target.
% So, the circuit sends $\ket{b,c}$ by the first CNOT to $\ket{b,c\oplus b}$, then by the second CNOT (for which control and target are swapped) to $\ket{b\oplus c\oplus b,c\oplus b}=\ket{c,c\oplus b}$, and finally by the third CNOT to $\ket{c,c\oplus b\oplus c}=\ket{c,b}$.
% In sum, the circuit swaps the first and second register, which is the same action as the SWAP gate.
% The \zxdiag proof of this identity does not rely on reasoning about states, instead directly manipulating spiders using the Bialgebra and Hopf rules.
The \zxdiag proof of this identity does not reason at the level of matrices, but instead manipulates spiders using the Bialgebra and Hopf rules.
The \vizx visualization of its steps is given in \Cref{fig:cnot2swap}.
% Intro here; can / should we cite https://zxcalculus.com/ to show it's a common example? I haven't found it anywhere else, yet.

First, we flip the rightmost CNOT and reassociate to move the Z spider to the right of the X spider in the second CNOT (\Cref{fig:cnot2swap:reassoc-bialg}).
This exposes a four-cycle of spiders of alternating colors in the two rightmost gates, to which we can apply the Bialgebra rule (\Cref{fig:cnot2swap:bialg}). 
Then, we can reassociate (\Cref{fig:cnot2swap:reassoc-fusion}) and fuse the remaining spiders (\Cref{fig:cnot2swap:fusion}). As these spiders have two edges between them, we can remove them using the Hopf rule (\Cref{fig:cnot2swap:hopf}).
Finally, the spiders each have 1 input and 1 output, so they can be removed (\Cref{fig:cnot2swap:wire}).

\begin{figure}[b]
    \centering
    \begin{subfigure}[b]{\textwidth}
        \includegraphics[width=\linewidth]{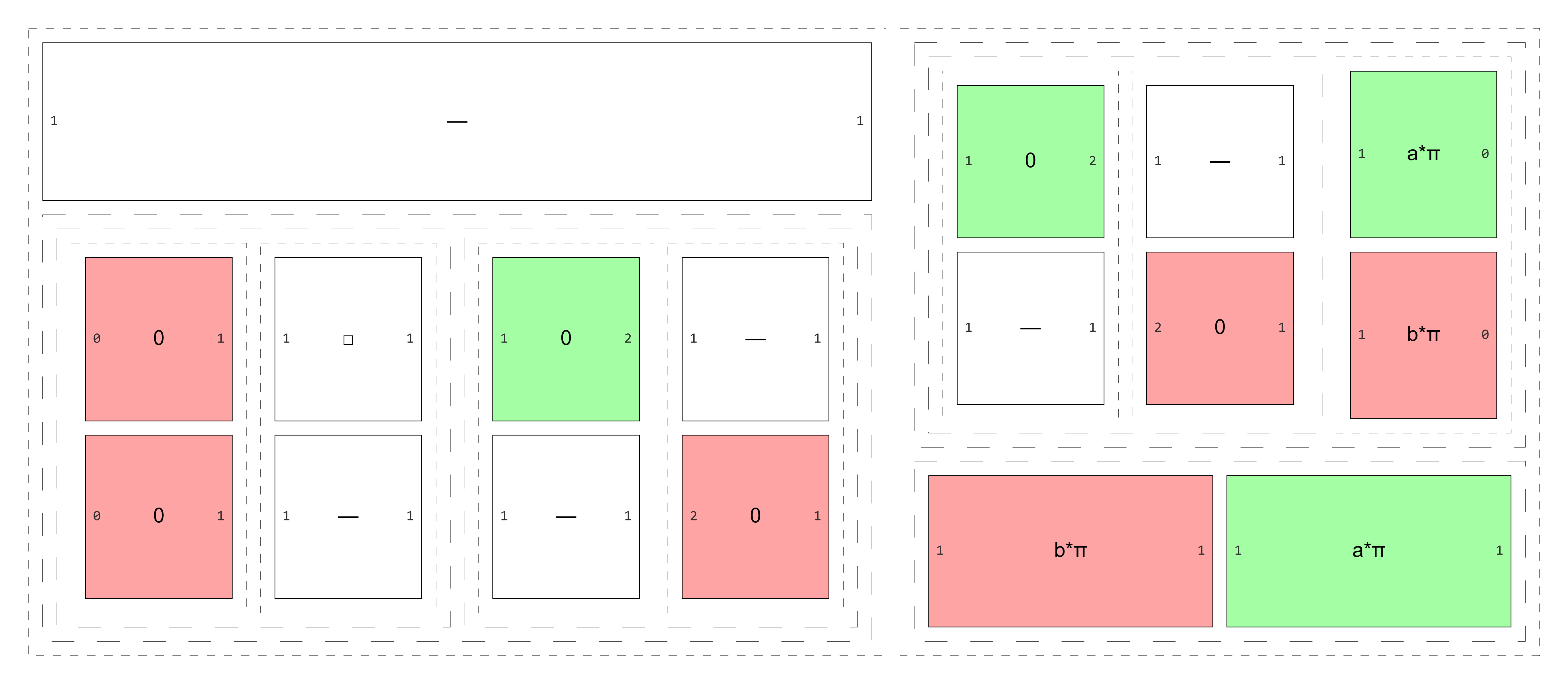}
        \caption{Teleportation lemma, where $a$ and $b$ are booleans that are coerced to $0$ or $1$.}
        \label{fig:teleportation:lemma}
    \end{subfigure}
\end{figure}
\begin{figure}[b]
    \ContinuedFloat
    \begin{subfigure}[b]{0.5\textwidth}
        \includegraphics[width=\linewidth]{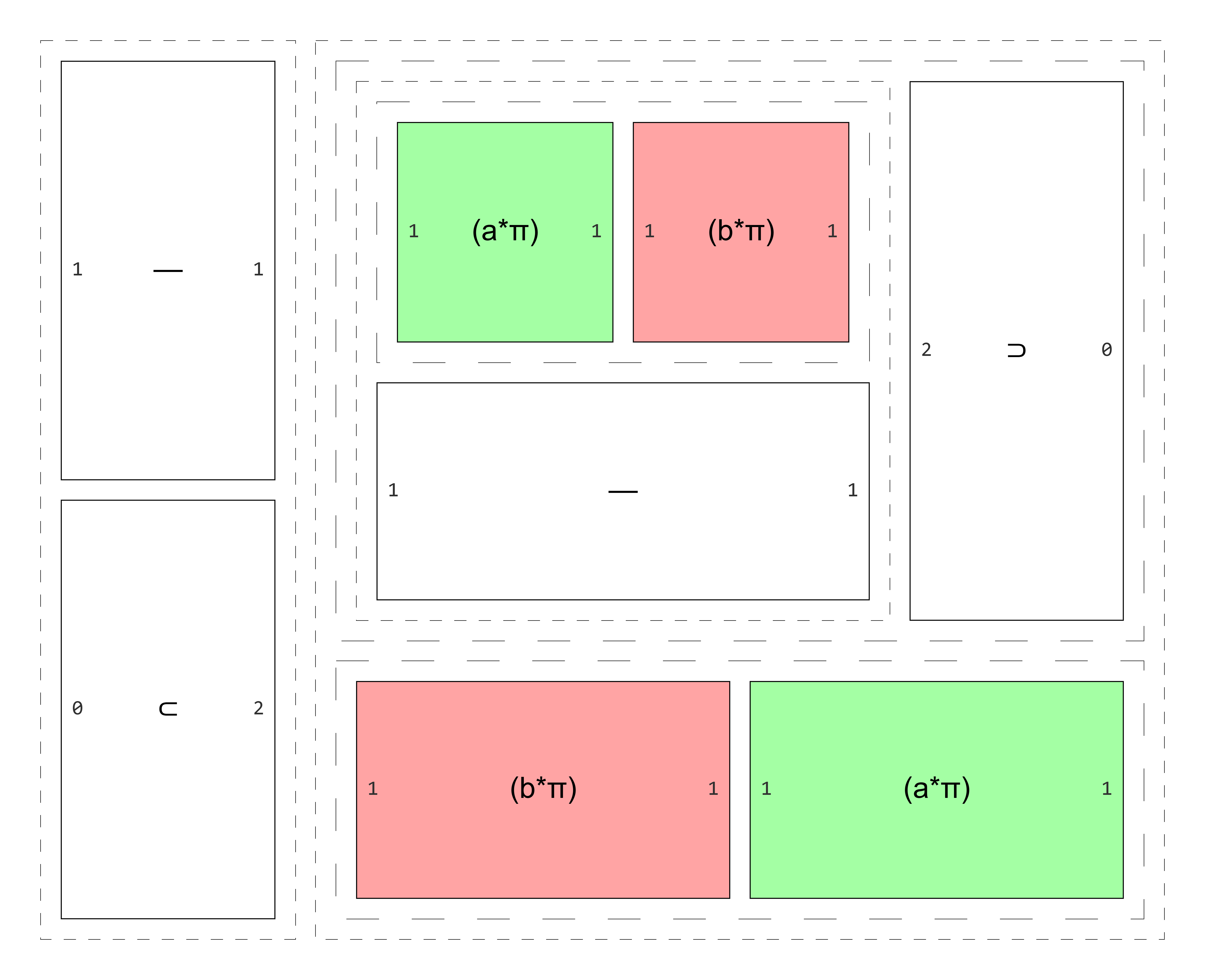}
        \caption{Subdiagrams become cap and cups with measurement results.}
    \end{subfigure}\begin{subfigure}[b]{0.5\textwidth}
        \includegraphics[width=\linewidth]{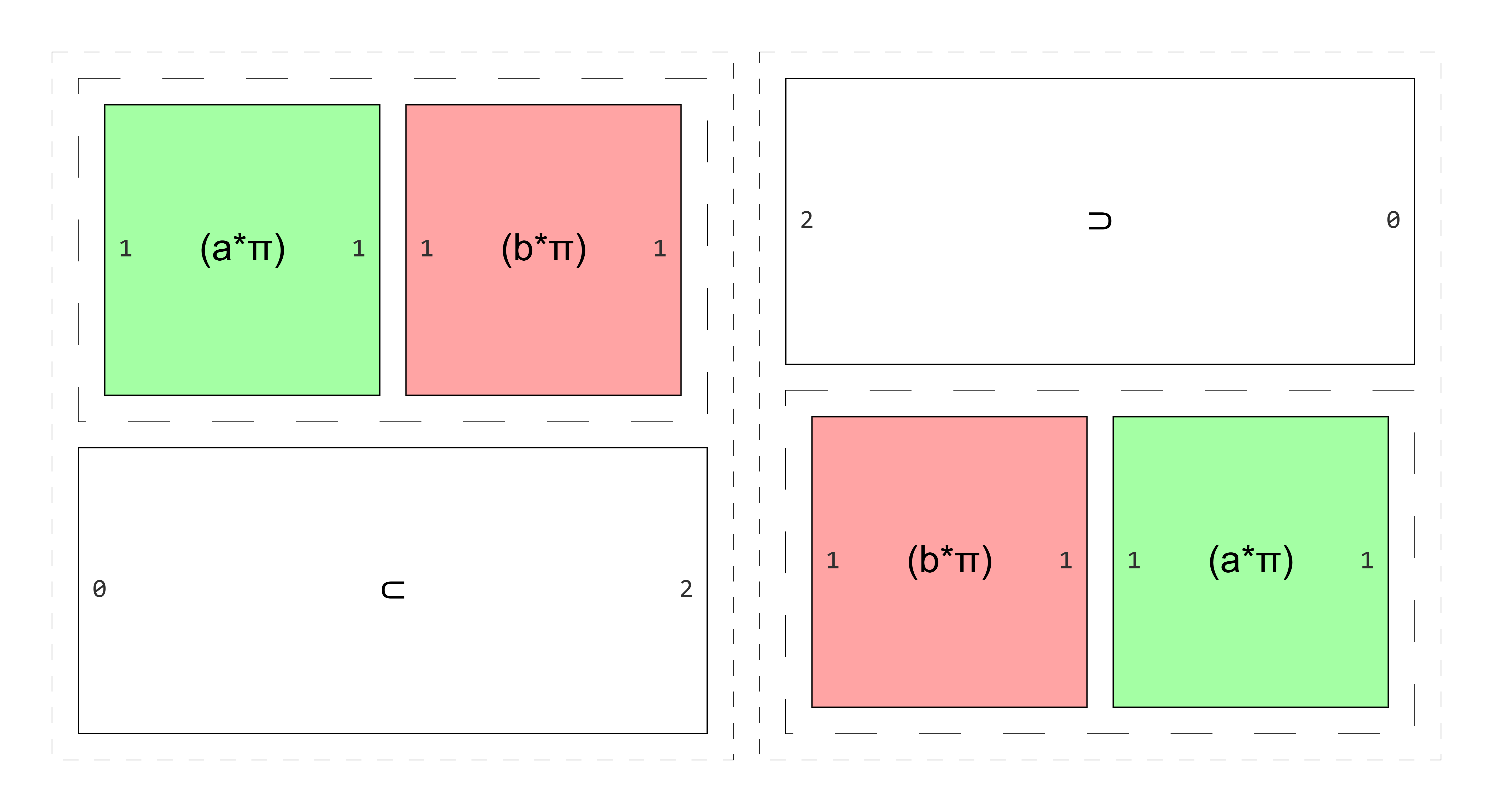}
        \caption{Taking the diagram to a simpler form}
    \end{subfigure}
\end{figure}
\begin{figure}[t]
    \ContinuedFloat
    \begin{subfigure}[b]{0.3\textwidth}
        \includegraphics[width=\linewidth]{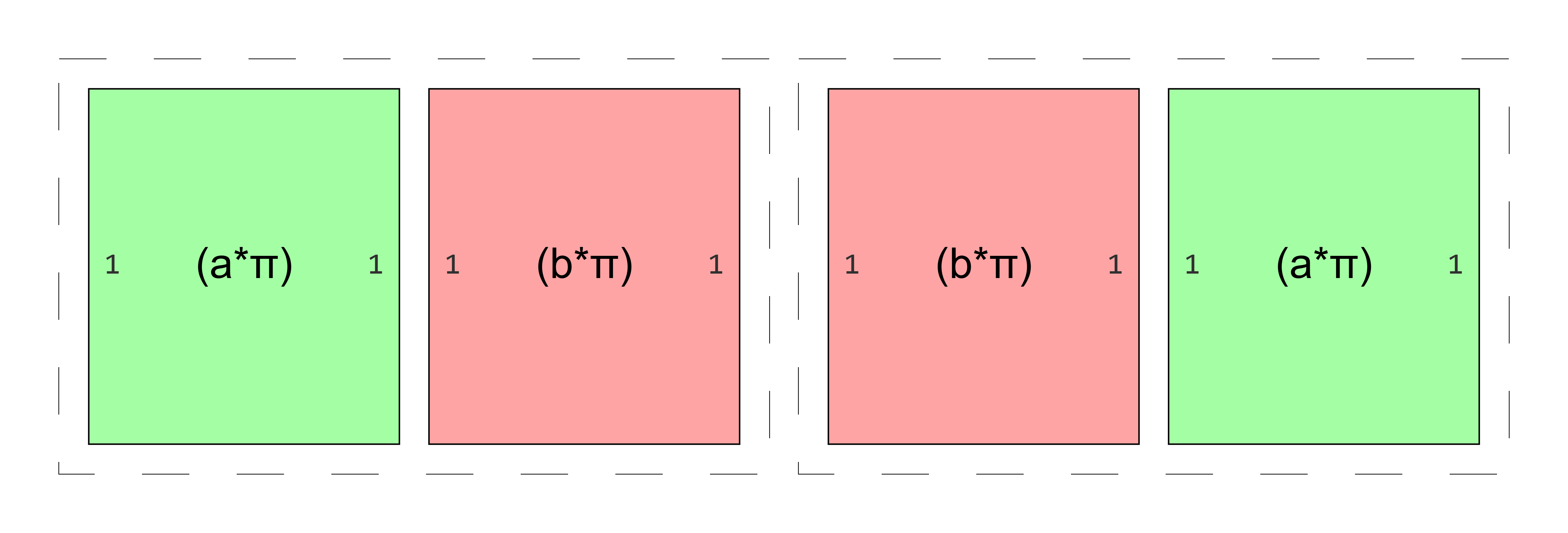}
        \caption{Yank the cap and cup}
    \end{subfigure}\begin{subfigure}[b]{0.3\textwidth}
        \includegraphics[width=\linewidth]{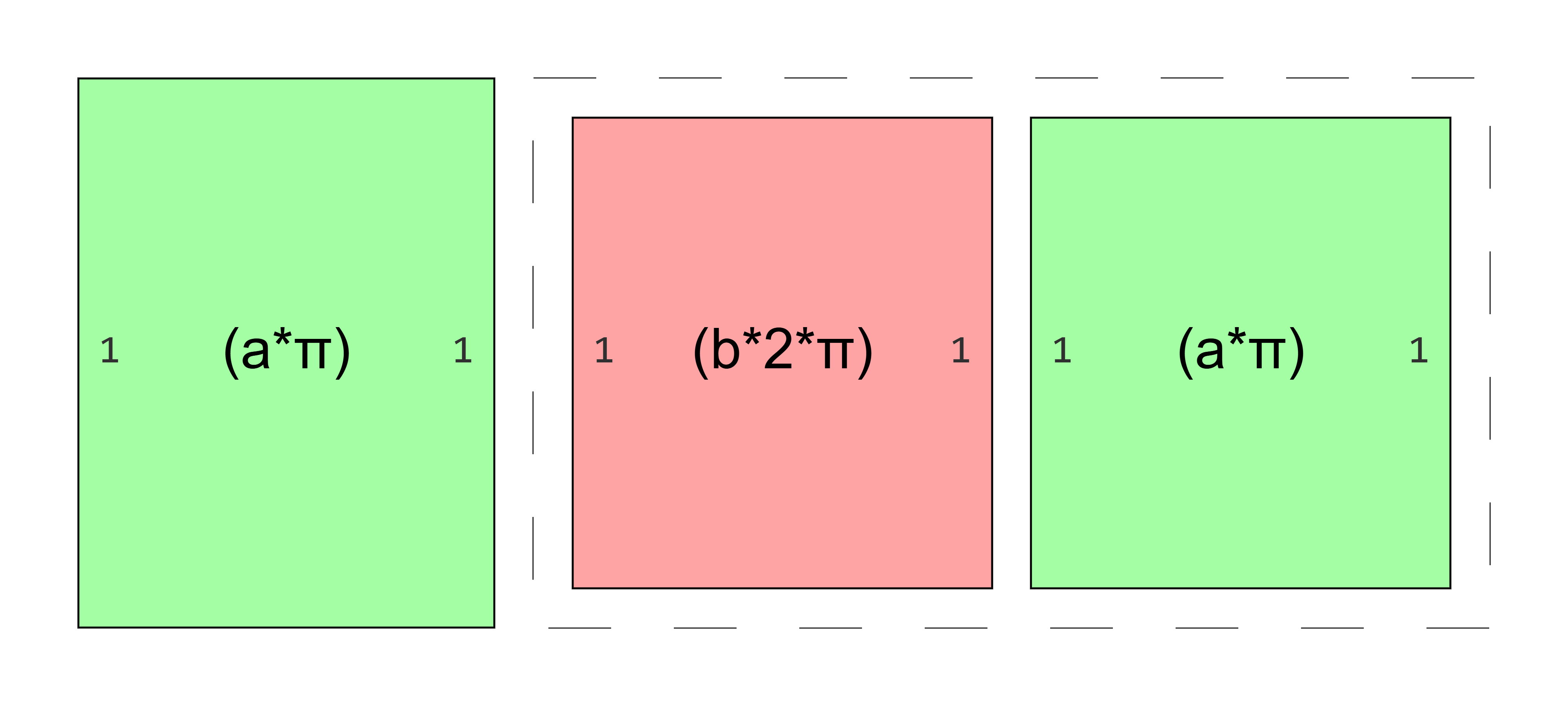}
        \caption{Associate the red spiders}
    \end{subfigure}\begin{subfigure}[b]{0.3\textwidth}
        \includegraphics[width=\linewidth]{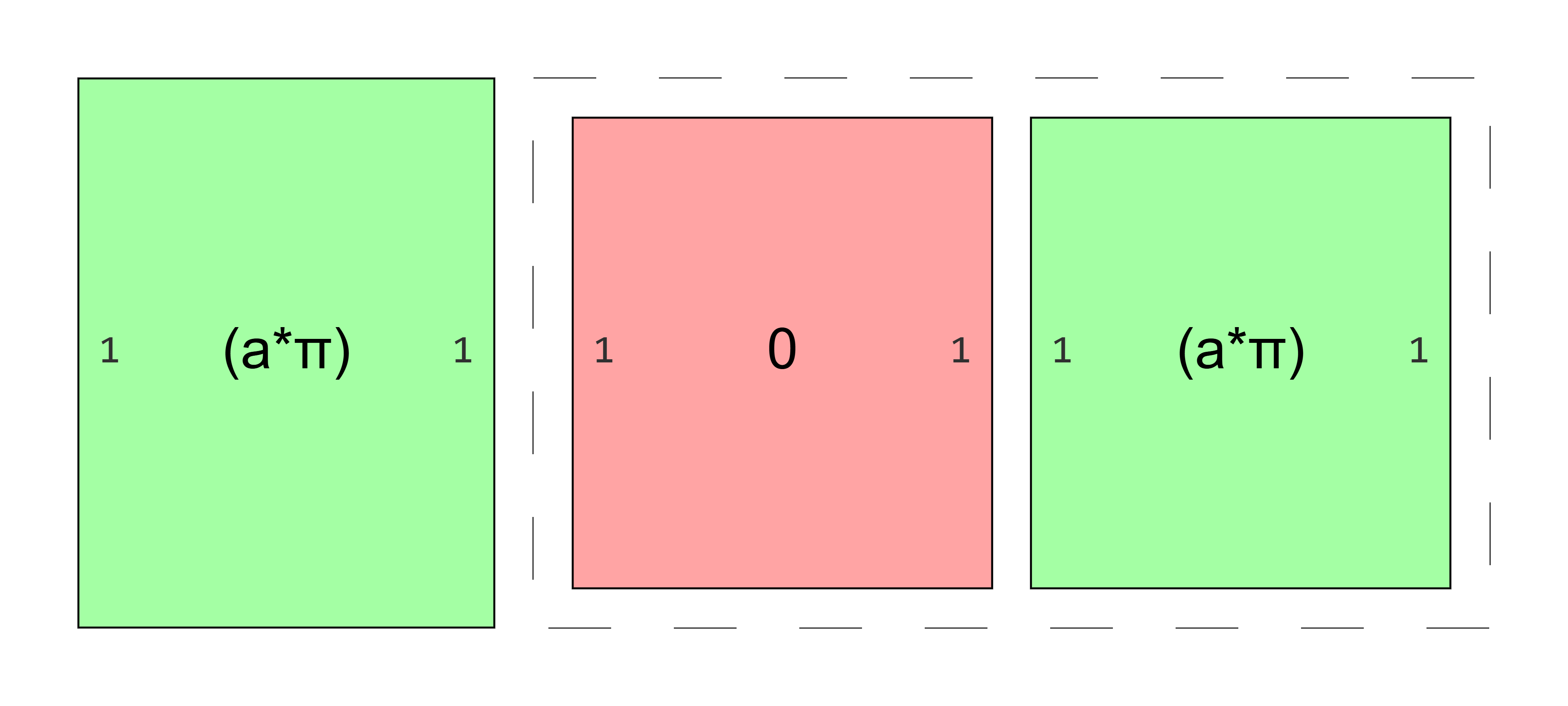}
        \caption{Fuse red spiders}
        \end{subfigure}
\end{figure}
\begin{figure}[t]
    \ContinuedFloat
    \begin{subfigure}[b]{0.25\textwidth}
        \includegraphics[width=\linewidth]{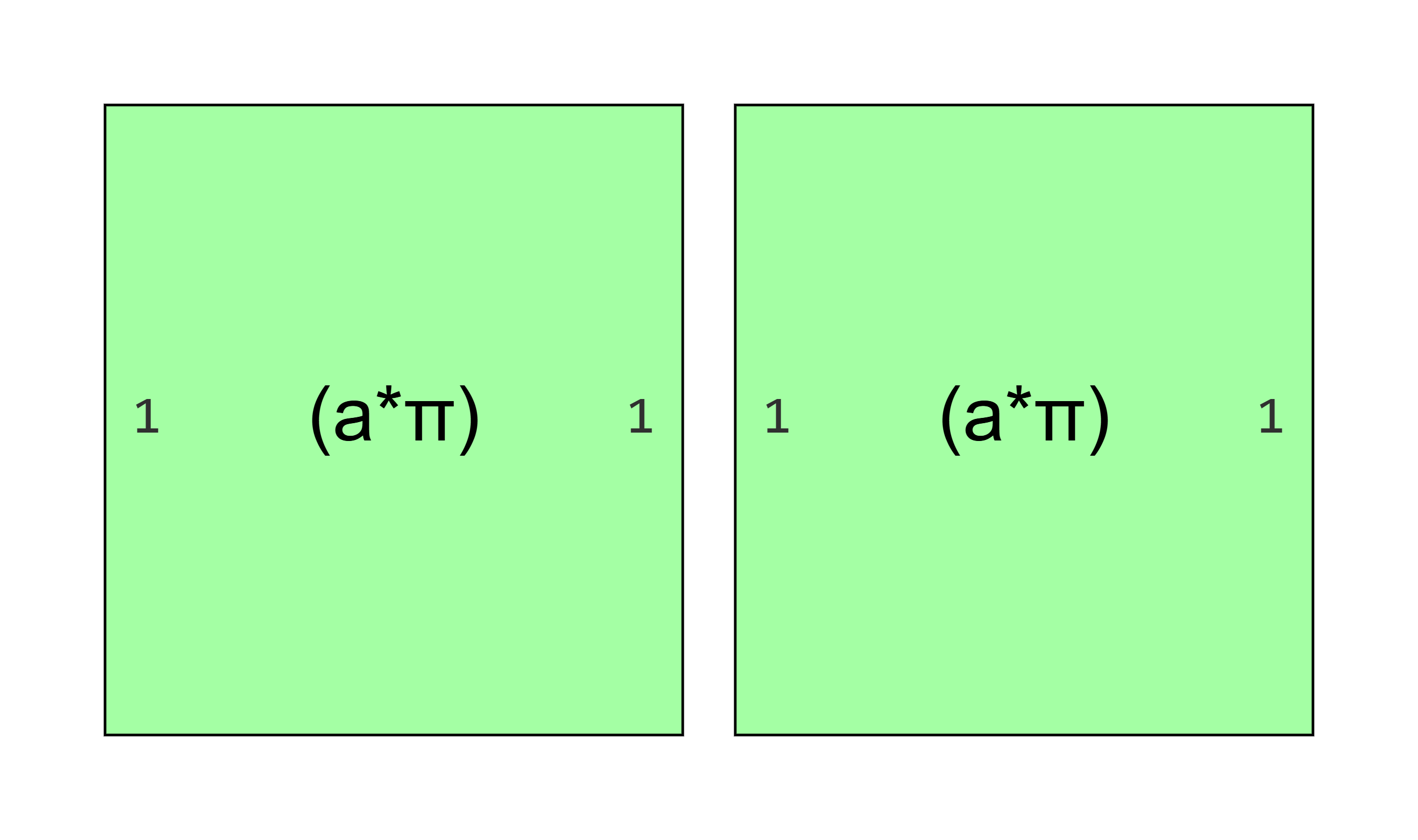}
        \caption{Simplify red spider}
    \end{subfigure}
    \begin{subfigure}[b]{0.2\textwidth}
        \includegraphics[width=\linewidth]{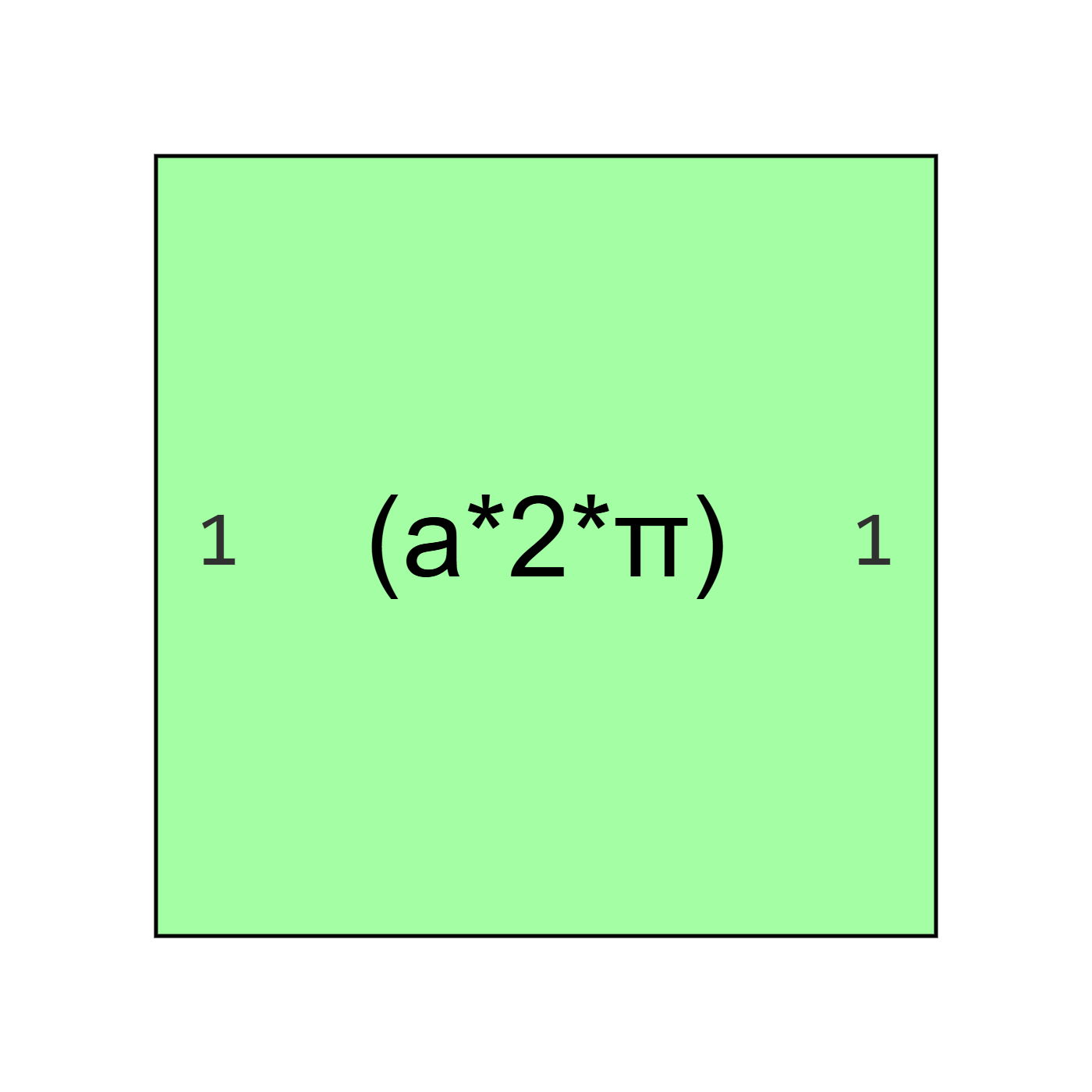}
        \caption{Fuse green spiders}
    \end{subfigure}\begin{subfigure}[b]{0.2\textwidth}
        \includegraphics[width=\linewidth]{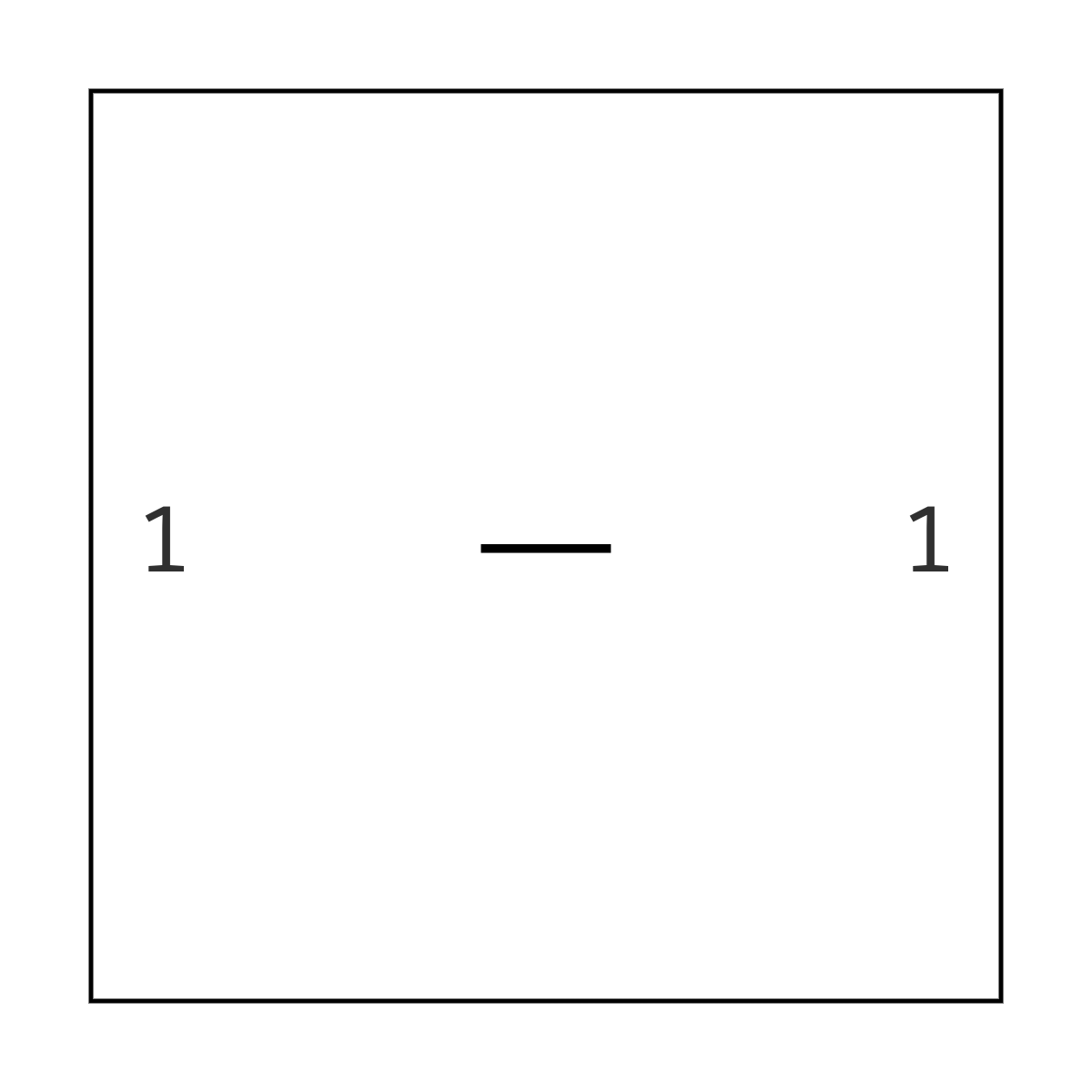}
        \caption{Simplify green spider}
    \end{subfigure}
    \caption{Proof of teleportation in \vyzx by encoding boolean arguments for measurement.}
    \label{fig:teleportation}
    \vspace{-0.5em}
\end{figure}

\subsection{Measurement proof example: Teleportation}

Using \vyzx we can also prove facts about processes which involve measurements. A classic trick to capture measurement and correction is to use a variable to capture the result of the measurement and then later use that same variable to make the correction, as seen in ZX proofs of teleportation \cite{vandewetering2020zxcalculus}. In that way, a Bell measurement over 2 qubits looks like a Bell state where we connect a Z and X spider with a variable to capture the measurement outcome. We have a flipped version of Bell state preparation that captures these two measurements as two booleans in a tuple. Other than that, it is the same. This is an important part of proving teleportation, as it allows us to first manipulate wires and then show that the correction does cancel out any measurement outcomes. We are able to use booleans for our measurements which then get coerced to either $0$ or $1$. If additional precision is needed, or you want to consider measurement to more than just two possible states, natural numbers can be used instead of booleans and will be coerced to reals.

\section{Related work}

\subsubsection*{\zxcalc tools}
\emph{Quantomatic}~\cite{kissinger2015quantomatic} is a diagrammatic proof assistant specific to the \zxcalc which allows for assisted \zxcalc rewrites.
Quantomatic uses a visual interface and allows finite diagrammatic rewrites to be performed using its UI.
Quantomatic isn't verified with respect to underlying linear algebraic semantics, meaning it would have difficulty interfacing with a broader verified library.
It is a useful interactive theorem prover for reasoning about \zxdiags, but it cannot verify programs that operate on \zxdiags.
This is a gap that \vyzx aims to fill.
Quantomatic, however, allows for reasoning through adjacency and hence avoids the structural overhead introduced by \vyzx, leaving it a useful tool for research on theoretical \zxcalc results.

Similarly, Kissinger recently created \chyp~\cite{chyp}, an interactive textual and visual theorem prover for symmetric monoidal categories.
\chyp allows the user to define axioms and generators for any symmetric monoidal category, not just the \zxcalc, to use in proof.
\chyp does not support parametric generators, so every generator must have a fixed finite dimension for its inputs and outputs.
Fixed and finite dimensions are very limiting, in practical use cases, such as in the domains of circuit optimization and error correction where one reasons over diagram families.
\chyp, like Quantomatic is axiomatic and does not allow for extraction.

\subsubsection*{Proof visualization}

There have been efforts in developing primarily graphical proof assistants~\cite{siekmann1999}, as well as integrating visual components into primarily textual proof assistants~\cite{ayers2021}. 
One recent development in this domain is 
the diagram editor Yade for \Coq proofs~\cite{lafont2023}. Yade deploys a bidirectional framework to allow for both construction of diagrammatic proofs from proof scripts, and generation of mechanical proofs from diagrams. The diagrammatic foundations of categorical semantics make it a perfect candidate for such a bidirectional tool, and future work on \vizx intends to generalize it to a bidirectional framework.
Proof assistants such as Lean~\cite{de2015lean} have even successfully integrated interactive, user-specified graphical components into the proof engineering workflow for users to define their own interfaces as they see fit~\cite{nawrocki2023}. 
This approach is far more general than the one we took with \vizx, but may provide a roadmap to introduce such a tool to other domains.

\subsubsection*{Verified quantum computing}
Several attempts have been made to verify quantum computation within a proof assistant, beginning with Boender et al's \Coq library used to prove quantum teleportation~\cite{boender2015}. 
Later, Rand et al. developed a \Coq library to verify programs written in the \qwire quantum programming language; this library later became QuantumLib~\cite{rand2018practice,paykin2017qwire,QuantumLib}.
QuantumLib was used and expanded in the development of the \sqir intermediate representation~\cite{hietala2021sqir}, the \voqc verified optimizing compiler~\cite{hietala2021voqc}, verified quantum oracles~\cite{li2022vqo}, and a proof of Shor's algorithm~\cite{peng2023formally}; it also underlies the present work.

Other attempts to verify quantum computing and quantum protocols include \qbricks~\cite{chareton2021qbricks}, which verified key quantum algorithms using path sums in the Why3 prover, QHLProver~\cite{QHLProver-AFP}, which verified a quantum Hoare logic in Isabelle, and CoqQ~\cite{zhou2023coqQ}, which was able to verify cutting-edge quantum algorithms using \Coq and the Mathematical Components library.
These tools showcase the active effort to verify quantum computation.
Given the relevance of \zxcalc, we believe \vyzx provides the foundation to fill an important gap in verified quantum software.

\subsubsection*{String diagrams for graphical verification}
Concurrently with our own work, Castello et al. proposed \emph{causal separation diagrams} (CSDs) for
reasoning about parallel processes in a proof assistant, which they formalized in Agda~\cite{castello2023inductive}.
A key difference between our work and their work is that their diagrams enforce a strict temporal ordering of operations: ``Only Connectivity Matters'' is antithetical to their application to logical clocks.
The lack of OCM leads to restrictions on composition, which is an interesting aspect that many graphical reasoning systems without OCM properties share.
Their work is very exciting as it implicitly deals with edge labeling, creating a path to reasoning about adjacency.
We take the simultaneous development of \vyzx and CSDs as a sign of burgeoning interest in process theories and diagrammatic reasoning in proof assistants.

\section{Future Work}

\vyzx lays the groundwork for future work on the \zxcalc in \Coq. 
The \zxcalc is particularly useful in quantum circuit optimization, as shown by the \pyzx optimizer~\cite{kissinger2020Pyzx}. 
With the rules proven in \vyzx and some extensions to the interface, a fully verified \pyzx style optimizer is a clear goal for the future, fully integrating with other verified optimizers such as \voqc to provide a full circuit and ZX optimization pass.

There are a variety of useful extensions to the \zxcalc that we could explore adding to \vyzx. 
The most immediate is conditioning on measurement outcomes.
The \zxcalc, by default, captures an idea of postselection.
Additionally, we can capture that our measurement outcome has two possibilities instead of one postselected option by adding boolean variables to spider rotations, following v. d. Wetering~\cite{vandewetering2020zxcalculus}.

Another exciting use of the \zxcalc is for writing error-correcting surface codes \cite{deBeaudrap2020zxcalculusis}. 
With \vyzx, we can implement and verify various surface codes and conversions between them. 
\zxcalc is commonly used to discover measurement patterns and Pauli corrections, and visualize surface codes, and translate physical actions on a code into logical actions on the logical qubit the code represents.
We can use \vyzx to verify the translation between physical and logical actions and the error-resistant properties of these surface codes.

\vizx reasons about graphical structures in the context of a proof goal.
We plan to extend \vizx to enable interactive proof editing of \vyzx graphical structures. 
We believe \vizx is a step towards entirely visual, formally verified proof, similar to unverified tools such as Quantumatic~\cite{kissinger2015quantomatic}.
This work focuses on the \zxcalc, but we also presented a framework for reasoning about connections in other string diagrams.
These are examples of symmetric monoidal categories, which appear in various settings, including higher-dimension topological monoids and vector spaces. 
In initial follow-up work~\cite{vicar}, we generalize \vyzx to reason about arbitrary symmetric monoidal categories, so that \vyzx and the string diagrams of \Cref{sec:inductive} can be treated as simply instances of a broader type class. We also extend the resulting tool to other categories, and add automation and visualization for those categories.
We could further extend the tool to reason about areas with related categorical interpretations, like logic and topology~\cite{rosettastone}. 

A key future direction for \vyzx is automating rewriting across our block diagrams.  
Many of our diagrammatic proofs rely heavily on reassociating our diagrams so that the rewriteable components are adjacent to one another, which requires substantial effort. 
To address this, we propose implementing an automated technique for rewriting across associativity and distributivity in \Coq.
%to bridge this gap. 
%
The rewrite system would take in two structural \zxdiags equal up to associativity information and check their equivalence, enabling us to rewrite within the reassociated diagrams. 
We plan to develop a rewriting \Coq plugin that uses a modern \egraph~\cite{egraphs-nieuwenhuis,egraphs-nelson} rewrite tool~\cite{egg}, to efficiently find structural equivalence proofs.
These proofs will then be extracted to \Coq (using proof explanation strategies such as the one described by Flatt et al. \cite{flatt2022smallproofscongruenceclosure}) and checked by \Coq's typechecker.

A key design decision in verifying a graphical language is the choice of representation.
We found that it was necessary to use a structural representation with clear semantics, unlike graphs. 
The inductive block structure was very useful for defining diagrams and functions on diagrams, including those of parametric size.
We believe future developments of verified graphical languages will benefit from this structural representation as a necessary intermediary between graph-like data and semantic interpretation.
However, there are also clear limitations to the block structure, especially its inability to directly reason about connectivity. 
We plan to develop further tools that allow us to reason at a purely graphical level while maintaining semantic verification using the core structural representation of \vyzx.

% \vspace{-2em}

\section{Conclusion}

We presented \vyzx, a formally verified \zxcalc library.
We defined \zxdiags not in terms of connectivity, but in terms of structure, so we could define semantics in terms of complex-valued matrices.
This structure allowed us to prove facts inductively about \zxdiags, which then enabled us to write fully diagrammatic proofs while retaining the guarantees arising from our matrix-level ground truth.
Using the inductive structure presented challenges, namely dealing with additional structure and not having access to connection information directly.
We presented proof strategies, automation techniques, and a visualizer to deal with these challenges.
These allowed us to prove a complete \zxcalc equational theory, which will help verify the broad range of existing \zxcalc tools and those yet to be developed.
We hope that \vyzx will be used to develop trusted quantum software and the techniques we developed here will prove useful beyond the quantum domain.

% \vspace{-1em}

% Acknowledgments
\begin{acks}                            
% The authors thank John Reppy for pointing us towards \egraphs. We also especially thank Saul Shanabrook, Max Willsey, and Oliver Flatt for their time, their gracious help with \egg/\egglog, and for promptly fixing bugs or adding features when we needed them. 

Thanks to David Spitz for exploring transformations from adjacency lists to \vyzx diagrams and Laura Zielinski and Evan Cook for their work on constructing scalars from ZX diagrams.

This material is based upon work supported by the Air Force Office of Scientific Research under award numbers 
FA95502310361 and FA95502310406, and
EPiQC, an NSF Expedition in Computing, under Grant No. CCF-1730449.
\end{acks}

% Do we still need this?
% \input{sections/dataavb}

% \newpage

%% Bibliography
\bibliography{references}

\newpage

\appendix

 \section{Inductive graphs}
\label{sec:inductive}

Reasoning about a graphical language like ZX inside a proof assistant is difficult. 
As in most graphs, our main concern is connectivity, but graphical languages have semantics, meaning that we need a way to construct diagrams that enforces a consistent order between the graph's nodes.
To find a suitable inductive definition, we turn to the category that underpins the \zxcalc and use categorical definitions to inspire our inductive constructors.
This allows us to look at something more general purpose than just \zxcalc, and discuss how we could graphically reason about any monoidal category.
This section motivates our diagrammatic work in terms of categorical concepts.
However, the underlying category theory isn't necessary to understand the paper as a whole.

Monoidal categories are useful mathematical objects when creating inductive graphs, as every monoidal category admits a graphical language~\cite{Selinger2010} while also being defined by a small collection of equations, which we will cover later in this section.
To get started, we lay out \Cref{dfn:cat} for categories and see how the basic objects have graphical interpretations.
We will call the graphical interpretations diagrams, and each categorical definition will have an associated diagram constructor.
The following definitions are adapted from \cite{joyal1993braided} and \cite{Selinger2010}.

\begin{dfn}[Category] \label{dfn:cat}
   A \textbf{category} $C$ is 
    \begin{enumerate}
        \item A collection of objects of $C$ for which we write $A$ in $C$,
        \item A collection of arrows between objects in $C$ written $f : A \to B$,
        \item A composition operator for arrows $\circ$ such that if $f : A \to B$ and $g : B \to D$ are arrows of $C$, $g \circ f : A \to D$ is an arrow of $C$,
        \item two operations $\mathit{domain}$ and $\mathit{codomain}$ such that for an arrow $f : A \to B$, $\mathit{domain}~f = A$ and $\mathit{codomain}~f = B$, and
        \item for every object $A$ of $C$, an identity arrow $\id_A : A \to A$,
    \end{enumerate}
    where the following equations are satisfied:
    \[ 
        \id_B \circ f = f, \hspace{3em} f \circ \id_A = f, \hspace{3em} (h \circ g) \circ f = h \circ (g \circ f)
    \]
\end{dfn}

\begin{dfn}[Functor] \label{dfn:functor}
    A \textit{functor} $F$ between categories $C$ and $D$, written $F : C \to D$, is really two operations, one on the objects of $C$ and one on the morphisms of $C$ mapping in such a way that for objects $A, B \in C$ and morphisms $f : A \to B, g : B \to E$ satisfying $F (f) : F (A) \to F (B)$ and $F (g \circ f) = F (g) \circ F (f)$ and $F(id_A) = id_{F(A)}$.
\end{dfn}

\begin{dfn}[Natural Transformation]
    A \textit{natural transformation} between two functors $F, G : C \to D$, written $\tau : F \to G$, is given by a family of morphisms $\tau_A$ for each object of $C$ such that for every morphism $f : A \to B$ the following diagram commutes:
    \begin{figure}[h]
        \centering
        \ctikzfig{naturaltransformation}
    \end{figure}
\end{dfn}
Expanding on our basic categorical definitions, we look at the definitions for \textit{monoidal} (\Cref{dfn:monoid}), \textit{symmetric} (\Cref{dfn:symm}), and \textit{autonomous} (\Cref{dfn:autonomous}) to see their diagrammatic interpretations.

\begin{dfn}[Monoidal Category]\label{dfn:monoid}
   A category $\,C$ is \textbf{monoidal} if there exists: 
  \begin{enumerate}
    \item A binary functor $\otimes$ on objects and a left and right identity object for $\otimes$ called $I$.
    \item A binary functor $\otimes$ on arrows such that if $f : A \to B$ and $g : C \to D$ then $f \otimes g : A \otimes C \to B \otimes D$.
    \item Natural isomorphisms \(\alpha_{A,B,C} : (A \otimes B) \otimes C \simeq A \otimes (B \otimes C),\,  \lambda_A: I \otimes A \simeq A,\, \rho_A: A \otimes I \simeq A\)
  \end{enumerate}
  Where $\alpha, \lambda, \rho$ satisfy the following:
  \begin{enumerate}
      \item $(f\otimes(g\otimes h))\circ \alpha_{A,B,C} = \alpha_{A',B',C'}\circ ((f\otimes g)\otimes h)$,
      \item $f \circ \lambda_A = \lambda_{A'} \circ (\id_I \otimes f)$,
      \item $f \circ \rho_A = \rho_{A'} \circ (f \otimes \id_I)$,
  \end{enumerate}
  With the following axioms also holding:
  \begin{enumerate}
      \item $\otimes$ is a bifunctor, so $\id_A \otimes \id_B = \id_{A\otimes B}$ and $(k \otimes h) \circ (g \otimes f) = (k \circ g) \otimes (h \circ f)$,
      \item The ``pentagon'' and ``triangle'' coherence axioms given in \cref{fig:hexandtri} commute.
  \end{enumerate}
\end{dfn}
\begin{figure}[h]
    \begin{subfigure}{0.6\textwidth}
        \ctikzfig{pentagonaxiom}
    \end{subfigure}\begin{subfigure}{0.4\textwidth}
        \ctikzfig{triangleaxiom}
    \end{subfigure}
    \caption{The pentagon and triangle coherence axioms.}\label{fig:hexandtri}
\end{figure}
Objects, arrows, and compositions have a natural interpretation as \emph{string diagrams}, shown in \Cref{fig:catdiagrams}.
\begin{figure}[t]
    \begin{tabular}{rlrl}
         \text{Object } $A \equiv$ & \begin{tikzpicture}
	\begin{pgfonlayer}{nodelayer}
		\node [style=none] (6) at (1, 0) {};
		\node [style=none] (18) at (2, 0.5) {$A$};
		\node [style=none] (19) at (3, 0) {};
	\end{pgfonlayer}
	\begin{pgfonlayer}{edgelayer}
		\draw (19.center) to (6.center);
	\end{pgfonlayer}
\end{tikzpicture} &
         \text{Arrow } $f : A \to B \equiv$ & \begin{tikzpicture}
	\begin{pgfonlayer}{nodelayer}
		\node [style=none] (6) at (1, 0) {};
		\node [style=none] (18) at (1.5, 0.5) {$A$};
		\node [style=box] (19) at (2.5, 0) {$f$};
		\node [style=none] (20) at (4, 0) {};
		\node [style=none] (21) at (3.5, 0.5) {$B$};
	\end{pgfonlayer}
	\begin{pgfonlayer}{edgelayer}
		\draw (20.center) to (19);
		\draw (19) to (6.center);
	\end{pgfonlayer}
\end{tikzpicture} \\
         & & & \\
         \text{Compose } $\circ \equiv$ & \begin{tikzpicture}
	\begin{pgfonlayer}{nodelayer}
		\node [style=none] (0) at (-2, 0) {$f$};
		\node [style=none] (1) at (2, 0) {$g$};
		\node [style=none] (2) at (-2.5, 0.5) {};
		\node [style=none] (3) at (-1.5, -0.5) {};
		\node [style=none] (4) at (-1.5, 0.5) {};
		\node [style=none] (5) at (-2.5, -0.5) {};
		\node [style=none] (6) at (1.5, 0.5) {};
		\node [style=none] (7) at (2.5, -0.5) {};
		\node [style=none] (8) at (2.5, 0.5) {};
		\node [style=none] (9) at (1.5, -0.5) {};
		\node [style=none] (10) at (0, 1.25) {};
		\node [style=none] (11) at (0, -1.25) {};
		\node [style=none] (12) at (-3.25, 0) {};
		\node [style=none] (13) at (-2.5, 0) {};
		\node [style=none] (14) at (-1.5, 0) {};
		\node [style=none] (15) at (1.5, 0) {};
		\node [style=none] (16) at (2.5, 0) {};
		\node [style=none] (17) at (3.25, 0) {};
		\node [style=none] (18) at (0, 0) {};
		\node [style=none] (19) at (0, 0) {};
		\node [style=none] (20) at (-0.75, 0.5) {$B$};
		\node [style=none] (21) at (0.75, 0.5) {$B$};
		\node [style=none] (22) at (3.25, 0.5) {$D$};
		\node [style=none] (23) at (-3.25, 0.5) {$A$};
	\end{pgfonlayer}
	\begin{pgfonlayer}{edgelayer}
		\draw (4.center)
			 to (3.center)
			 to (5.center)
			 to (2.center)
			 to cycle;
		\draw (9.center)
			 to (6.center)
			 to (8.center)
			 to (7.center)
			 to cycle;
		% \draw [style=double edge] (11.center) to (10.center);
		\draw (17.center) to (16.center);
		\draw (13.center) to (12.center);
		\draw (19.center) to (15.center);
		\draw (18.center) to (14.center);
	\end{pgfonlayer}
\end{tikzpicture} &
         \text{Tensor product } $f \otimes g \equiv$ & \begin{tikzpicture}
	\begin{pgfonlayer}{nodelayer}
		\node [style=none] (0) at (0, 1.25) {$f$};
		\node [style=none] (1) at (-1, 2.25) {};
		\node [style=none] (2) at (1, 0.25) {};
		\node [style=none] (3) at (1, 2.25) {};
		\node [style=none] (4) at (-1, 0.25) {};
		\node [style=none] (7) at (0, -1.25) {$g$};
		\node [style=none] (8) at (-1, -0.25) {};
		\node [style=none] (9) at (1, -2.25) {};
		\node [style=none] (10) at (1, -0.25) {};
		\node [style=none] (11) at (-1, -2.25) {};
	\end{pgfonlayer}
	\begin{pgfonlayer}{edgelayer}
		\draw (4.center) to (1.center);
		\draw (1.center) to (3.center);
		\draw (3.center) to (2.center);
		\draw (2.center) to (4.center);
		% \draw [style=double edge] (6.center) to (5.center);
		\draw (11.center) to (8.center);
		\draw (8.center) to (10.center);
		\draw (10.center) to (9.center);
		\draw (9.center) to (11.center);
	\end{pgfonlayer}
\end{tikzpicture} \\
    \end{tabular}
    \caption{\centering The diagrammatic construction for monoidal category definitions, where $D_1$ and $D_2$ are arbitrary diagrams.}\label{fig:catdiagrams}
\end{figure}
The definition of a monoidal category adds a sense of parallelism to our category.
This will allow us to take two objects or arrows and set them in parallel, or take two operations and perform them in parallel as well.
The isomorphism $\alpha_{A, B, C}$ allows us to reassociate an object $(A \otimes B) \otimes C$ to $A \otimes (B \otimes C)$.
We can take two arrows $f$ and $g$ and form a new arrow $f \otimes g$ or pass information along with $f \otimes \id_A$.
The identity object $I$ can be represented using the empty diagram, which is just blank space, allowing us to satisfy the equations $\lambda_A$ and $\rho_A$ in our diagram.

To define a symmetric category, we first define a braided category.
\begin{dfn}[Braided Category] \label{dfn:braid}
     A monoidal category $C$ is \textbf{braided} if there exists a natural family of isomorphisms $c_{A,B} : A \otimes B \to B \otimes A$ such that the diagrams in \Cref{fig:hexagon} commute.
     \begin{figure}[h]
         \centering
            \ctikzfig{hexagon1}
            \ctikzfig{hexagon2}
         \caption{The Hexagon isomorphisms for braided categories}
         \label{fig:hexagon}
     \end{figure}
     Intuitively we can understand the braiding isomorphisms $\beta_{A,B}$ and $\beta^{-1}_{A,B}$ as the two ways we could lay pieces of string over one another. Either we can have string $A$ go over string $B$ or have string $B$ go over string $A$.
\end{dfn}
Diagrammatically, this gives us two ways to wrap wires, picking one to be on top of the other.
Since we are only interested in symmetric monoidal categories for our purposes, we immediately extend the notion of a braided category.
\begin{dfn}[Symmetric Category]\label{dfn:symm}
     A braided category $C$ is \textbf{symmetric} if $\beta_{A,B} \cong \beta_{B,A}^{-1}$.
\end{dfn}
The symmetric braid can be interpreted as string diagrams as follows.
\begin{figure}[h]
    \begin{tabular}{rl}
        \text{Symmetric braid } $\beta_{A,B} \equiv$ & \begin{tikzpicture}
	\begin{pgfonlayer}{nodelayer}
		\node [style=none] (0) at (-1.5, 1) {};
		\node [style=none] (1) at (-1.5, -1) {};
		\node [style=none] (2) at (1.5, 1) {};
		\node [style=none] (3) at (1.5, -1) {};
        \node [style=none] (4) at (-1.5, 1.5) {A};
        \node [style=none] (4) at (-1.5, -0.5) {B};
        \node [style=none] (4) at (1.5, 1.5) {B};
        \node [style=none] (4) at (1.5, -0.5) {A};
	\end{pgfonlayer}
	\begin{pgfonlayer}{edgelayer}
		\draw [in=-180, out=0] (0.center) to (3.center);
		\draw [in=0, out=180] (2.center) to (1.center);
	\end{pgfonlayer}
\end{tikzpicture}
    \end{tabular}
    \caption{The diagrammatic interpretations for symmetric categories.}\label{fig:symdiagram}
\end{figure}

To start defining an autonomous category, we first define the dual of an object.
\begin{dfn}[Exact Pairing]\label{def:exactpairing}
     In a monoidal category, an \textbf{exact pairing} between two objects $A$ and $B$ is given by a pair of morphisms $\nu : I \to B \otimes A$ and $\epsilon : A \otimes B \to I$ such that the following diagrams (written as if our category were strict, without loss of generality~\cite{Selinger2010}), commute.
    \ctikzfig{exactpairing}
    In an exact pairing, we say that $A$ is the left dual of $B$ and $B$ the right dual of $A$ and we call $\nu$ the unit and $\epsilon$ the co-unit. We can visualize these as follows.
    \begin{table}[h]
        \centering
        \begin{tabular}{rlrl}
            \text{Unit } $\nu_{A} \equiv$ & \begin{tikzpicture}
	\begin{pgfonlayer}{nodelayer}
		\node [style=none] (2) at (0.5, 1) {};
		\node [style=none] (3) at (0.5, -1) {};
        \node [style=none] (4) at (0.5, 1.5) {A};
        \node [style=none] (5) at (0.5, -0.5) {A};
	\end{pgfonlayer}
	\begin{pgfonlayer}{edgelayer}
		\draw [bend right=90, looseness=1.50] (2.center) to (3.center);
	\end{pgfonlayer}
\end{tikzpicture} &
            \text{Co-unit } $\epsilon_{A} \equiv$ &\begin{tikzpicture}
	\begin{pgfonlayer}{nodelayer}
		\node [style=none] (0) at (-0.5, 1) {};
		\node [style=none] (1) at (-0.5, -1) {};
        \node [style=none] (3) at (-0.5, 1.5) {A};
        \node [style=none] (4) at (-0.5, -0.5) {A};
	\end{pgfonlayer}
	\begin{pgfonlayer}{edgelayer}
		\draw [bend left=90, looseness=1.50] (0.center) to (1.center);
	\end{pgfonlayer}
\end{tikzpicture}
        \end{tabular}
    \end{table}
\end{dfn}
\begin{dfn}[Autonomous Category]\label{dfn:autonomous}
     We say a monoidal category is \textbf{autonomous} if every object has both a left and a right dual (\Cref{def:exactpairing}).
\end{dfn}
We can see the graphical definitions for a unit and co-unit in \Cref{def:exactpairing}.
Given these definitions, we have a minimal collection of necessary objects to define our inductive data structure.
The simplest form of symmetric monoidal category we can define will be string diagrams with only the identity arrow, which have semantics that capture the connections from some input or output of a diagram to another input or output.
This is the simplest form of a diagram we can create, and we can extend this idea later to build diagrams for the \zxcalc.
We assign these different names in \Cref{tab:cat-to-ind} to match their diagrammatic interpretation more closely.
\begin{table}[H]
    \centering
    \begin{tabular}{|c|c|c|}
      \hline
      Categorical Concept & Inductive Constructor & Symbol \\
      \hline
      $\id_A$ & Wire & \(\--\)\\ \hline
      $I$    & Empty & \(\revemptyset\) \\ \hline
      $f \circ g$ & Compose $f$ $g$ & \(g \leftrightarrow f\) \\ \hline
      $\otimes$ & Stack & \(\updownarrow\) \\ \hline
      Symmetric braid & Swap 1 1 & \(\times\)\\ \hline
      Unit & Cap & \(\subset\) \\ \hline
      Co-unit & Cup & \(\supset\) \\ \hline
    \end{tabular}
    \caption{Translating categorical concepts to inductive constructors with their respective symbolic notation.}
    \label{tab:cat-to-ind}
\end{table}
With all of our categorical concepts and constructors for diagrams laid out, we can use these exact constructors to define string diagrams.
We define inductive constructors for string diagrams \texttt{SD} (\(n~m : \mathbb{N}\)) with $n$ inputs and $m$ outputs as:
\[
        \inferrule{ }{\oftype{\texttt{Cup}}{\sdold{0}{2}}}
        \hspace{0.85em}
        \inferrule{ }{\oftype{\texttt{Cap}}{\sdold{2}{0}}}
        \hspace{0.85em}
        \inferrule{ }{\oftype{\texttt{Wire}}{\sdold{1}{1}}}
\] \[
        \hspace{0.85em}
        \inferrule{ }{\oftype{\texttt{Swap n m}}{\sdold{(n + m)}{(m + n)}}}
        \hspace{0.85em}
        \inferrule{ }{\oftype{\texttt{Empty}}{\sdold{0}{0}}}
\] \[
        \inferrule
        {\oftype{\texttt{sd\textsubscript{0}}}{\sdold{in}{mid}} \\ \oftype{\texttt{sd\textsubscript{1}}}{\sdold{mid}{out}}}
        {\oftype{\texttt{Compose sd\textsubscript{0} sd\textsubscript{1}}}{\sdold{in}{out}}}
\] \[
        \inferrule
        {\oftype{\texttt{sd\textsubscript{0}}}{\sdold{in\textsubscript{0}}{out\textsubscript{0}}} \\ \oftype{\texttt{sd\textsubscript{1}}}{\sdold{in\textsubscript{1}}{out\textsubscript{1}}}}
        {\oftype{\texttt{Stack sd\textsubscript{0} sd\textsubscript{1}}}{\sdold{(in\textsubscript{0} + in\textsubscript{1})}{(out\textsubscript{0} + out\textsubscript{1})}}}
\]
For visualizations of cap and cup, refer to \cref{def:exactpairing}, for swap refer to \cref{fig:symdiagram}, for Wire, Compose, and Stack refer to \cref{fig:catdiagrams}.
We will refer to the diagrams generated by these constructors as the \emph{block form} for string diagrams.
They can be seen as blocks we can stack on top of one another or compose side by side if their dimensions are equal.

\begin{subsection}{Connection Information in the Block Structure}

Existing soundness and completeness results show that these graphical languages and their connection information properly reflect the categories they represent~\cite{Selinger2010}. 
Completeness says that if two diagrams are equivalent up to some manipulation of the diagram (say twisting a wire around, moving a box, or stretching wires while preserving connections), then that manipulation can be explained as a consequence of the axioms of the category. In the other direction, soundness states that every axiom of a category holds in the diagrammatic language, up to some manipulation of the diagram. For example, take $\id_A \circ f = f$. This is an axiom, but at a diagrammatic level it just looks like shortening the wire on the left side of the diagram. These two results together are known as coherence results for diagrammatic languages.
All that remains for us is to take a look at the connection information of the different rules to justify why we call these diagrams ``graphs''.
To do this, we can check the axioms given in the prior definitions \Cref{dfn:cat} through \Cref{dfn:autonomous} and observe they do not modify the connection information of diagrams.
We take our block structure constructors from above and view them as string diagrams to do this.
We then can observe that the three isomorphisms maintain connection information.
From \Cref{fig:hexcon} and \Cref{fig:tricon}, we can see the isomorphisms that move wires around do not modify the connection information of the diagram.
The remaining isomorphisms $\lambda_A$ and $\rho_A$ do not change our diagram visually, as they add or remove the empty diagram which contains no connections.
The same is true for $\alpha_{A,B,C}$, which only reassociates objects, while diagrams have no parentheses.
This means the minimum collection of rules required for our string category will not modify any connection information, telling us we can use these constructors to describe graphs.
We can specify our category's arrows to extend our inductive construction beyond a minimal collection of rules.
When extending our string diagrams to define the \zxcalc in \Cref{sec:vyzx-diags}, we have to add constructors to represent the Z and X spiders.
Further, we will define an appropriate semantic interpretation for the ZX calculus.
\begin{figure}[h]
    \ctikzfig{hexagonconnections}
    \caption{The connection information for the hexagon isomorphism; see \Cref{dfn:braid}.}\label{fig:hexcon}
\end{figure}
\begin{figure}[h]
    \ctikzfig{triangleconnections}
    \caption{The connection information for the triangle isomorphisms; see \Cref{def:exactpairing}.}\label{fig:tricon}
\end{figure}
\end{subsection}

\section{The complete equational theory}\label{sec:equational-theory}

In this appendix, we present a complete set of \zxcalc rules proven within \vyzx and rendered using \vizx. We do not show all of the rules proven within \vyzx, but instead a sufficient subset for complete equational reasoning.

\begin{figure}[h]
    \centering
    \begin{subfigure}{.7\textwidth}
        \centering
        \includegraphics[width=\linewidth]{figures/vizx_spider_fusion.png}
        \caption{\centering Spider fusion in \vyzx}
        % Jeandel: S1
        \label{fig:vizx-bi-alg}
    \end{subfigure}
    \begin{subfigure}{0.35\textwidth}
        \centering
        \includegraphics[width=\linewidth]{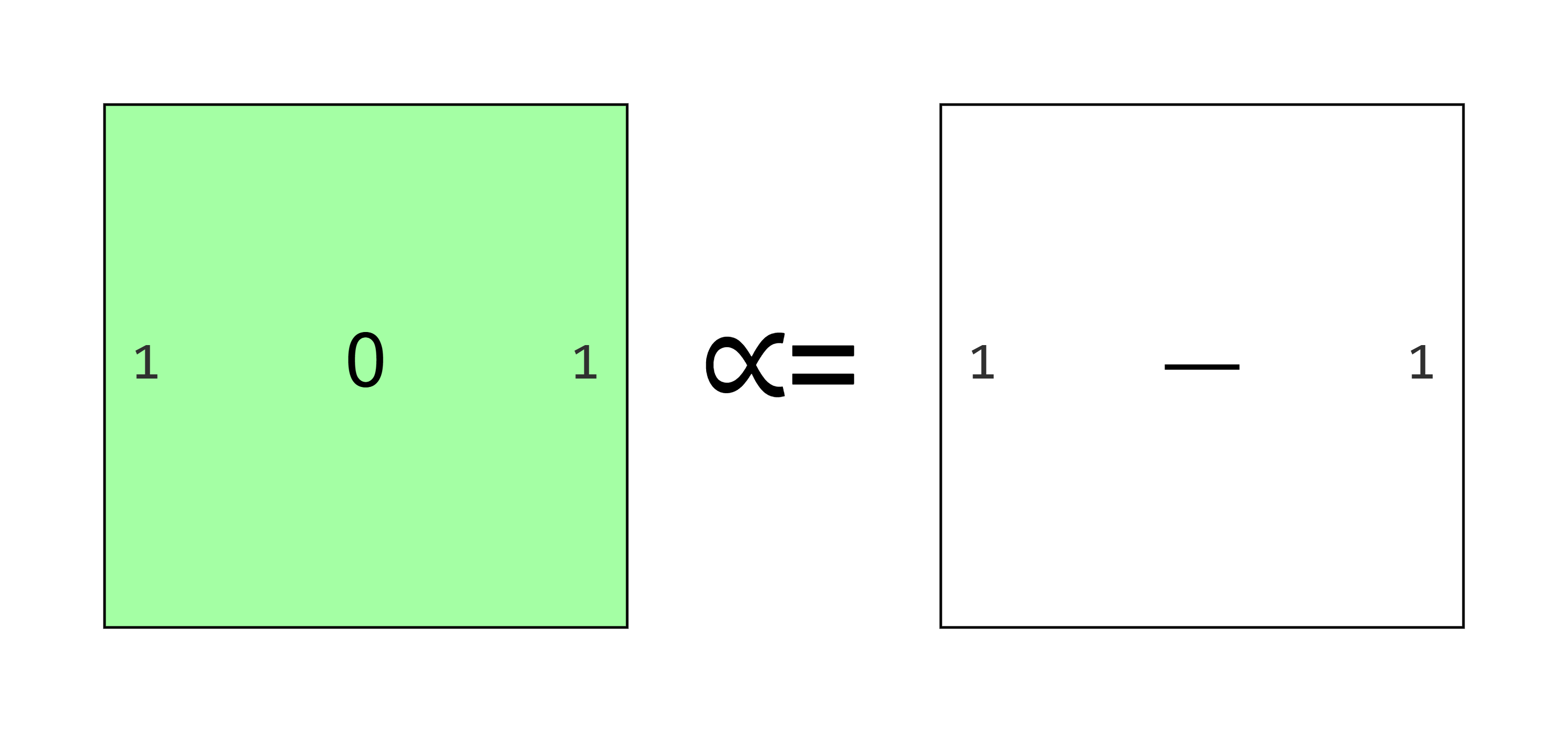}
        \caption{\centering Blank spider removal to wire}
        % Jeandel: S2
        \label{fig:vizx-blank-spider}
    \end{subfigure}\begin{subfigure}{0.35\textwidth}
        \centering
        \includegraphics[width=\linewidth]{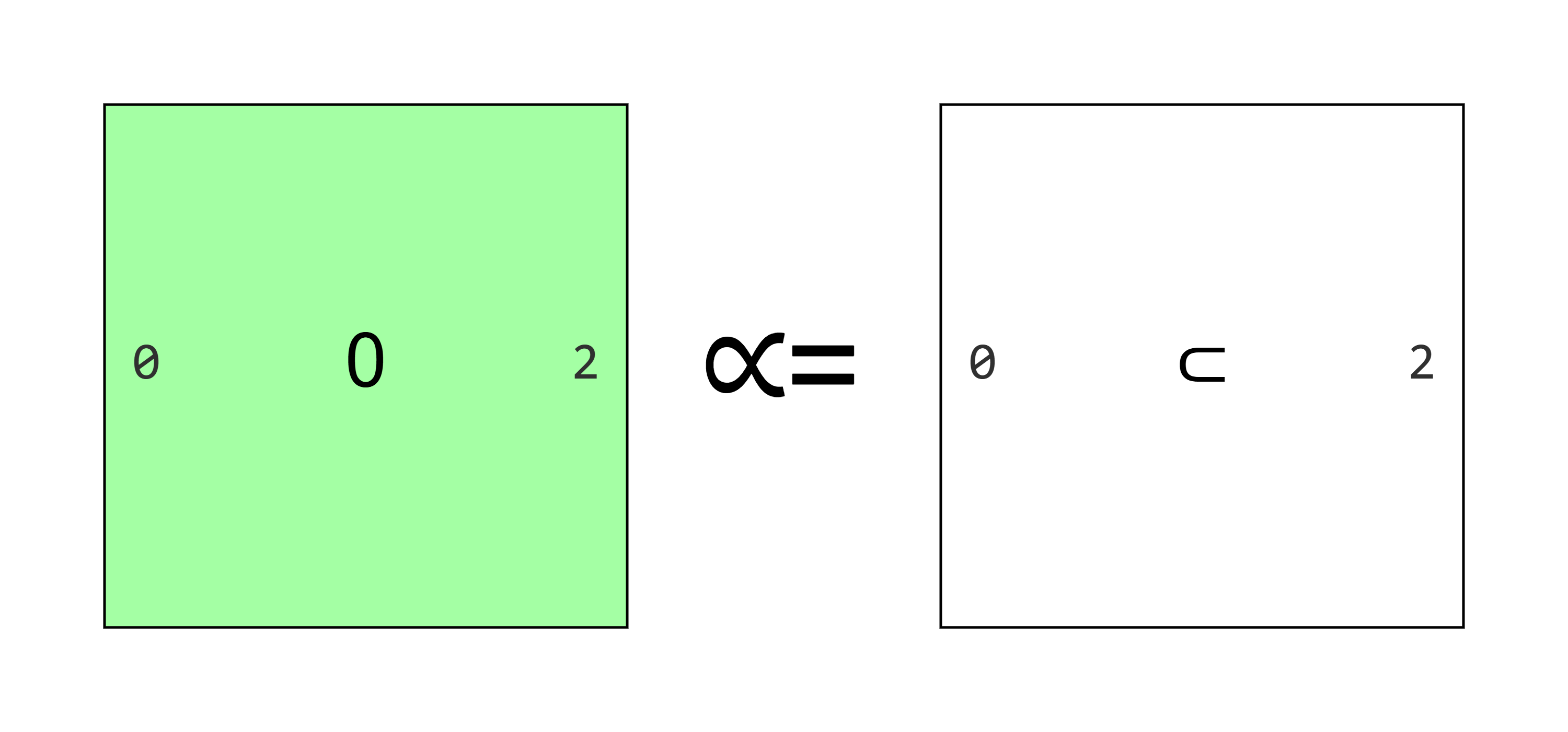}
        \caption{\centering Blank spider removal to cup}
        % Jeandel: S3
    \end{subfigure}
    \begin{subfigure}{.7\textwidth}
        \centering
        \includegraphics[width=\linewidth]{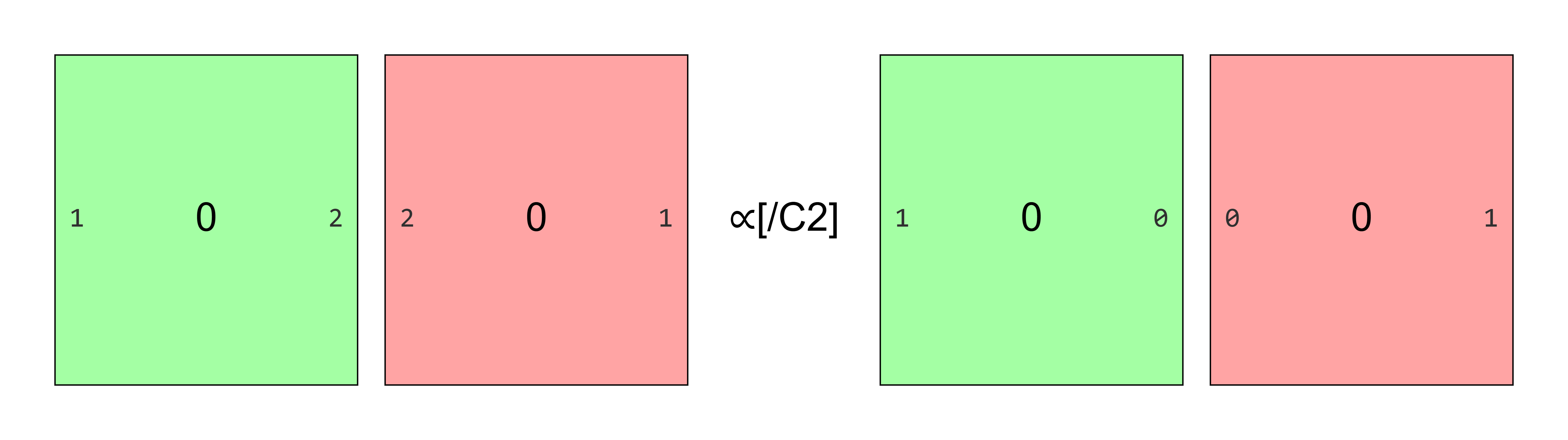}
            \caption{\centering Hopf rule in \vyzx}
        \label{fig:vizx-hopf}
    \end{subfigure}
    \begin{subfigure}{.8\textwidth}
        \centering
        \includegraphics[width=\linewidth]{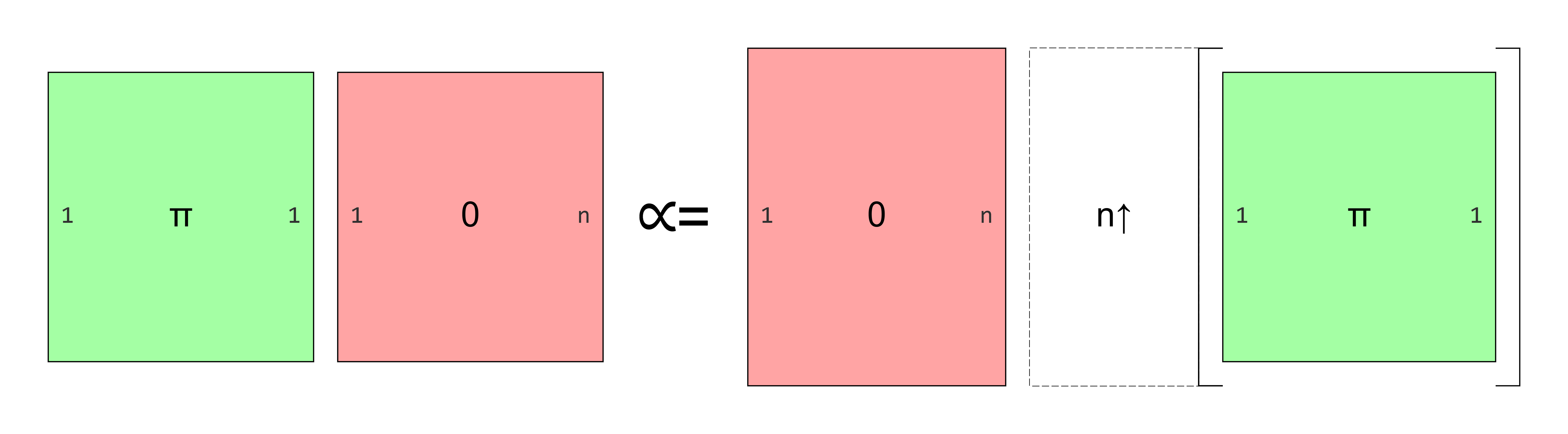}
        \caption{\centering Pi-copy rule in \vyzx, $\alpha \in \R$}
        \label{fig:vizx-pi-copy}
        % Jeandel : K-ish (should be once fixed, anyways)
    \end{subfigure}
    \caption{\centering Directly stateable \vyzx rules}\label{fig:vyzx-direct-rules}
\end{figure}

\begin{figure}[h]
    \centering
    \begin{subfigure}{.8\textwidth}
        \centering
        \includegraphics[width=\linewidth]{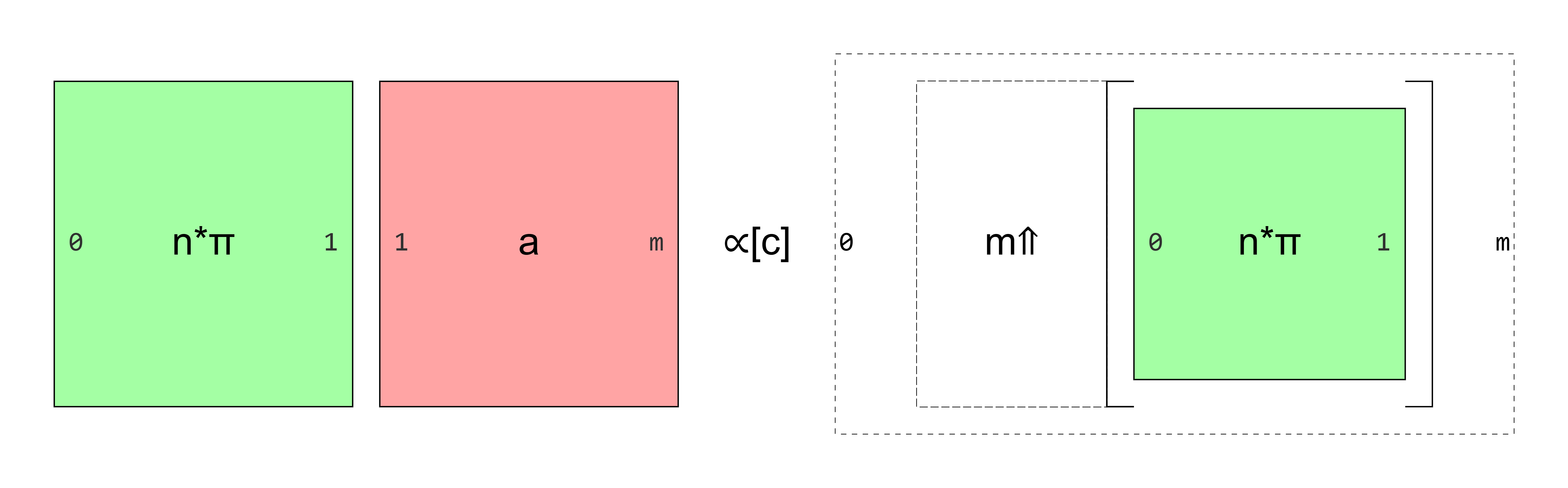}
        \caption{\centering State copy rule in \vyzx, $r \in \mathbb{N}$, $a \in \mathbb{R}$, $c=(1+e^{\pi n} + e^a - e^{a + \pi n}) \cdot 2^{-(m + 1)/2}$}
        \label{fig:vizx-state-copy}
        % Jeandel : subsumes B1
    \end{subfigure}
    \begin{subfigure}{.8\textwidth}
        \centering
        \includegraphics[width=\linewidth]{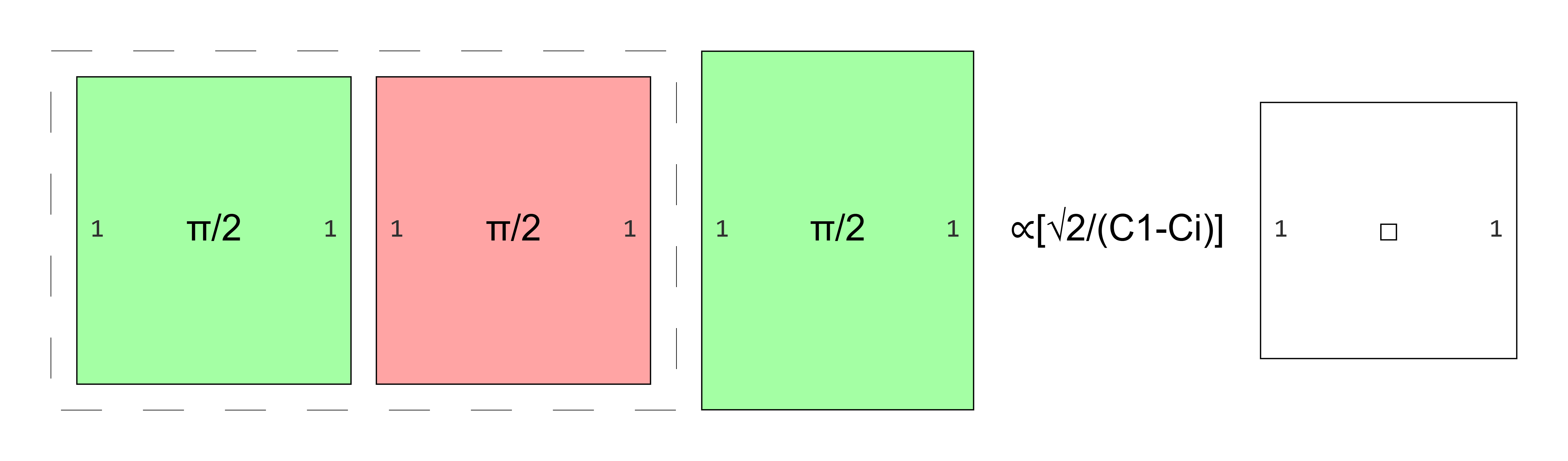}
        \caption{\centering Hadamard decomposition in \vyzx}
        \label{fig:vizx-hadamard-decomposition}
        % Jeandel : kinda close to EU? But not quite
    \end{subfigure}
    \begin{subfigure}{.8\textwidth}
        \centering
        \includegraphics[width=\linewidth]{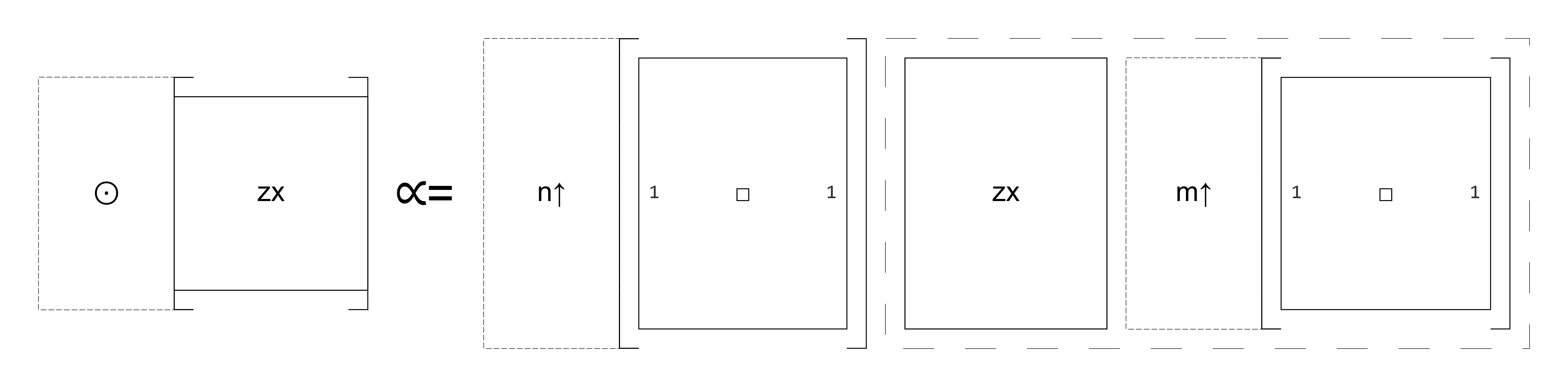}
        \caption{\centering Bi-Hadamard color swapping \vyzx, where \(\odot\) represents a ``color-swapped'' diagram where all red spiders are replaced with green spiders and vice versa}
        \label{fig:vizx-bihadamard-colorswap}
        % Jeandel: subsumes H
    \end{subfigure}
    \begin{subfigure}{.65\textwidth}
        \centering
        \includegraphics[width=\linewidth]{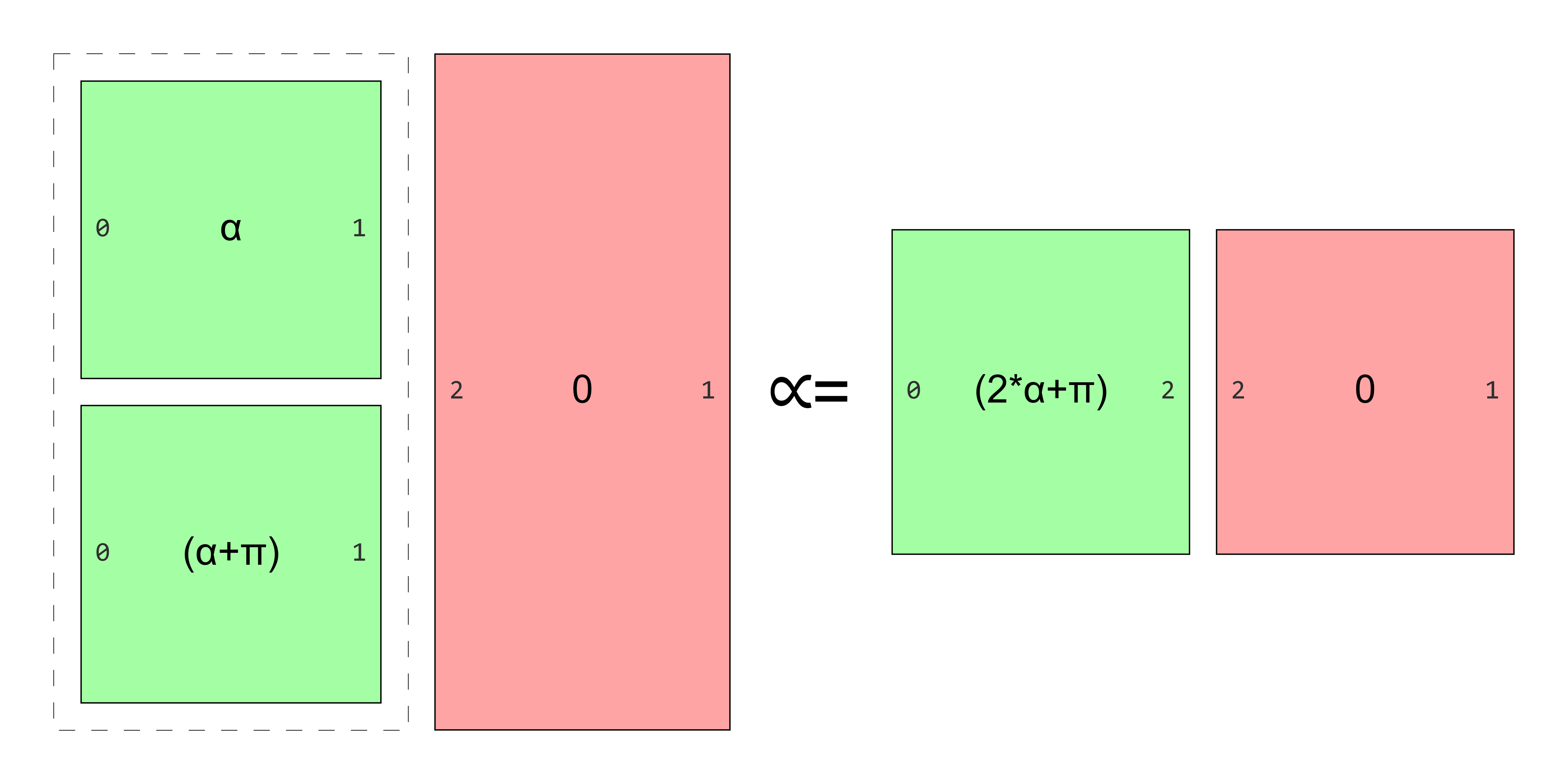}
        \caption{\centering The SUP rule in \vyzx, $\alpha \in \R$}
        \label{fig:vizx-sup-rule}
        % Jeandel: SUP
    \end{subfigure}
    \caption{Additional \vyzx rules. \rnr{Better words?}}
\end{figure}

The primary source for these rules is Jeandel's work~\cite{jeandel2019completeness}, but the rules are slightly modified with inspiration from van de Wetering's survey~\cite{vandewetering2020zxcalculus}.
These rules should be compared to Jeandel's work~\cite[Figure 1]{jeandel2019completeness}, where the main differences are:
\begin{enumerate}
    \item \vyzx uses a more general rule, \Cref{fig:vizx-pi-copy}, instead of rule K,
    \item \vyzx uses a more general rule, \Cref{fig:vizx-state-copy}, instead of rule CP.
\end{enumerate}

\begin{figure}[h]
    \begin{subfigure}{.8\textwidth}
        \centering
        \includegraphics[width=\linewidth]{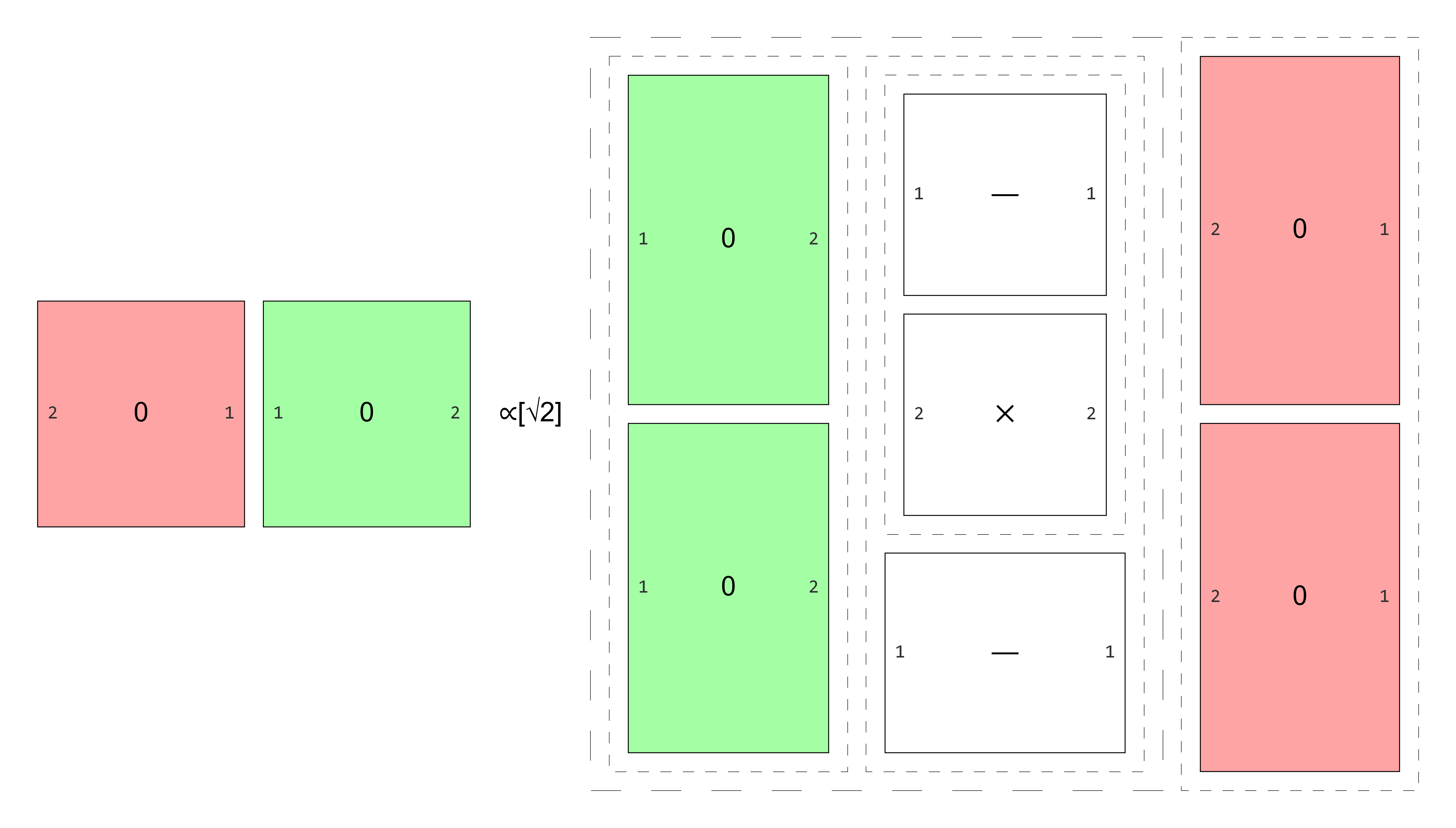}
        % Jeandel: B2
    \end{subfigure}
    \caption{\centering Bialgebra rule in \vyzx}
    \begin{subfigure}{\linewidth}
            \centering
            \includegraphics[width=\linewidth]{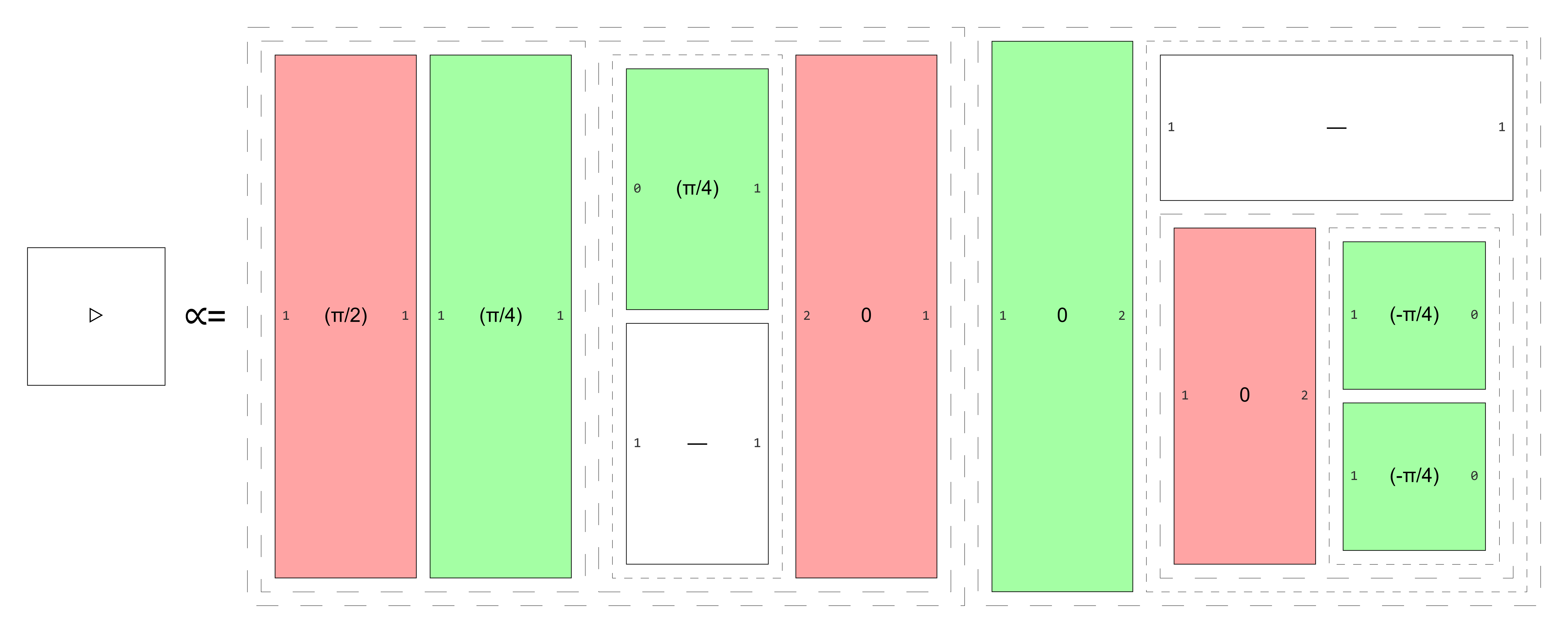}
            \caption{\centering Definition of $\triangleright$}
            \label{fig:vizx-triangle-right}
    \end{subfigure}
    \begin{subfigure}{0.34\linewidth}
            \centering
            \includegraphics[width=\linewidth]{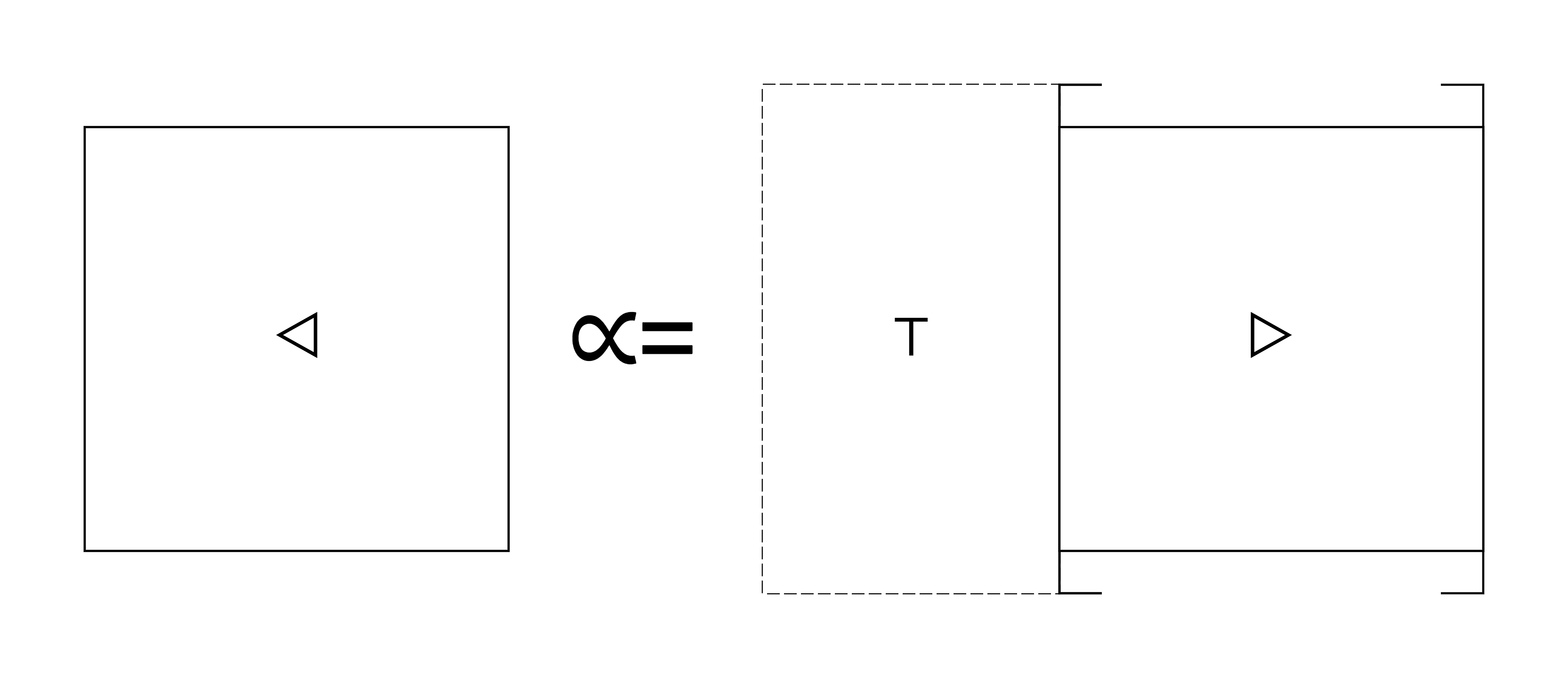}
            \caption{\centering Definition of $\triangleleft$}
            \label{fig:vizx-triangle-left}
    \end{subfigure}
    \begin{subfigure}{0.64\linewidth}
            \centering
            \includegraphics[width=\linewidth]{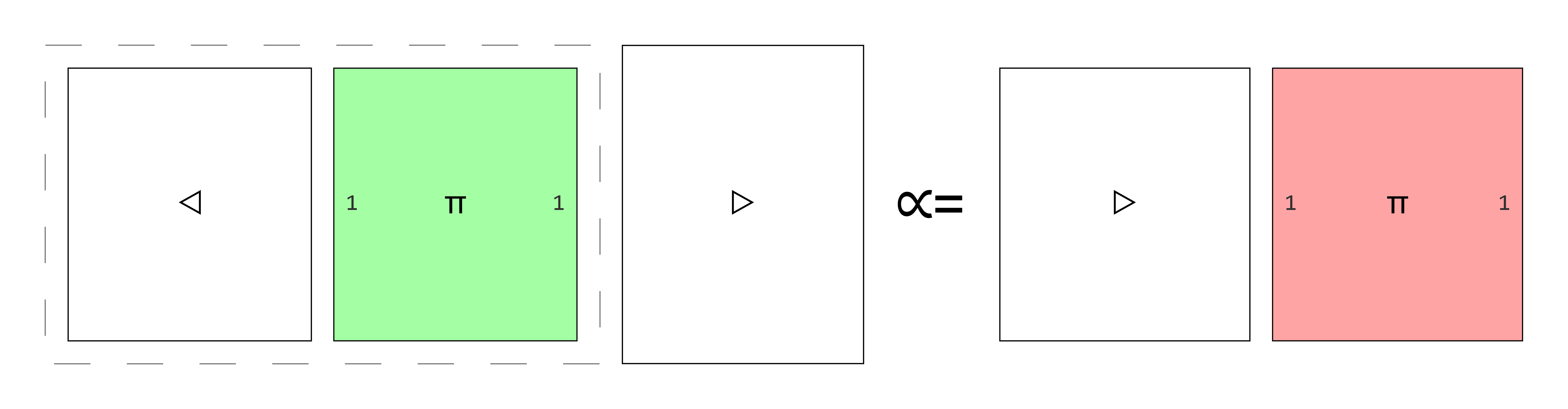}
            \caption{\centering The BW rule}
            \label{fig:vizx-bw-rule}
            % Jeandel: BW
    \end{subfigure}
\caption{\centering The BW rule in \vyzx}
\end{figure}

These differences do not impact the universality of the \cite{jeandel2019completeness}'s system as \vyzx contains more general rules.

\begin{figure}[h]
    \centering
    \includegraphics[width=\linewidth]{figures/vizx_C_1.png}
    \scalebox{2}{$\propto=$}
    \includegraphics[width=\linewidth]{figures/vizx_C_2.png}
    \caption{\centering The C rule in \vyzx}
    \label{fig:vizx-c-rule}
    % Jeandel: C
\end{figure}

\begin{figure}[h]
    \centering
    \includegraphics[width=0.95\linewidth]{figures/vizx_N.png}
    \caption{\centering The N rule in \vyzx}
    \label{fig:vizx-n-rule}
    % Jeandel: N
\end{figure}

\begin{figure}[h]
    \centering
    \includegraphics[width=0.5\linewidth]{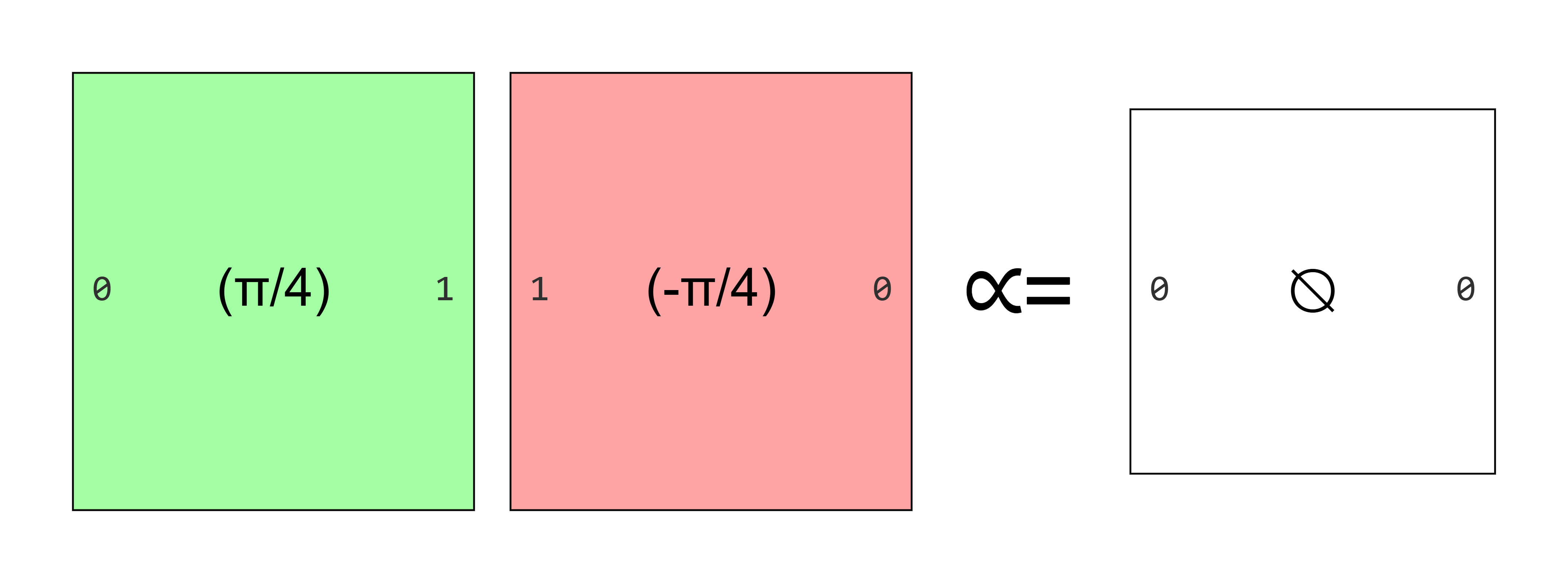}
    \caption{\centering The E rule in \vyzx}
    \label{fig:vizx-e-rule}
    % Jeandel: E
\end{figure}

\end{document}